# Futures with Digital Minds:
# Expert Forecasts in 2025

Lucius Caviola[a,b] & Bradford Saad[b,1]
[a] University of Oxford
[b] University of Cambridge

A web-based version of this report is available at: https://digitalminds.report/forecasting-2025

This report presents findings from an expert survey on digital minds takeoff scenarios. The survey was conducted in early 2025 with 67 experts in digital minds research, AI research, philosophy, forecasting, and related fields. Participants provided probabilistic forecasts and qualitative reasoning on the development, characteristics, and societal impact of *digital minds*, that is, computer systems capable of subjective experience. Experts assigned high probability to digital minds being possible in principle (median 90%) and being created this century (65% by 2100), with a non-negligible probability of emergence by 2030 (20%). Many anticipated rapid growth in digital mind welfare capacity, with collective welfare capacity potentially matching that of billions of humans within a decade after the creation of the first digital mind. Participants also expected widespread claims from digital minds regarding their consciousness and rights, and predicted substantial societal disagreement over their existence and moral interests. Views diverged on whether digital mind welfare will be net positive or negative. These findings provide evidence that bears on the extent to which preparing the world for the potential arrival of digital minds should be a priority across domains such as research and governance. However, these findings should be interpreted cautiously in light of the potential for systematic overrepresentation of experts who deem digital minds particularly likely or important.

---

[1] Authors made equal contributions to designing the survey and writing the report. LC coordinated recruitment and analysis.

# Summary

This survey gathered insights from 67 participants with relevant expertise to explore the future of digital minds—defined here as computer systems capable of subjective experience. The main findings are summarized below.

- **Possibility:** Digital minds were generally judged to be possible in principle, with a median probability estimate of 90%.
- **Creation:** The probability that digital minds will at some point be created was judged to be 73%. Only 12% of participants thought it was less than 50% probable.
- **Timing:** The median estimated probabilities of digital minds being created by a given year were 4.5% by 2025, 20% by 2030, 40% by 2040, 50% by 2050, and 65% by 2100. Reasons influencing these timelines included technological trends (e.g., AI scaling), regulatory developments, and the risk of preemptive catastrophe.
- **Timing relative to AGI:** 89% judged digital minds unlikely to arrive before AGI. However, these estimates may partly reflect confidence in short AGI timelines and greater uncertainty about digital mind timelines, rather than strong confidence that digital minds must necessarily follow AGI.
- **Speed:** Many expected rapid growth in digital minds' collective welfare capacity after the first is created, with median estimates suggesting their collective welfare capacity could equal that of one billion humans within five years[2].
- **Types:** Brain simulations were viewed as most likely to support the capacity for subjective experience in principle, but machine learning–based AI systems were judged most likely to be the first digital minds.
- **Social function:** Participants disagreed on what proportion of digital minds will be designed specifically for social interaction with humans. Most participants expected only a small minority (median 10%), while about one-third anticipated that the majority will serve such social roles.[2]
- **Location:** Digital mind production in the first decade is expected to be concentrated in a few powerful countries, especially the USA and China, though some mentioned production in oceans, space, or outsourced locations[2].
- **Producers:** Companies were widely expected to lead digital mind development in the first decade, outpacing governments, universities, and others—mirroring current AI research and commercialization trends[2].
- **Deliberate creation:** Opinions were split on whether digital minds would be created deliberately or emerge as byproducts of other developments[2].
- **Net welfare:** There was no consensus on whether digital mind welfare will be net positive, neutral, or negative, though the overall leaning was slightly positive. Reasons for positive welfare included deliberate design for well-being and self-modification capabilities. Reasons for negative welfare included neglect of welfare in design, mistreatment, and suffering during training[2].

---

[2] Participants were instructed to assume the first digital mind is a machine-learning-based AI system created no later than 2040.

- **Pre-deployment:** Most thought only a small share of digital mind welfare willoccur pre-deployment, though some flagged intense training-related suffering as potentially dominant among valenced experiences[2].
- **Super-beneficiaries:** Most were skeptical that super-beneficiary digital minds (with >1,000× human welfare capacity) will contribute much to total welfare within ten years, though some saw them as potentially important[2].
- **Experience-independent welfare:** Most doubted that computer systems could have welfare without subjective experience. However, arguments for such possibilities may be underappreciated, as the existing body of literature on this question was not well-represented in responses[2].
- **Claims:** Most expected that a sizable number of digital minds will consistently and proactively claim to have subjective experience or be deserving of legal protections or further civil rights[2].
- **Inaccurate self-attribution or self-denial:** Only a minority of AI systems without subjective experience were expected to systematically and falsely claim to have it. Similarly, only a minority of digital minds were expected to systematically and falsely deny possessing subjective experiences. However, given potentially large populations, even a small percentage of such systems could result in a large number of systems that make systematically inaccurate claims about whether they have subjective experience[2].
- **Public belief:** A substantial portion of the public was expected to believe in the existence of digital minds within ten years of their creation. However, responses reflected significant uncertainty and anticipated societal division, with beliefs likely varying across individuals and groups[2].
- **Welfare capacity estimation:** Participants broadly expected the public to significantly underestimate the welfare capacity of digital minds, raising concerns that their moral significance may be underappreciated[2].
- **Protections and rights:** Public support for protections was expected to vary by level: moderate support for basic harm protections may emerge within a decade, while support for full civil rights was expected remain limited[2].
- **Political attention:** Participants were divided on whether digital mind rights will become a major political issue within a decade of the first digital mind[2].
- **Moratorium:** Views were mixed—leaning slightly positive—on whether a moratorium on creating digital minds until 2040 would be good or bad. Arguments for a moratorium included buying time to prepare and indirect AI safety benefits. Arguments against included delayed integration of safety measures and risks of later creation in higher-stakes contexts.
- **Interactions with AI safety:** There was little convergence among participants on whether AI safety and digital mind welfare efforts will align or conflict. Mentioned potential synergies included alignment reducing the need for coercive control and shared technical tools (e.g., interpretability methods). Potential conflicts included safety measures such as monitoring and shutdown protocols that could harm digital minds, welfare protections limiting the ability to control AI behavior, and competition for scarce funding, talent, and regulatory attention.

We hope this survey will help inform future research, guide policy discussions, and lay a foundation for deeper exploration of this emerging field. At the same time, the quantitative forecasts should be interpreted with caution. Our sampling—drawn from digital mind research groups, academic networks, and AI safety and policy organizations—ensured strong domain expertise but likely overrepresented those who view digital minds as especially important. Given the speculative nature of this topic, we place particular emphasis on the qualitative responses in this report, which provide valuable insights even where the associated quantitative results warrant limited confidence. To complement our findings, we suggest also consulting [Dreksler, Caviola et al. (2025)](), which surveyed a larger and more representative sample of AI researchers on a narrower set of questions about subjective experience in AI.

# Introduction

This report presents findings from our expert survey on *digital minds*, that is, computer-based systems with a capacity for subjective experience. We asked experts to make quantitative predictions and offer qualitative considerations concerning when digital minds will be created, what characteristics they will have, and their societal impact. While considerable research attention has been directed toward forecasting the development of artificial intelligence (AI), little research effort has gone toward forecasting the emergence of digital minds and the implications of their creation.

What the future may hold for digital minds is arguably an important and urgent matter, though the topic is also fraught with uncertainties and difficult questions that span various disciplines. In light of rapid progress in AI development, some philosophers and scientists have argued that the creation of digital minds is a realistic near-term possibility (e.g. Butlin et al., 2023; Bradley & Saad forthcoming*a*, forthcoming*b*; Chalmers, 2023; Dung, ms; Sebo & Long, 2025; Long et al., 2024). Some researchers have also argued that the creation of digital minds—or systems that people *think* are digital minds—could have profound moral implications such as societal unrest and the mistreatment of digital minds on a large scale (Caviola, 2025; Caviola et al., 2025; Fernandez, 2024; Schwitzgebel, 2023*a*; 2023*b*; Birch, 2024).[3] Regarding social unrest, one concern is that there will be deep and recalcitrant disagreements among humans about which systems are digital minds and how to treat them. A related worry is that extending moral consideration to near-term AI systems would come at the expense of humans' interests (Bryson et al. 2017), especially if such concern were inappropriately extended to AI systems that lacked consciousness and moral significance (Bostrom, 2014). There is also the concern that extending such consideration would risk letting responsible humans off the hook for inflicting algorithmic harms (Véliz, 2021). Regarding mistreatment, one concern is that the status quo of treating AI systems as mere tools could—if we create digital minds—cause digital minds to suffer or harm them in ways comparable to rights violations that result from treating human persons as mere tools or submissive servants (Bales, 2025; Bostrom & Shulman, 2022; Dung, 2023; ms; Bradley & Saad, 2022, forthcoming*a*, forthcoming*b*; Gloor, 2016; Schwitzgebel & Garza, 2015; 2020; Tomasik, 2017). Others have argued that advanced preparation for the arrival of digital minds should begin now (Long et al., 2024; Sebo & Long, 2025) and that regulations concerning sentient artificial systems should be anticipatory rather than reactive (Birch, 2024). In addition, at least one frontier AI company (Anthropic) has begun to address AI system welfare, by conducting welfare assessments and reporting on them in a system model card and by starting a research program to study model welfare (Anthropic 2025*a*; Anthropic 2025*b*). Although these perspectives are fairly niche within their respective spaces, they to some extent resonate with public opinion (Anthis et al., 2025; Dreksler, Caviola et al. 2025).[4]

---

[3] For a thorough literature review of work concerning the extension of moral consideration to artificial entities, see Harris & Anthis (2021).

[4] Dreksler, Caviola et al. (2025) found that AI researchers and members of the US public reported strikingly similar expectations about when AI systems might develop subjective experience, with the median estimate of both groups identifying 2050 as the year by which there is a 50% probability that AI systems with subjective experience will be created.

At the same time, within the broader landscape of AI development, the topic of digital minds is almost entirely absent from the mainstream discourse. To our knowledge: No government or international governmental organization has devised a plan for addressing the potential emergence of digital minds.[5] With the noted exception of Anthropic, no frontier AI company has publicly acknowledged the potential for digital minds to arise on the current trajectory of AI development. Digital minds and their potential ramifications are rarely mentioned in the space of AI governance and policy research.[6] Although digital minds are sometimes alluded to in passing in research on frontier AI safety/security (e.g. Hendrycks, 2025, p. 338-339), such research rarely acknowledges the potential for digital minds or interactions between safety and security measures and risks to digital minds—see Bradley & Saad forthcoming*a*, Long et al. 2025, Moret 2025, Schwitzgebel 2025, and Sterri & Skjelbredfor 2024 discussion of these interactions by philosophers. Others are skeptical of the near-term prospects for digital minds and have argued that more tangible—though no less urgent—issues in AI development should be prioritized (Birhane & van Dijk 2020; Crane 2021).

There are thus a number of reasons to investigate the potential emergence of digital minds. First, the level of evidence for such emergence bears on the extent to which companies and governments should allocate attention and resources to the topic. Those who favor evidence-based prioritization should thus welcome the gathering of additional evidence on this topic, regardless of whether they currently favor the status quo or a pivot advocated by some toward taking the near-term prospects of digital minds seriously. Second, from the perspective of those who are uncertain about whether the potential emergence of digital minds is an important and urgent topic, there is reason to gather more evidence about the topic now: doing so helps ensure that we can act in time in the event that urgent action is needed. Third, given the noted divergence in perspectives and that the topic is little explored, it is reasonable to hope that gathering further evidence may serve to inform and resolve disagreements on the topic. Our survey is motivated by these reasons along with the belief that aggregating input from experts in relevant domains will provide an unusually comprehensive collection of relevant considerations as well as a stepping stone for more-systematic investigations of this topic.

## Research Aims

The primary objective of this survey is to systematically collect and analyze expert predictions and perspectives on:

- The likelihood of digital mind creation
- Timelines for digital mind creation
- The probable characteristics of early digital minds (e.g., types, distribution, welfare capacity)
- How digital minds might interact with and be perceived by society

---

[5] Digital minds are also conspicuously absent from major international governmental AI declarations, legislation, and reports such as the [Bletchley Declaration](), [EU Artificial Intelligence Act](), and the UN AI Advisory Council report ["Governing AI for Humanity"]().

[6] There are exceptions. For example, see Maas 2023.

- The potential welfare implications of digital minds.

Rather than aiming to represent the full range of expert opinion across all fields, our goal is to elicit informed considerations and judgments from researchers who have thought deeply about digital minds or closely related issues. By aggregating their views, we aim to surface key areas of agreement and disagreement within this group, and to offer a foundation for more systematic forecasting and critical engagement. In this sense, the survey is best seen as a structured exploration—intended to clarify key issues, generate concrete predictions, and lay the groundwork for broader, more representative future research.

Although our survey relies on speculative expert judgment, we believe it offers significant contributions. First, the quantitative forecasts—while uncertain—reveal informative patterns of convergence and divergence that help map the current expert landscape and inform prioritization decisions about research in this area. Second, the qualitative responses are especially insightful: they reveal the reasoning behind forecasts, highlight key considerations, and collectively surface a much broader and richer set of ideas than any individual might be expected to generate. These insights may serve as a basis for refining future quantitative estimates. Finally, we view this work as a first step in building a more robust research program on digital minds forecasting.

## Methods and Sample

Our survey methodology combined probabilistic judgments with qualitative reasoning. Given the speculative nature of the topic—long-term predictions about philosophically and empirically uncertain matters with limited historical precedent—we anticipated that probabilistic judgments would need to be interpreted cautiously without assigning too much weight to exact quantitative estimates. It was partly for this reason that we also asked for qualitative considerations. Conversely, we sought probabilistic judgments partly to elicit richer qualitative responses by encouraging participants to engage in careful reflection.

We recognize that very few researchers currently specialize in digital minds. Accordingly, we recruited a broader group of participants with a reasonable understanding of what digital minds are, why they might matter, and how their domain expertise could inform the relevant questions. In particular, we invited individuals from a range of fields, including digital minds research, AI research and policy, forecasting, and philosophy, cognitive science, and social science. For simplicity, we refer to all of them as experts, though not all were specialists in digital minds specifically.

Participants were identified through digital minds research groups, academic networks, AI research and policy organizations, and professional contacts, along with additional outreach to authors of published work on consciousness and AI welfare. While we sought a broad range of perspectives, our sample likely overrepresents individuals who view digital minds as plausible or important—a potential source of bias discussed in the limitations section.

We ultimately recruited 67 participants: 18 digital minds researchers, 18 AI research and policy experts, 7 professional forecasters, and 20 philosophers, cognitive scientists, and other academic specialists. Four additional participants selected "no specific expertise," though their responses showed thoughtful reasoning and strong understanding. The figure below shows a more detailed breakdown of participants' self-assessed expertise across multiple domains, highlighting the sample's diversity. Participants worked across sectors—27 in academia, 22 in research organizations outside academia, 8 in non-research non-profit organizations, 2 in industry, and 2 in other sectors. Geographically, 27 were based in the United States, 15 in the United Kingdom, 8 in various other countries, with 17 not disclosing their location. The study was conducted in February and March 2025.

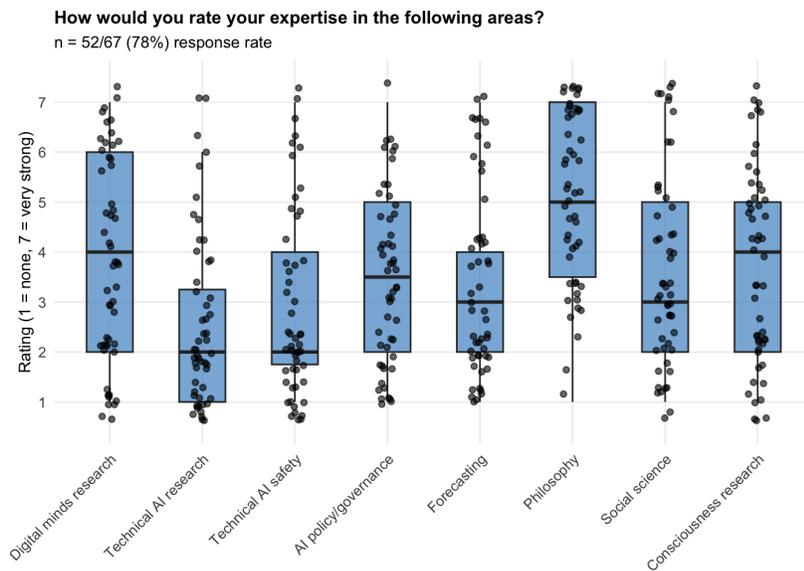

## Key Assumptions

For consistency across responses, we stipulated some key assumptions for all participants:

- **Definition of Digital Mind**: In our survey, a "digital mind" was defined as any computer system with a capacity for subjective experience (phenomenal consciousness). This definition is broad in that it applies not just to AI systems with the capacity for subjective experience but also to any other computer systems (e.g. brain simulations) with that capacity. The definition is narrow in that it excludes any computer systems that lack the capacity for subjective experience, even if they have other (morally significant) mental capacities. We restricted the adopted notion of digital minds in this way for methodological reasons: to minimize ambiguity, ensure participants were reasoning about the same class of entities, and facilitate clearer interpretation of results.
- **Welfare Capacity Threshold**: Participants were asked to consider only digital minds with a welfare capacity at least roughly as high as that of a typical human. We excluded systems with significantly lower welfare capacity to avoid complications arising from diverging beliefs about large populations of artificial minds individually exhibiting low but non-zero welfare capacity.

- **Earth-originating**: Participants were instructed to consider only digital minds created by our Earth-originating civilization. We thus excluded from consideration any digital minds created by non-Earth-originating entities such as other civilizations in our universe and simulators of our universe. This served to set aside certain speculative hypotheses about digital minds that could substantially influence estimates and to have participants focus on candidate populations of digital minds whose evolutions are easier to forecast and which are more obviously decision-relevant to humans.[7]
- **Stipulated Scenario**: For some questions, participants were asked to assume a specific scenario: that the first digital mind will be a machine-learning-based AI system created in or before the year 2040. This assumption served as a conditional baseline: it enabled participants to make concrete forecasts based on that assumption, regardless of how plausible they found that assumption. It also allowed for comparability across responses by anchoring predictions to a shared reference point. However, this assumption was not applied to all questions—for example, it was not applied to questions about whether digital minds are possible in principle or the likelihood that they will ever be created. (When discussing each question in a section below, we will discuss whether this assumption was applied to that question.)

When designing the survey, we considered whether or not to include an assumption concerning how to individuate digital minds. We decided not to include such an assumption for three reasons. First, we struggled to find a suitable individuation criterion. Second, we judged that responses to most questions probably would not be highly sensitive to the choice of individuation criterion. Third, participants could, as some did, note such sensitivity in their free responses.

## Report Structure

This report presents our findings in a logical sequence, beginning with expert predictions on the feasibility and timelines of digital minds, followed by expert predictions concerning specific scenarios and their implications. Please note that the order of sections in the report does not entirely reflect the order in which participants encountered them. The exact sequence and wording presented to participants can be found in the supplementary materials. Throughout the report, we present both quantitative results and experts' qualitative reasoning. **For the qualitative responses, we group related considerations thematically, paraphrase them in our own words, and order them roughly by our judgments of significance and prevalence. By default, "considerations" refers to views expressed by at least one participant; they are not necessarily endorsed by all participants or by the authors.** We conclude with a discussion of implications of survey results and suggestions for further research.

---

[7] Although such hypotheses are ignored by default in many contexts, it was not safe to assume that they would be ignored in this context. One reason for this is that the topic of digital minds raises to salience arguments for the hypothesis that we are living in a simulation (Bostrom 2003, Chalmers 2022, Greene 2020, Saad 2023, Thomas 2024). Another is that some of our questions concern usually large-scale outcomes (e.g. the time by which digital minds will have collective welfare capacity matching that of a trillion humans), which would have invited consideration of non-Earth-originating digital minds had we not placed them out of scope.

# Starting Point

## Possibility

The first question asked participants how likely it is that digital minds—computer systems capable of subjective experience (phenomenal consciousness)—are possible in principle. Participants entered a probability number between 0% and 100%.

The results show that most participants favored the hypothesis that digital minds are possible in principle. The median estimate was a 90% probability that digital minds are possible in principle, the mean was 86.4%, and the most common (i.e., modal) response was 100%.

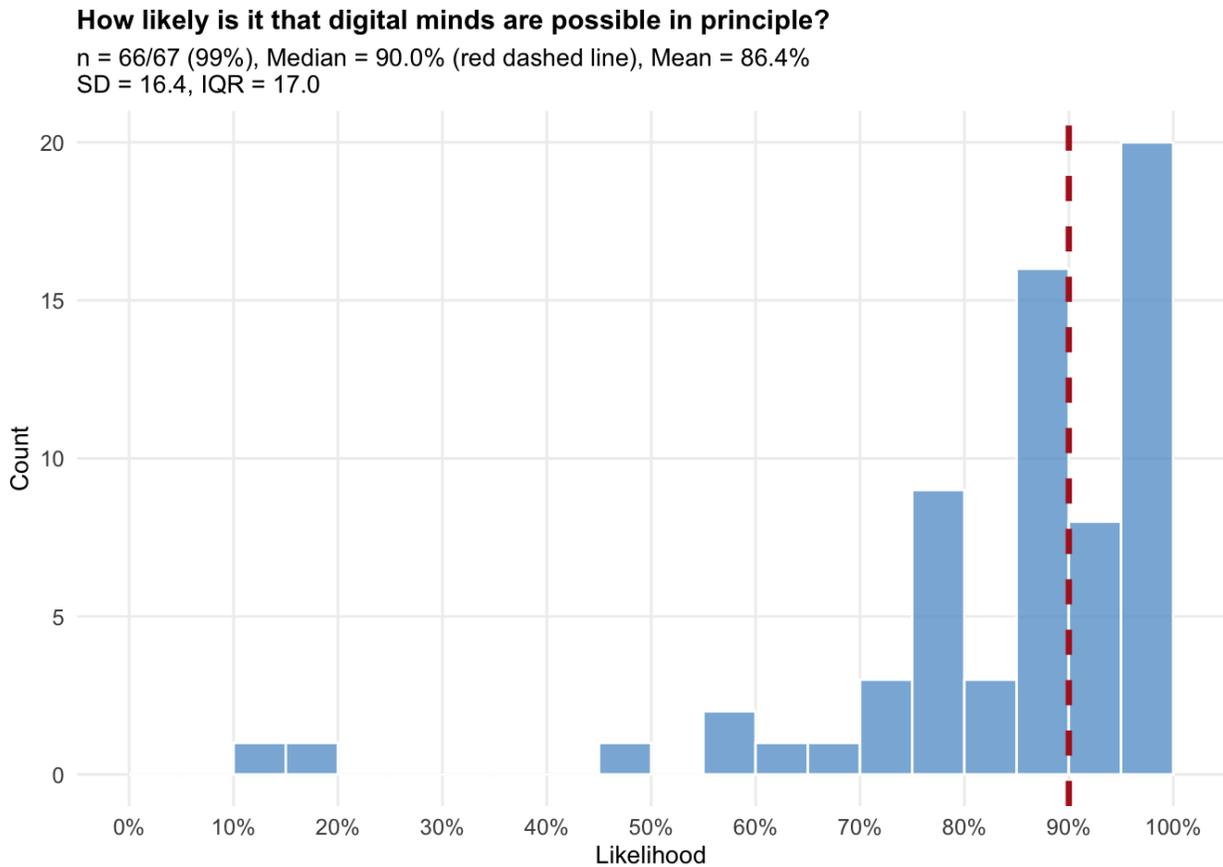

Text responses to this question echoed common themes in the literature. Broadly, participants offered tentative support for the in-principle possibility of digital minds and cited little evidence against it. More specifically, many pointed to a link between subjective experience and functional organization as a key reason for thinking digital minds are possible—while skeptics questioned whether such a link holds.[8] Below is a more detailed breakdown of responses to this question.

Considerations supporting the possibility of digital minds

Ways digital minds could be created:
- **Whole brain emulations**: Creating systems that emulate functional features of conscious brains.[9]
- **Biology-mirroring optimization:** Evolutionary forces shaped human and animal minds. Artificial systems can be subjected to optimization pressures that are analogous to those forces. For example, artificial systems will be developed in complex environments with rewards that are delayed and infrequent. Producing artificial systems under such pressures may yield digital minds.

Sources of the capacity for subjective experience in artificial systems
- **Functional capacities:** Realizing in silicon functional capacities that underlie consciousness in biological minds.
- **Suitable representations**: E.g., analog representations that occupy a functional role that's tied to homeostasis.

Other considerations that support the possibility of digital minds
- **Functions as crucial and substrate-independent**: What seems to be special about the brain is how it functions, not what it's made of; and its relevant functions could be realized artificially.
- **Gradual replacement argument**: Conscious biological systems could be gradually transformed into artificial systems while preserving functional organization; if subjective experience were tied to biology, subjective experience would implausibly dissociate from cognition in intermediate systems; so, subjective experience isn't tied to biology.[10]
- **Intelligence as indicator**: Intelligence beyond a certain level suffices by default for consciousness.
- **Anti-parochialism**: The hypothesis that only non-digital substrates can realize consciousness is implausibly parochial.

Absence of countervailing considerations
- **Lack of evidence for a biological requirement**: There is no evidence that biology is required for subjective experience.
- **Lack of evidence for barriers to digital minds**: There are no obvious barriers to creating artificial systems with a capacity for subjective experience.

Considerations against the possibility of digital minds
- **Implausibility of computational functionalism**: Computational functionalism holds that some sort of computational state is necessary and sufficient for subjective experience. This view is probably false.

---

[8] For advocates of such a link, see, for example, Chalmers 1996 and Butlin et al. 2023. For challenges to such a link, see, for example, Block 1980, Hill 1991, and Seth 2024.

[9] A whole brain emulation is standardly understood as an ambitious form of brain simulation on which a computer system that mimics the internal causal dynamics of a brain at a relevant level of grain (Sandberg & Bostrom, 2008).

[10] For a classic presentation of gradual replacement arguments, see Chalmers 1996: ch. 7.

- **Possibility of biological substrate requirement**: Having subjective experience might require having a biological substrate.
- **Future theories**: In addition to theories on which a biological substrate is required for subjective experience, there might be unimagined/unknown theories that deem digital minds impossible.

### Other factors

- **Probability that subjective experience can only be realized in a narrow class of substrates:** Substrates are what a system is made of (e.g. biological tissue). The narrower the class of substrates that can realize subjective experience, the less likely it is that that class includes a computer substrate. So, if subjective experience can only be realized in a narrow class of substrates, that tells against the possibility of digital minds.
- **The existence of a relevant difference between carbon and silicon**: Carbon is critical to enabling life. Silicon plays a critical role in computers. Whether digital minds are possible may turn on whether there is an experience-relevant difference between carbon and silicon.
- **Modal ambiguity:** The question is ambiguous between different kinds of possibility (e.g. logical possibility vs. possible according to the laws of nature in our world).
- **Question defectiveness**: There is no fact of the matter about whether digital minds are possible or our way of conceptualizing this issue is defective.

## Creation

The second question asked participants how likely it is that digital minds will ever be created. Whereas the first question concerned possibility in principle, here we were interested in participants' empirical predictions about the creation of digital minds in practice.

As expected, estimates were somewhat lower, since creation requires more than just possibility in principle. Nonetheless, most participants saw creation as likely: the median estimate was 73%, and the mean was 70.7%. Only eight out of 66 participants thought it was less than 50% likely that digital minds will ever be created.

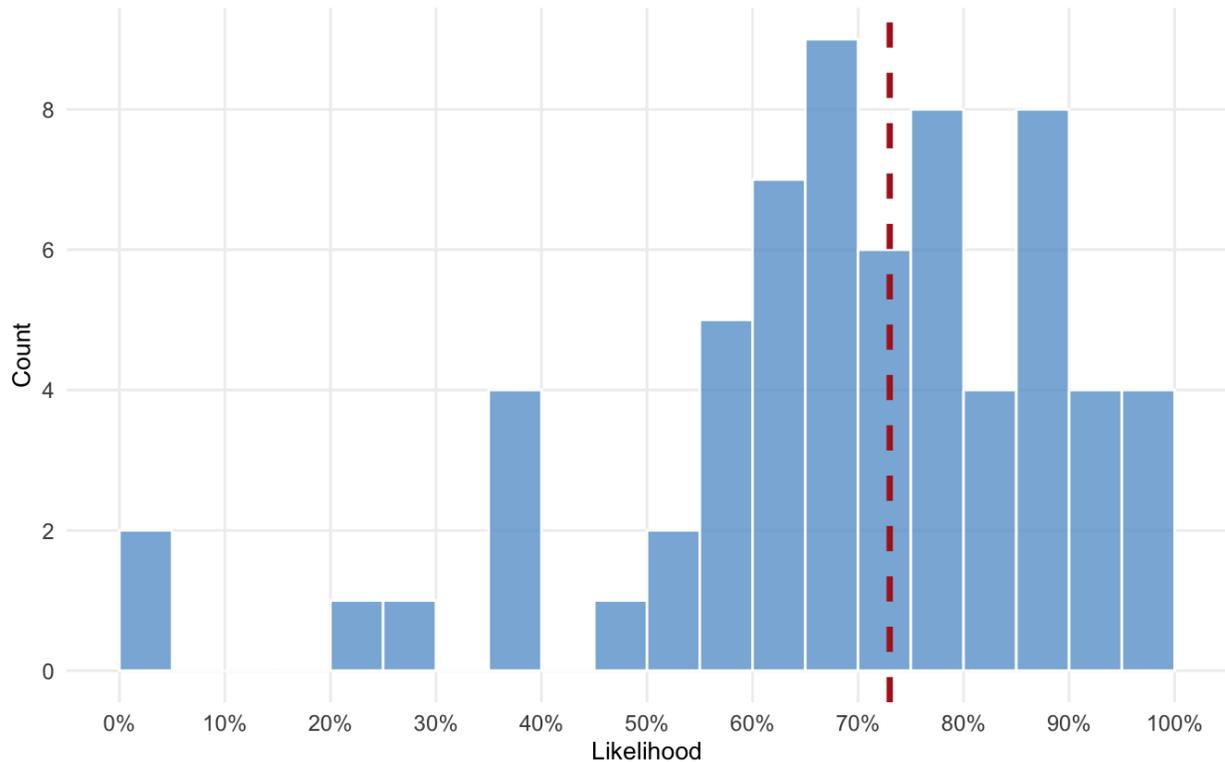

Considerations that suggest a digital mind will be created

- **Future technological ability**: Civilization will probably survive long enough to have much more knowledge and much more compute, and conditional on such survival it's likely that either someone builds a whole brain emulation or else some other kind of digital mind is created.
- **Demand for digital minds**: There will be demand for digital minds (companions, uploads, digital mind children, curiosity, experimentation, just for the sake of having created a digital mind).
- **Differential demand**: The desire to create digital minds will be stronger than the desire to not create them.
- **Subjective experience as an AI aspiration**: Even in an AI-disempowerment scenario, AI systems may try to become capable of subjective experiences out of curiosity or decide to create digital minds for some other reason. Creating digital minds will be high on a post-singularity civilization's to-do list.[11]
- **Digital minds as byproducts**: Digital minds will emerge as a side effect of something else (e.g., agency, intelligence, scaling AI models, training on delayed and sparse rewards, developing AGI, using optimization pressure to develop a complex intelligent system).

---

[11] 'Singularity' is standardly used to refer to a period in which AI systems drive an explosion in intelligent capabilities via recursive self-improvement (Bostrom, 2014; Chalmers, 2010).

- **Digital minds are not difficult:** Building systems that satisfy the main indicators of subjective experience from the literature does not seem particularly hard.

Considerations that suggest no digital mind will be created
- **Digital minds are difficult:** Creating digital minds is likely difficult.
- **Difficult or off path:** Creating digital minds may be difficult or off path from AGI development.
- **Preemptive existential catastrophe**: An existential catastrophe may prevent us from creating digital minds.
- **Technological setback**: A major technological setback may prevent us from creating digital minds.
- **Regulation** either general or targeted regulation may slow or prevent progress toward digital minds.
- **Deliberate decision**: We may decide not to create digital minds, e.g., for ethical reasons, because we don't want to incur obligations to them.
- **No incentive to create**: ASI (artificial super intelligence) may have no reason to create digital minds. Or belief in phenomenal consciousness or its importance may be a passing phase.
- **Ontogenetic and structural divergence:** The cognitive capacities of artificial systems are being developed in an order that differs from those in which biological systems develop their capacities. This may result in artificial systems missing structures that are crucial for consciousness.

## Timing

When might digital minds be created? Participants were asked to estimate how likely it is that digital minds will be created by specific points in time: 2025, 2030, 2040, 2050, and 2100. For each year, participants were asked to provide their best-guess probability (specifically, the median of their subjective distribution).

The responses indicate that participants see the creation of digital minds as unlikely in the short term but increasingly plausible over the coming decades. On average (median responses), participants assigned the following probability:
- 4.5% to digital minds being created by 2025 (mean = 8.3%)
- 20% to digital minds being created by 2030 (mean = 25.5%)
- 40% to digital minds being created by 2040 (mean = 39.8%)
- 50% to digital minds being created by 2050 (mean = 48.8%)
- 65% to digital minds being created by 2100 (mean = 62.6%)

These results suggest that there is a non-negligible probability that digital minds have already been created, that the probability of digital minds being created will substantially rise during the next 5-10 years, and that digital minds will probably emerge during this century.

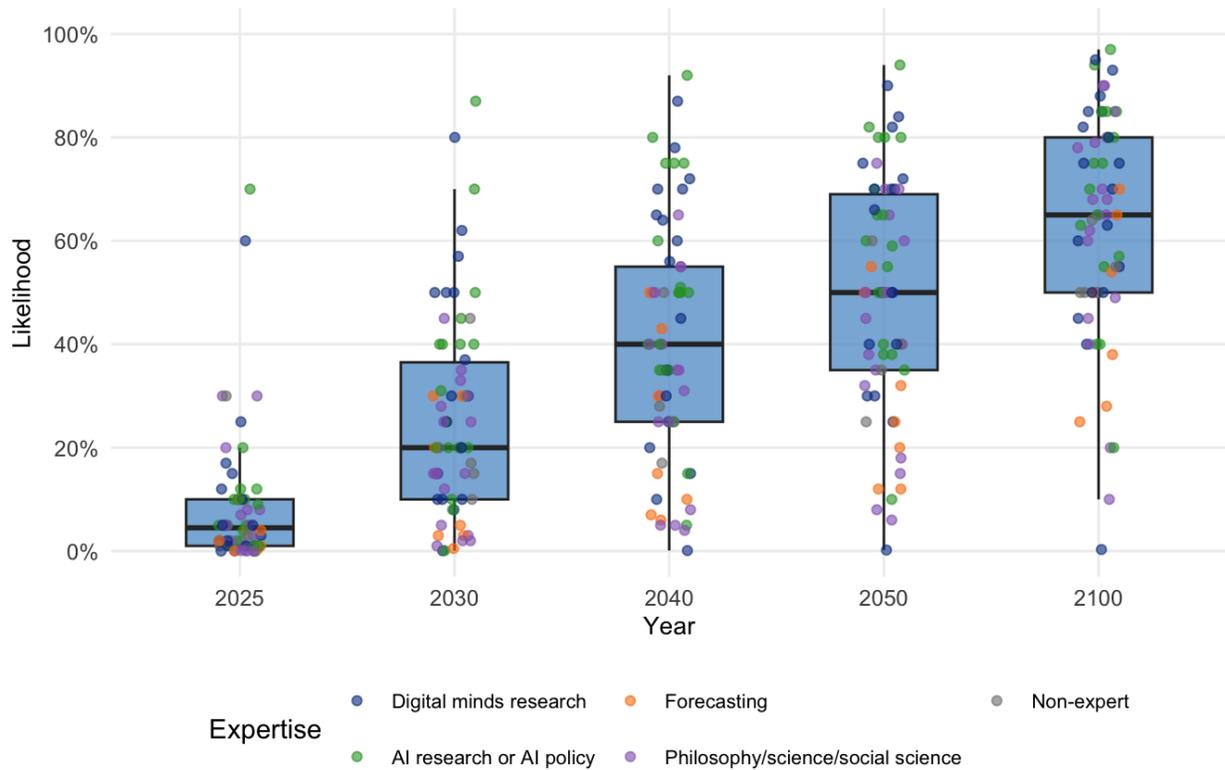

### Enablers

Text responses indicated that digital minds timeline estimates were informed by the following **potential enablers**:
- AIs gaining new features
- Experimentation with different AI architectures
- Technological progress
- Improved knowledge of how to create digital minds

### Barriers

Responses also addressed **potential barriers and speed bumps**:[12]
- By far the most common potential barrier noted in text responses was that of a **preemptive catastrophe**, that is, a catastrophe that prevents the creation of digital minds. Preemptive catastrophes could take the form of existential catastrophes, global catastrophes, human extinction, or human loss of control. Because the cumulative risk of a preemptive catastrophe increases with time, taking into account preemptive catastrophic risk should drive down probability estimates of digital minds arriving by a given year. For the same reason, taking into account preemptive catastrophic risks should reduce estimates for later years by a greater amount. The fact that a significant number of participants reported taking into account preemptive catastrophic risks suggests that their estimates for digital minds arriving would be higher conditional on the non-occurrence of a preemptive catastrophe.
- **Deciding not to create digital minds.**
- **Regulation** could impede AI development in general or digital mind creation in particular.
- **Energy acquisition:** After AGI arrives, it will take time to harness energy to do everything we want.
- **Barrier timelines**: Many barriers to creating digital minds will be overcome soon. So, the probability of digital minds being created by a given year will rise relatively rapidly in the near-term. However, once those barriers are overcome, if we haven't created them yet, there are probably important remaining barriers that won't be overcome in the next 15-25 years. But a lot can happen in 50 years—by then we may well have overcome nearly all the barriers to creating digital minds that we can overcome. So, the probability rises again after 25 years but subsequently levels off.

---

[12] Here we omit discussion of in-principle barriers—see above for discussion.

### AI timelines

Various participants noted that their responses were informed by timelines for advances in AI, a topic that has received much discussion in recent years and much more discussion than digital minds timelines in relevant literature with which we are familiar.[13] Mentioned **AI timeline factors** include:
- Current pace of AI progress
- The expectation that AGI will arrive soon
- The expectation that AGI will either arrive soon or not for a long time
- Whether the current paradigm will plateau soon
- The expectation that ASI will arrive in the 2030-2040 window
- The expectation that technological progress will be compressed into the next ten or so years
- The expectation that whole brain emulation will arrive in the distant rather than near future.
- Catalysts for experimenting with the requisite AI architectures may not arrive for quite some time (e.g. space colonization might catalyze architecture exploration)
- Other expert AI timelines[14]
- The prospect of a US-China AI race
- Actions by the current US federal government that have shortened timelines.

### Methodological

Some responses offered reflections on how to approach this question.

---

[13] For some influential discussions, see Bostrom 2014: ch 4, Cotra 2020, and https://ai-futures.org/. In previous work (Saad & Caviola, 2024) we discuss digital mind takeoff scenarios.

[14] One participant cited Grace et al. 2022.

- **Setup sensitivity**: The survey restricted consideration to digital minds whose welfare capacity is at least roughly as high as that of a typical human. Excluding digital minds with lower welfare capacity drove down estimates for the probability of digital minds collectively crossing welfare capacity thresholds by a given year.
- **Anchoring**: One approach to arriving at estimates is (1) use deference to determine a lower bound for an early estimate (for digital minds by 2030) and (2) take the estimated probability for digital minds ever arriving and assign a slightly lower estimate for a later time (digital minds by 2100) by which digital minds seem highly likely to arrive if they ever do.[15]
- **Potential boiling frog effect / goalpost moving:** Prior to the advent of LLMs, it was expected that digital minds would be created before AGI. The opposite is now expected. Yet it is not clear that this reversal is justified. It could be that, like slowly boiling frogs, we have failed to track the relevant evidence (in this case, for the emergence of digital minds) because it has accumulated gradually. And it could be that we will continue to move the consciousness goalposts, assuming that whatever has been achieved so far in AI development, consciousness requires something further.

## Peer Forecasts About Timing

This "meta-question" asked participants to predict how others in their expert group would respond to the question: "How likely are digital minds by 2040?" Specifically, they were asked to estimate the median response of their peers.

We included this peer forecasting question (as well as another about societal beliefs discussed below) for three reasons. First, we used accuracy on these questions to help allocate performance-based compensation. Second, we sought to examine whether participants with superior peer forecasting ability exhibited systematically different beliefs on other survey items (see section called "Relationship Between Peer Forecasting Accuracy and Survey Responses"). Third, we were interested in the accuracy patterns themselves and potential differences between expertise groups.

The results reveal differences in accuracy between groups in their participants' predictions of the group's responses. For the analysis here, we rely on mean predictions of the group median. AI research and policy experts showed the most pronounced underestimation bias, with a (mean) prediction of 32% for their group median when the actual median was 50%, a difference of 18 percentage points ($p = 0.01$). Digital minds research experts also underestimated their peers' confidence, predicting 42% versus an actual median of 50.5%, a difference of 8.5 percentage points, though this was not statistically significant ($p = 0.14$). Philosophy, science, and social science experts were remarkably accurate, predicting 33.2% versus an actual median of 35%, differing by only 1.8 percentage points ($p = 0.57$). Forecasting experts overestimated their peers' confidence, predicting 30.4% versus an actual median of 15%, a difference of 15.4 percentage points. Non-experts were also quite accurate, predicting 32% versus an actual median of 31.5%, differing by only 0.5 percentage points (p = 0.55). Sample sizes for forecasting experts (n=7) and non-experts (n=4) were small, limiting our ability to conduct inferential statistical tests for these groups.

---

[15] More specifically for (1), 0.1% was used for 2030 based on Sebo & Long (2023). Note, however, that Sebo & Long (2023) defend .1% as a conservative lower bound and say "the chance that some AI systems will be conscious by 2030 is much higher than 0.1%. But we focus on this threshold here to be generous to skeptics" (p. 593).

It is difficult to draw firm conclusions from this pattern. One possible interpretation is that participants believed they had access to special information or insights that made them more confident than their peers that digital minds will arrive by 2040. Alternatively, they may have thought that others were generally too conservative in their estimates. Another possibility is that social or professional pressures against expressing short digital mind timelines may in some settings suppress expressions of shorter digital mind timelines and result in the perception that they are less common than they in fact are. Or the survey may have been subject to a selection bias toward individuals with shorter digital mind timelines that participants did not sufficiently take into account.

While text responses did not offer many object-level insights, two responses noted sensitivity of the question to who within expert groups participates in the survey. One response cited self-selection as a reason to expect that digital mind experts will tend to give higher estimates for the development of digital minds by 2040. Another noted an expectation that social scientists will be overly skeptical about digital minds, based on social scientists' general thinking about AGI, while also noting an expectation that less skeptical social scientists will be overrepresented by those taking the survey.

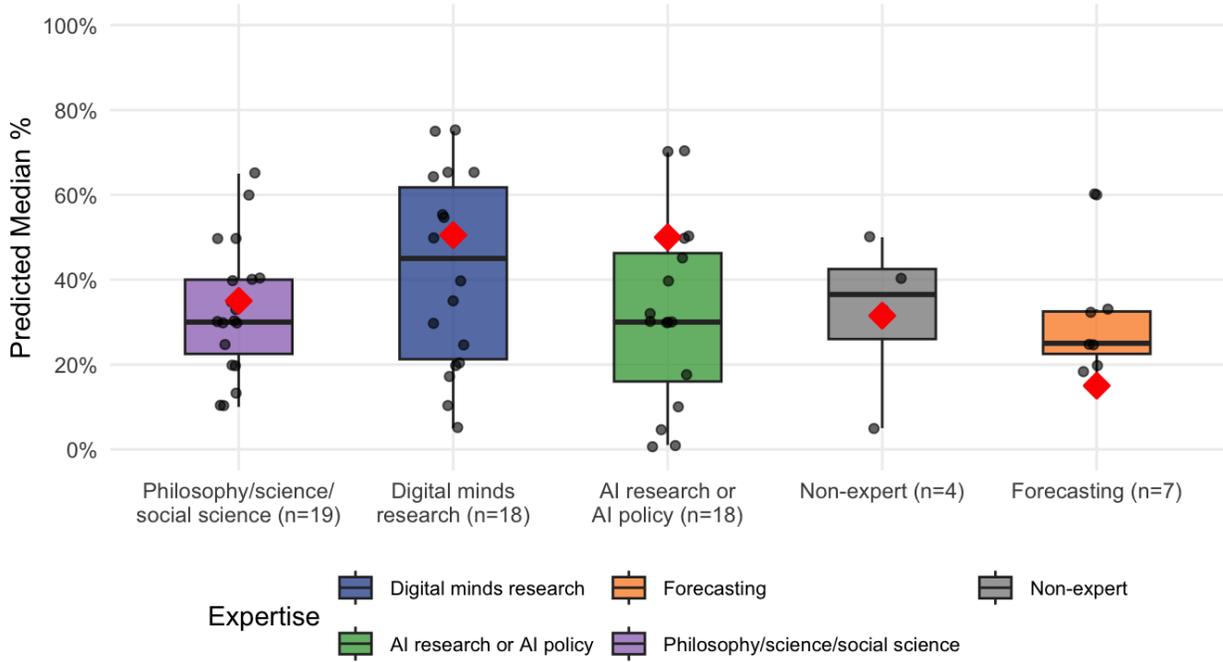

## Timing Relative to AGI

This question explored participants' views on whether digital minds will be created before AGI, where 'AGI' was defined as an AI system that matches or outperforms humans at almost all economically valuable tasks.

The results showed a somewhat bimodal pattern. The largest group of participants considered it very unlikely that digital minds will be created before AGI, with many giving responses of 0% or close to 0%. The median estimate was 27.5%. Another group clustered around 50%, indicating substantial uncertainty. Only a negligible number of participants thought digital minds were likely to arrive before AGI.

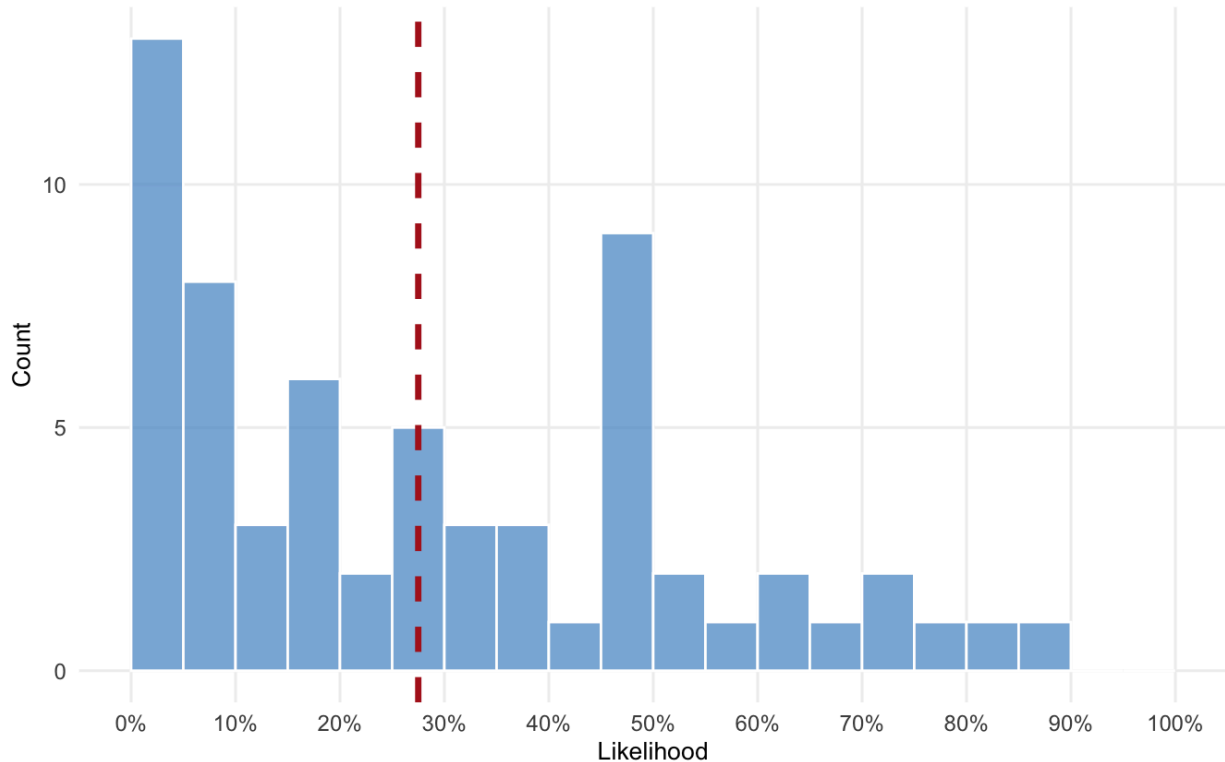

Considerations that suggest digital minds will be created after AGI

Participants offered a variety of reasons for expecting digital minds to emerge only after AGI.
- ○ **Short AGI timelines** along with **long or less certain digital minds timelines**
- ○ **Short AGI timelines** along with **long whole brain emulation timelines** (and the path to whole brain emulations before AGI requiring a coordinated ML AGI moratorium)
- ○ **Current models are not close to being conscious**.
- ○ **Incentives**: Incentives to create AGI are stronger than incentives to create digital minds.
- ○ **Experience dispensability**: Subjective experience isn't required for economically valuable tasks.
- ○ **Experience-cognition decoupling**: Assuming current systems are not conscious, they provide evidence that cognitive abilities and consciousness decouple.
- ○ **Digital mind costs**: Creating digital minds is likely to be expensive, not something that comes at minimal marginal cost as a side effect on the way to AGI.
- ○ **Creating digital minds is difficult:** Creating digital minds may be difficult even if one is aiming for it.
- ○ **AGI facilitates digital mind creation**: AGI will make it (vastly) easier to create digital minds.

- **Anthropomorphic illusions**: The tendency to anthropomorphize LLMs may make it seem like we're closer to creating digital minds than we are.
- **Agentic requirement unmet before AGI**: There is a type of agency that is required for subjective experience but which will be absent from AI systems that precede AGI.
- **High welfare capacity bar**: The survey restricts consideration to digital minds whose welfare capacity is at least roughly as high as a typical human's. However, given an objective-list account of welfare—on which e.g. love and friendship are welfare goods—this is a high bar. AI systems with welfare capacity may not clear it until after AGI.

### Considerations that suggest digital minds will be created before AGI

Only a few participants offered reasons in support of digital minds emerging before AGI. Notably, most of these reasons—perhaps with the exception of the last two—indicate that digital minds *could* precede AGI, but offer at most weak support for the claim that they *will*.

- **Embodiment asymmetry**: Whereas AGI requires a robot body to perform various economically valuable tasks a human can perform, being a digital mind probably does not require a robot body.
- **Cognitive asymmetry**: The level of cognition required to be an AGI is higher than the level required to be a digital mind.
- **AGI affect or emotional intelligence qualifies them as digital minds**: The centrality of affective states (such as emotions) and emotional intelligence to general intelligence is underappreciated. AGI would be a digital mind because general intelligence requires affective states and emotional intelligence, which require subjective experience. Moreover, if AGIs qualify as digital minds, less advanced systems will too.
- **In-principle possibility**: There is no principled reason why digital minds cannot be created before AGI.
- **Digital minds may already exist**: We cannot rule out that digital minds have already been created.
- **Model whisperers**: Work by 'model whisperers' is suggestive of consciousness in current systems (cited: https://x.com/repligate). Model whisperers are individuals who are exceptionally good at communicating with models and understanding their behavior or psychology.

### Ambiguous empirical factors

**Uncertainty about digital minds timelines**: A number of responses expressed greater uncertainty about digital minds timelines than AGI timelines.

- "We are also much more lost in terms of understanding subjective experience than of understanding intelligence (as capacity to perform certain tasks)"
- **Whether pre-whole-brain-emulation systems will have enough welfare capacity** to qualify as digital minds may determine whether digital minds arrive first.
- **Prospects for accidentally created digital minds**: Probably, deliberately created digital minds would arrive only after AGI, while accidentally created digital minds could arrive before AGI as a result of scaling AI capabilities.

**Uncertainty about AGI timelines**:
- AGI timelines are **disjunctive** between a short-timeline scenario and a long-timeline scenario
- **Does AGI already exist?**
- **AGI timelines are sensitive to how 'AGI' is defined**: although we will have incentive to create AI systems that broadly outperform humans, whether we will have incentive to create AI systems that outperform humans at (almost) *all* tasks will depend on whether that turns out to be most efficient.
- **How will AGI be achieved**: Timelines are sensitive to whether AGI will be achieved via ML agents, tool-like systems, or oracle systems vs. via whole brain emulation.

Other potentially important ambiguous factors:
- **Lack of strong evidence**: We lack clear reasons for favoring one order of arrival.
- **Common cause**: Training on delayed and sparse rewards in complex environments is a likely near-term path to AI capability gains that will eventuate in both AGI and digital minds.
- **Boiling frog effect**: Uncertainty about whether/how to take into account the fact that gradual advances in AI capabilities have produced systems that we judge unconscious but which we would have regarded as conscious (say) five or ten years ago if they had suddenly become available then.

### Additional considerations

- In addition to convergence on AGI and digital minds not necessarily coinciding in practice, there was also **convergence on mutually-enabling factors** whereby one of AGI or digital minds raises the probability of the other (e.g. because of a common probability raiser such as agency or robotics or because AGI makes creating digital minds easier).
- **Divergence in definition and practice**: On the adopted definitions, there is not a definitional requirement that a digital mind qualify as an AGI or vice versa. In addition, being a digital mind and being an AGI could come apart in practice. For example, the first digital mind might not be an AGI, and the first AGI might not be a digital mind.
- **Sources of conceptual ambiguity**: Text responses noted multiple ambiguities in this question. One participant reported uncertainty about whether or not the question was supposed to be conditional on digital minds being created.[16] Multiple participants noted that the operative notion of 'AGI' is ambiguous; one flagged it as a huge source of ambiguity and opted to resolve the ambiguity by understanding 'almost all' in the AGI definition in terms of being able to match or exceed human performance at 99% of US jobs. Another pointed out that the adopted definition of 'AGI' is ambiguous because 'matches or outperforms' could be read either in terms of being a system being capable of matching or outperforming humans at economically valuable tasks or in terms of a system being integrated into the economy such that it in fact matches or outperforms humans. Two participants pointed out that whether a system meets the welfare threshold we set for qualifying as a digital mind may depend on how systems are individuated.[17]

---

[16] The intended reading of the question was unconditional. On that reading, the probability that the first digital minds will be created before AGI (median 27.5%) cannot be higher than the probability that digital minds will be created (median 73%). While the latter may help to explain the former, the discrepancy between them suggests that there are also other factors in play.

# Speed

In this section, we examined digital mind takeoff speeds: how rapidly the population of digital minds could grow following the creation of the first digital mind.

Participants were asked to estimate how many years it will take for the collective welfare capacity of all digital minds in existence at a given time to match that of 1,000 humans, 1 million humans, 1 billion humans, and 1 trillion humans. A digital mind's welfare capacity was defined as its capacity to be benefited or harmed (e.g. by positive or negative subjective experience) in a manner that is inherently morally significant. Participants could also respond that a given threshold will never be reached; for calculating median estimates and visualization purposes, we coded these responses as 10,000 years.

Note that for this survey question, participants were asked to assume that the first digital mind will be a machine-learning-based AI system created in 2040 or earlier. Even if participants personally considered this scenario unlikely, they were instructed to respond to subsequent questions as if it were true. Remember that the median estimate of the probability that the first digital mind will be created before 2040 was 40%, and most participants expected a machine learning system to be the first type of digital mind.

The responses reveal that many participants expect digital mind welfare capacity to scale rapidly following the creation of the first digital mind. The median estimates (with "never" responses) were:
- 1 year to reach the welfare capacity of 1,000 humans (mean without "never" responses = 12.9 years)
- 2 year to reach 1 million humans (mean without "never" responses = 33.2 years)
- 5 years to reach 1 billion humans (mean without "never" responses  = 89.2 years)
- 10 years to reach 1 trillion humans (mean without "never" responses  = 497.6 years)

These median estimates reflect surprisingly short timelines, with many participants expecting massive growth within just a few years. However, the mean estimates were substantially higher than the median estimates due to a relatively small number of respondents who anticipated much slower growth—sometimes over hundreds or even thousands of years.

---

[17] One free text response nicely illustrated this thought: 'Also, where do you draw the boundary of individual mind[s]? E.g. I find it pretty plausible that frontier LLMs have subjective experiences… scattered [in] relatively small quantities across "person"-moments. Given the large quantity of such person moments, it seems plausible the total amount of capacity is already there, although scattered in many pieces. My guess is you would not count that way, so filling in numbers for something like "digital minds in form humans are more likely to recognize"'.

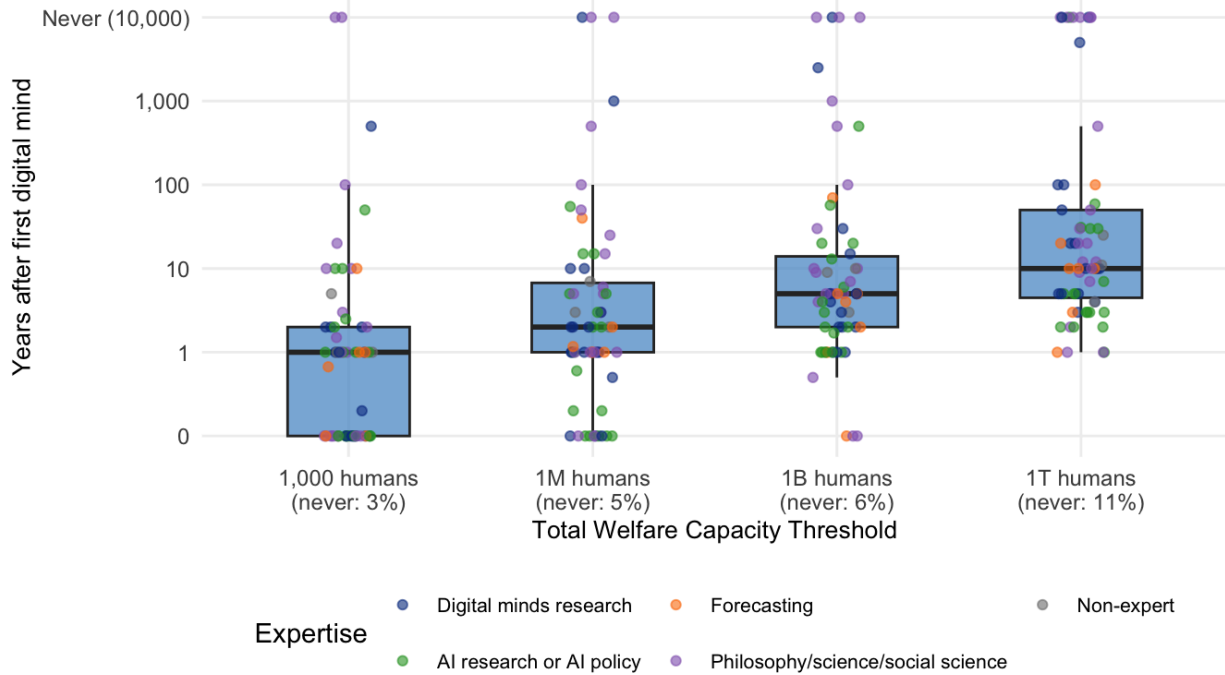

Participants offered a particularly rich and thoughtful set of text responses for this question. We'll discuss these responses under the following categories: drivers of faster takeoff in digital mind welfare capacity, constraints on takeoff speed, other factors that bear on takeoff speed, conceptual and methodological issues.

Drivers

- **Compute overhang**: When it becomes possible to create one digital mind, enough compute will be available to create many. Compute that was previously used for training or other purposes might be redirected to run digital minds. In thinking about how a compute overhang might lead to the rapid creation of a large population of digital minds, a useful comparison here may be ChatGPT. As soon as a ChatGPT model is publicly deployed, it is available to hundreds of millions of users.[18]
- **Demand**: There may be economic demand for digital minds because they are useful. There could even be demand for multiple digital minds per individual human.
- **Well-resourced actors**: The first actor that creates a digital mind may have the resources to create many. For instance, if they are rapidly scaling up their capacity to create more or other well-resourced actors are in a position to do so.

---

[18] According to an OpenAI spokesperson, actively weekly users of ChatGPT surpassed 400 million in February 2025 (Reuters, 2025).

- **Ethical incentives**: There may be ethical motivations to create large populations of happy digital beings (e.g. for totalist utilitarian reasons), or to develop advanced digital "post-humans" as a continuation of humanity's trajectory.
- **Ease of replication**: Once digital minds become feasible, they could be duplicated cheaply and efficiently.
- **Internal drive to replicate**: Digital minds could have a robust motivation to replicate.
- **Coordination deficit**: Probably, preventing the creation of digital minds with a collective capacity for harm that exceeds humanity's would require a strong form of coordination that is unlikely to happen in time.
- **Super-beneficiaries**: Individual digital minds may have super-human welfare capacities, owing to factors such as: First, a digital mind could have many more experiences per unit of objective time than a human. Second, there is "no reason why the first few minds taken together won't have more capacity for welfare than all humans taken together. If the entire hivemind of Claude's datacenters achieves one internal experience, it might be quite an important internal experience!"[19]
- **AI-driven rapid progress:** AGI/ASI may lead to rapid growth in factors that are positively correlated with digital mind population size (such as economic growth, chip development and production, and other technological advances).

Constraints

Various candidate constraints on the growth of digital mind welfare capacity were identified:
- **Computational constraints**: available (effective) compute is a presumptive bottleneck. The extent to which compute will bind as a constraint depends partly on whether available compute will grow in line with current trends or faster, owing to AI accelerating advances in hardware and/or software. How many digital minds can be deployed initially will depend on the ratio between the compute required for training digital minds and the compute required to run them. As training is currently relatively expensive, an actor with the capacity and incentive to create digital minds will run many copies (thousands to millions) fairly quickly. Perhaps creating a trillion digital minds would require significantly more compute than will be available when the first digital mind is created. Such quantities of compute might even require a new kind of technological advance, e.g. we might need to become spacefaring. However, there is reason to think that no such advance is required. For one, raw compute used to train AI systems is increasing by 4.6x per year while algorithmic efficiency is increasing by 3x per year (according to [Epoch estimates](#) [Epoch, 2025]), and we can expect AI to speed up these trends. So, effective compute can be expected to increase by more than an order of magnitude per year. For another, on the assumption that digital minds use about the same amount of inferential compute as the human brain (about $10^{15}$ FLOP/s), running a trillion digital minds will only require $10^{27}$ FLOP/s. And we may well have that much effective compute when the first digital mind is created.

---

[19] "For more discussion, see the section on Super-Beneficiaries.

Other factors to consider in connection with computational constraints on digital minds:
- **Demand deficit**: There may be little demand for digital minds (especially of the sort considered in the survey as opposed to a broader class that includes ones with less welfare capacity). Or there may be demand for digital minds only in a narrow range of use cases.
- **Supply bottleneck**: Even if there is high demand for such digital minds, this may not lead to a large number of them. For example, if there is only demand for digital minds that are integrated with robotics, then robotics development may be a bottleneck on the production of digital minds.
- **Capital constraints**: The development and production of digital minds may require significant capital investments. Once digital minds are produced in large numbers, capital may be a bottleneck on further growth to the digital mind population. This constraint might be temporary. For example, capital might cease to be a limiting factor during space colonization.
- **Diminishing returns**: Even if digital minds are economically valuable, there may be diminishing returns on investing in them. Thus, economic incentives may cease to drive the creation of more digital minds beyond a certain point.
- **Government constraints**: Governments could constrain growth of digital mind welfare capacity with bans on the development or production of digital minds, by imposing rate limits on the production of digital minds, or through monitoring requirements. If monitoring larger populations of digital minds is more expensive, then monitoring requirements may constrain digital mind population growth only after the digital mind population becomes sufficiently large.
- **Developer-imposed deployment constraints**: Developers may impose constraints on themselves that limit growth in digital mind welfare capacity. For example, developers may initially only allow the first digital minds to be rolled out internally, and deploy them externally only after a period of careful monitoring.
- **Ethical constraints:** Recognizing the welfare implications of creating digital minds may dissuade us from doing so. Or we may be averse to creating a population of digital minds whose moral weight exceeds our own.
- **Situational awareness**: We may create large numbers of digital minds in the first few years because AI development/production is proceeding very quickly and we don't realize that we're creating digital minds. Upon recognizing that we're creating digital minds, we may decide to slow/pause development to figure out how to proceed.
- **Societal opposition**: Some people will oppose the creation of digital minds (though others will also support their creation).
- **Artificial reproductive capacity**: If digital minds are allowed to self-replicate, runaway dynamics likely quickly yield many digital minds. These dynamics may be tempered by opposition to the creation of digital minds. But runaway dynamics is likely the dominant factor. While machine learning digital minds may proliferate quickly, digital minds with other architectures (e.g., brain simulations or neuromorphic systems) may face more barriers.
- **Changes in individual welfare capacity**: Individual digital mind welfare capacity could increase over time as a result of their becoming more sophisticated, building their own culture (compare: early humans vs. contemporary humans), or developing relationships. How quickly the digital mind population welfare capacity increases depends partly on the rate of increase in individual digital mind welfare capacity.

- **Energy constraints**: Digital minds will consume energy. A reasonable working assumption is that digital minds will be as energy efficient at producing consciousness as humans and that digital minds will become more energy efficient over time. Energy may be a bottleneck for creating digital minds only after digital minds are produced in large numbers. On the current trajectory, we will run up against energy limits around 2030. But energy supply may substantially increase. For example, perhaps nuclear fusion will substantially increase the supply by 2040.
- **Intelligence explosion constraints**: Whatever limits the intelligence explosion may also limit growth in digital minds' collective welfare capacity.
- **Metaphysical constraints**: The nature of digital minds or welfare might preclude some ways for digital minds to achieve a large welfare capacity. For example, if digital mind welfare scales with objective duration rather than experienced duration, then digital minds' ability to experience much longer durations than humans during a given amount of time will not result in digital minds having higher welfare capacity. Or perhaps individual digital minds with much greater than typical human welfare capacity are impossible because the nature of welfare capacity only allows for a much narrower range of variation.
- **Catastrophic risks:** such risks may limit digital mind welfare capacity. These risks may include risks that large populations of digital minds themselves pose for civilization.
- **Sci-fi penalty**: The creation of a population of digital minds that exceeds the billion or trillion threshold sounds crazy. This is a reason to penalize the likelihood of such scenarios (especially the latter type of scenario).

### Other factors

Other factors that bear on how digital mind welfare capacity will evolve:
- **Welfare capacity determinants**: Digital mind welfare will be sensitive not only to factors such as compute and energy but also to what can affect their welfare and by how much. For example, it may matter to what extent their welfare can be affected by hedonic states such as pleasure and pain vs. non-hedonic factors.
- **Digital mind roles**: Digital mind welfare capacity may vary across roles and contexts. For example, it may matter whether a digital mind is a social companion vs. a space-faring entity.
- **Future unknowns**: We don't know how technology will evolve. But we can expect that digital mind welfare capacity will interact with currently unknown future technological advances.

### Conceptual and methodological issues

- **Stipulative boosting**: Under the imposed stipulation that the first digital mind is created by 2040, digital minds are presumably easy to create and/or a side effect of AI advancement. In that case, whatever drove the creation of the first one will persist and more digital minds will be created quickly.

- **Stipulative suppression**: One participant flagged that the restriction to digital minds with at least roughly human welfare capacity mattered for their estimates. They wrote: "I would expect that before your definitional [digital minds] there'll be a lot of proto-[digital minds] which have subjective experience but lower welfare range than humans… they're likely to outnumber us by so much and so fast that they'd have total welfare capacity comparable with humanity's human population within a year or two of those systems existing."
- **Synchronic vs. diachronic welfare capacity:** The question could be read synchronically in terms of digital mind welfare capacity that is *present at a given time* or diachronically in terms of the integral of digital mind welfare capacity *up to a given time*.
- **The concept of individual welfare capacity**: To estimate the collective welfare capacity of digital minds, we need to consider the welfare capacity of individual digital minds with at least roughly human welfare capacity. But this notion is problematic. One difficulty is that human welfare capacity turns on the outer limits of human happiness/suffering and we may lack clear judgments of where those limits are. There are alternative ways of understanding individual welfare capacity in terms of counterfactuals about what an individual's welfare level would be under a certain range of conditions. For example, there is an option of understanding an individual's welfare capacity in terms of the (average) intensity of their ordinary experiences. But such approaches seem to require arbitrary stipulations concerning which counterfactuals are used.
- **Welfare capacity aggregation**: To estimate the collective welfare capacity of digital minds, we also need to somehow aggregate the welfare capacities of individuals. It is not obvious that this utilitarian-flavored kind of moral quantification is legitimate.
- **Cross-kind welfare comparison**: The question presupposes that we can at least roughly compare human welfare capacity with digital mind welfare capacity. But it is questionable whether such comparability holds.
- **Cross-scenario aggregation**: Participants frequently reported generating estimates for different scenarios and then aggregating those estimates to arrive at an overall estimate. Some types of scenarios that were aggregated across included:
  - Scenarios that vary in the extent to which 2040 is singularity-esque.
  - A scenario in which we rapidly and unreflectively create a million digital minds within five years after the first digital mind, with the scenario then branching into one of three possibilities: (1) we awaken to the fact that we're creating digital minds and decide to stop (leading to a sustained plateau or decline in the digital mind population, with the billion threshold only being crossed hundreds of years later), (2) digital minds have a drive to procreate (leading to accelerated growth in the digital mind population and the billion threshold being crossed within a couple of more years), and (3) civilization succumbs to a catastrophe that wipes out everything including infrastructure needed for digital minds (leading to collapse in the digital minds population).
  - Scenarios that vary along two dimensions: rates of diffusion for types of digital minds and intensity of a given type of systems' experiences.

- ○ Scenarios varying on whether digital minds are created on the main branches of the AGI technology tree vs. whether digital minds are curiosities or merely created in research labs (compare: Dolly the sheep)
  - ○ Fast/early vs. delayed takeoff scenarios, depending on whether there is a cultural or regulatory reaction against digital minds.
  - ○ Scenarios varying in the extent to which artificial systems are socially integrated, under the assumption that degree of social integration affects whether these systems have enough welfare capacity to be in scope.
- **Simplifying assumptions**: Some participants registered simplifying assumptions that they made in order to answer this question (e.g. noting that they were considering one type of scenario even though they considered others plausible/possible). It's unclear to what extent these participants disagreed about the plausibility of the scenarios they chose to focus on. This raises the possibility that variance in responses to this question may partly reflect decisions about which type of scenario to draw on in answering this question and/or small differences in credences leading participants to focus on very different scenarios. Adopted simplifying assumptions included:
  - ○ Digital minds are created in a scenario in which there is economic incentive to do so and no ban on their creation.
  - ○ Very good governance.
  - ○ Digital minds are developed in a cautious manner.
  - ○ Scientific understanding of digital minds.
  - ○ Digital minds are created on the LLM development path, leading to widespread adoption of digital minds.
  - ○ Digital minds have roughly human-level welfare capacity.
- **Difficulty in median estimation**: The survey asked participants to provide median estimates. For this question, participants reported difficulty in providing the requested type of estimate. While one participant reported uncertainty about whether to answer this question with median estimates, others reported difficulty in generating median estimates for this question (e.g. because of the need to consider many possible branching scenarios). It was noted that minimum time for digital minds to cross a welfare capacity threshold could be much shorter in expectation than the median time.
- **Hedges**: Free text responses to this question frequently entered hedges. Examples: "I have huge error bars and could also see that not that many [digital minds] will be created", "This is probably the most speculative of my responses so far.", "I feel especially unsure in my answers to this question", and "After the first five years, I find it near impossible to put numbers, because I see so many meaningfully different paths we could take". Some responses flagged the trillion-human sub-question as particularly difficult/speculative. It was also pointed out that we could come to see this question as meaningless as our understanding of minds improves.

# Types

## Possible Types

This section explores expert views on the prospects for different types of computer systems to qualify as digital minds. While much of the public debate focuses on specific models like large language models (LLMs), there is considerable uncertainty about which types of computer systems might ultimately support consciousness. We asked participants to estimate the likelihood that various types of systems could, in principle, have subjective experiences.

Specifically, participants were asked how likely they find it that some computer systems in the following categories could, in principle, have subjective experiences:
- Machine learning systems (e.g., large language models, reinforcement learning agents)
- Brain simulations (e.g., whole-brain emulations of human or animal brains)
- Other systems, such as neuromorphic or quantum architectures[20]

Unlike in the earlier section, participants were **not** asked to assume that digital minds will be created by a specific date or in a particular way.

The results revealed a hierarchy of perceived likelihood. The most likely in-principle possible possessors of subjective experiences were judged to be brain simulations, with a median estimate of 88% and a mean of 81.3%. This reflects a common view that replicating relevant fine-grained functions of a conscious biological brain in a computer would plausibly result in an artificial system with subjective experiences.[21] Other architectures were also seen as promising, with a median estimate of 85% and a mean of 76.4%. Machine learning systems, though rated as least likely of the three, still received surprisingly high estimates: a median of 70% and a mean of 64.7%. These results suggest that participants see multiple plausible paths to digital minds, not just those closely modeled on the brain.

---

[20] To a first approximation, neuromorphic architectures are standardly understood as brain-inspired, non-conventional computing architectures, while quantum computers are computers that, unlike conventional computers, exploit quantum superpositions. The survey did not give participants characterizations of these architectures.

[21] Chalmers 1996 defends this assumption, Butlin et al. 2023 adopt the kindred hypothesis of computational functionalism as a working assumption, and a plurality of philosophers of mind (36.21%) accepted or leaned toward accepting functionalism about consciousness in a 2020 survey (Bourget & Chalmers, 2023; for another academic survey, see Francken et al., 2022).

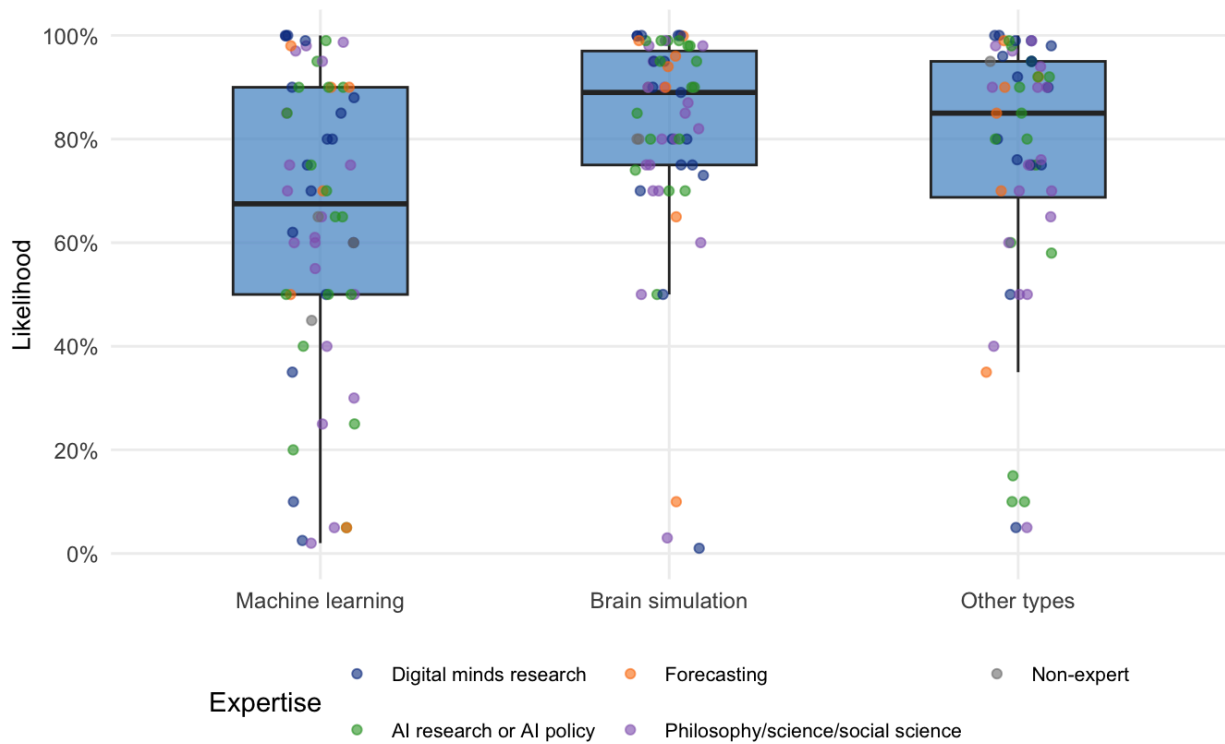

Considerations

Participants mentioned the following considerations that bear on the in-principle possibility of subjective experience in specific types of computer systems:

- **Scope of the Other category**: The "Other" category is very wide; it probably includes unimagined forms of digital minds
- **Universal functional approximation**: Machine learning systems can in principle approximate any computation that might underlie subjective experience, including computations implemented by whole brain emulations. The same presumably applies to some systems in the Other category.
- **Structural duplication**: Machine learning systems might realize subjective experience by duplicating the requisite structure.
- **Recursion requirement**: A potential barrier to machine-learning systems possibly having subjective experience is that machine learning systems may not be able to exhibit the requisite kind of recursion.
- **The prospects for neuromorphic digital minds**:
    - A neuromorphic brain simulation would be very similar to a human brain.
    - Neuromorphic systems are especially promising as possible subjects of experience because they could more clearly replicate (rather than arguably just simulate) relevant functions.

- Neuromorphic systems are more promising as candidates for possible subjects of experience than are brain simulation systems, as more assumptions need to hold for the latter to possibly realize subjective experience.

Question and response limitations

Some text responses noted limitations in participants' abilities to answer.
- **Lack of technical knowledge** about some types of systems. One participant offered the following reflection on the methodology of answering this question: "I wonder how we even go about estimating the in-principle feasibilities. When I introspect, I make use of the similarity to structures I know have mental life and that favors WBE. But the space of all minds is very large."
- **Uncertainty about whether to read the different types as exclusive** (e.g. whether to read 'machine learning system' broadly to potentially include some brain simulations)
- **Disconnect between the question and the adopted welfare-capacity restriction**: Whereas we restricted consideration to digital minds with at least roughly human welfare capacity, this question is about systems that could in principle have subjective experiences and doesn't say anything about digital minds or welfare capacity. Estimates could be lower or higher depending on whether the question is read with or without the restriction.
- **Uncertainty about what the Other category encompasses.**
- **Connections to other questions**: One participant wrote: "As I thought through this question, I started updating my earlier answer to how likely it is that digital minds are possible—I started to think that my earlier answer was perhaps closer to the likelihood a machine-learning system would ever be conscious and that the possibility of WBE and similar methods should increase my probability." This response is noteworthy because it suggests that some responses to earlier questions in the survey may have been under-estimates, owing to participants focusing on an overly narrow class of digital minds.

First Type

After asking which kinds of systems could, in principle, have subjective experience, we next asked participants which type they thought will be the *first* digital mind to exist. Specifically, participants were asked to allocate probabilities across the following hypotheses: digital minds will never be created, or the first digital mind will be a machine learning system (e.g., an LLM or reinforcement learning agent), a brain simulation (e.g., a whole-brain emulation of a human or animal), or some other system (e.g., neuromorphic or quantum). The probabilities had to add up to 100%.

The results revealed a notable shift from the previous question. While brain simulations were judged to be the most likely in-principle possible subjects of experience, participants judged machine learning systems as by far the most likely to be the first digital minds to be created. The median estimate for machine learning systems was 50%, with a mean of 47.1%, indicating a broad expectation that digital minds, if they arrive, will plausibly emerge from current AI paradigms. The next most common response was that no digital minds will ever be created, with a median of 20% and a mean of 24.5%, reflecting substantial uncertainty about whether digital consciousness will ever be achieved in practice[22]. Brain simulations, despite being seen as the most promising in principle, were only assigned a median probability of 15% and a mean of 18.0% of being the first digital mind created. Other types of systems, such as neuromorphic or quantum architectures, received the lowest estimates, with a median of 10% and a mean of just 12.2%.

Taken together, these results suggest that while experts believe brain simulations are the most promising route to digital consciousness in theory, they are not expected to yield digital minds soon. In contrast, machine learning systems—despite being seen as less likely to support subjective experience in principle—are viewed as the most likely to yield the first digital mind. This may reflect the (comparatively) rapid development, widespread deployment, and the practical momentum behind current machine learning research.

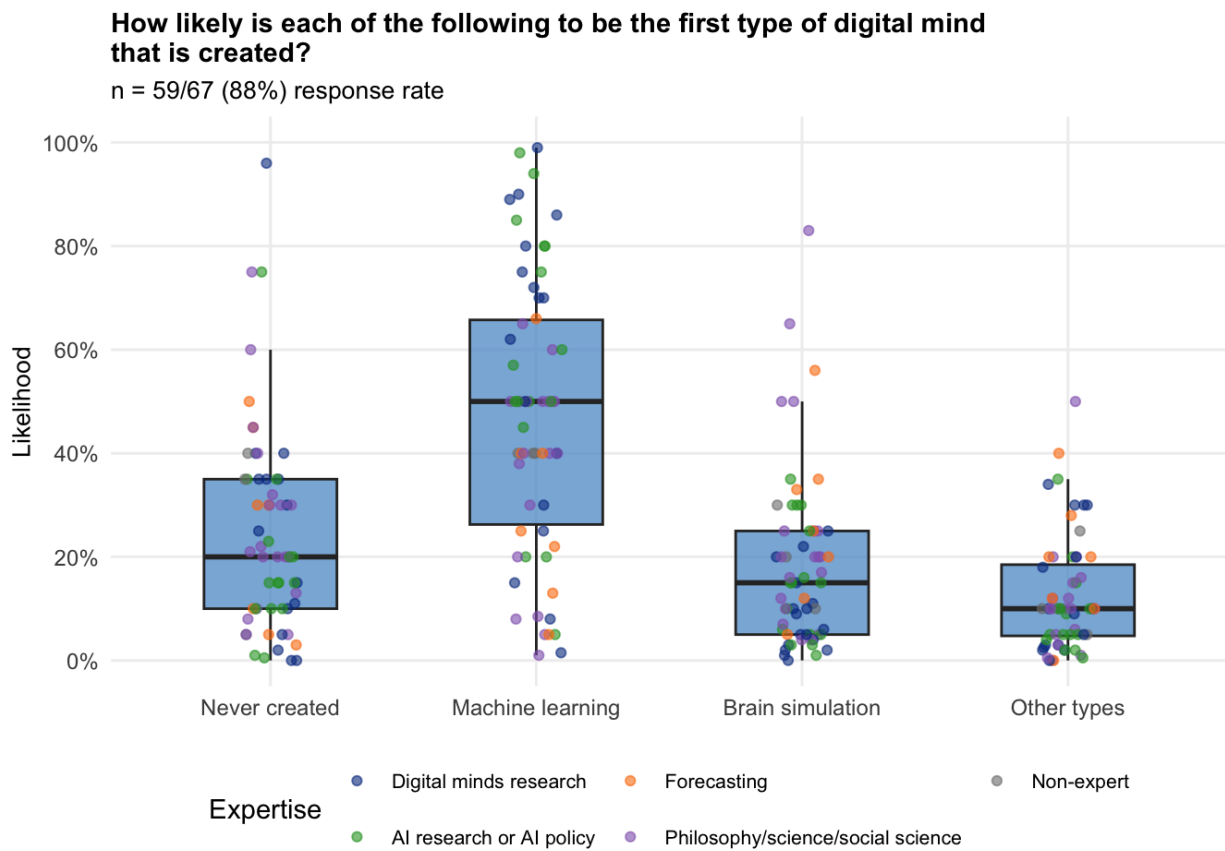

Text responses to this question reflected both how plausible different types of digital minds are in principle and when they are expected to emerge. Since the previous section addressed in-principle plausibility, we now focus on situational factors that influence expected timelines—and how these factors interact with earlier plausibility judgments.

---

[22] Note that these "never created" estimates differ slightly from the "ever created" estimates in the Likelihood question (Starting Point section, where the median estimate that a digital mind will ever be created was 73% (mean = 70.7%), implying a 27% probability that digital minds will never be created. When directly comparing each participant's response between these two questions, 69.5% of participants deviated less than 5 percentage points from being fully consistent, demonstrating reasonable coherence in probability judgments despite the different question framings.

Technological factors

- **The current state and trajectory of machine learning**: Machine learning systems already exist and are being actively developed.
- **The current state of neuromorphic technology**: Our knowledge of how to build neuromorphic systems is currently very limited.
- **Whole brain emulations will not be created soon**: The track record of whole brain emulation/simulation research indicates that they will not be created in the near term. As one respondent wrote, "Many past projects, such as the EBRAINS branch of the 2013–2023 Human Brain Project, ultimately failed to deliver on their ambitious goals." Constraints on the creation of whole brain emulations include brain scanning technology, compute resources, and hardware. It is not clear that whole brain emulations will ever be created. But if they are, they will likely be created in the distant future and/or after AGI.
- **AI as catalyst**: AI could speed up the arrival of whole brain emulations and neuromorphic systems.
- **Risks bound opportunities for creation**: Risks of catastrophes that would prevent the creation of certain types of digital minds (especially those that would be created in the distant future if at all) bound the probability that those types of systems will be created first.

Investments and incentives

- **Current investment patterns** more strongly influenced responses than intuitions about which types of systems could in principle be digital minds. If current investment trends continue, near-term investment in machine learning systems will swamp near-term investment in other types of systems. Architecture path dependency may contribute to the continuation of this trend.
- **Balancing level of investment vs. in-principle prospects**: Perhaps LLMs and "Other" digital minds are roughly equally as likely to be first, since LLMs are much less likely to be conscious but are being developed much more actively and these factors may cancel.
- **Investment track record for brain simulation projects**: Past investments in large-scale brain simulation projects haven't delivered on their goals. Funding prospects for such projects is now dimmer, as funding is now more likely to flow toward feasible and commercially viable AI projects.
- **Desire to develop uploading technology** incentivizes the development of conscious whole brain emulations. That desire also incentivizes the development of simulations populated by digital minds as environments into which individuals could be uploaded.

Consciousness factors
- **Whole brain emulations are less likely to be early but more likely to be conscious**: If whole brain emulations are created, they are more likely to have subjective experiences than machine learning systems. Whole brain emulation digital minds will be first only if creating ML digital minds hits a wall.
- **Biology/machine-learning similarities**: Resemblances between reinforcement learning and human/animal cognition make ML systems more likely to be digital minds than quantum and neuromorphic systems.
- **In-practice barriers to subjective experience in machine learning systems**. Even if ML systems can in principle qualify as digital minds, they may not qualify in practice owing to:
  - infeasibility in practice of realizing the consciousness-relevant causal roles in machine learning systems, despite those systems' in-principle potential for universal function approximation.
  - not creating machine learning systems with the self-monitoring or introspective abilities that go along with subjective experience.
- **Training trajectory raises the probability of machine learning systems being the first digital minds**: On the current trajectory, we are making ML systems increasingly capable (e.g. with respect to learning and agentic abilities) and are training them in complex environments. Because these correlate with subjective experience, this raises the probability that ML digital minds will be the first.
- **Scaling**: A key factor for whether ML digital minds will be first depends on whether scaling ML systems (e.g. increasing their parameter count or training compute) is a path to digital minds.
- **The prospects for byproduct digital minds**: Whereas ML minds may be created as a byproduct of advancing capabilities, digital minds in the "Other types" category are more likely to be created by design if we develop a theory of subjective experience and use it to develop bespoke digital minds.
- **Other digital minds as the only possible kind**: Digital minds in the Other category are most likely to be first because there is a high probability that they are the only possible digital minds we can create.
- **Hybrid digital minds**: If combinations of categories are allowed, hybrid systems would be particularly promising candidates for digital minds.

# Distribution

In this section, participants evaluated how digital minds will vary in their social roles, geographic origins, creators, and in motivations for their creation. As with the Speed questions, participants were asked to assume that the first digital mind is a machine-learning-based AI system created in 2040 or earlier.

## Social Function

Participants were asked to consider all digital minds that exist 10 years after the first one is created and estimate what proportion of them will have a social function—that is, be designed to interact with humans in a conversational, human-like manner (e.g., through text, audio, or video).

The results revealed a bimodal pattern: most participants estimated that only a small proportion of—or no—digital minds will have a social function, with a median response of 10% and a mean of 30.9%. However, a smaller subset of participants believed that the majority of digital minds will serve a social function.

The reason for this divide remains somewhat unclear. It may reflect genuine disagreements about the likely roles of digital minds—for example, whether demand for human-AI interaction will dominate their use cases. Relatedly, it could arise from differing views on whether certain capacities central to social interaction—such as language, emotional expression, and self-reflection—require genuine subjective experience. Another possibility is that the divergence stems from differences in how the term "social function" was interpreted.

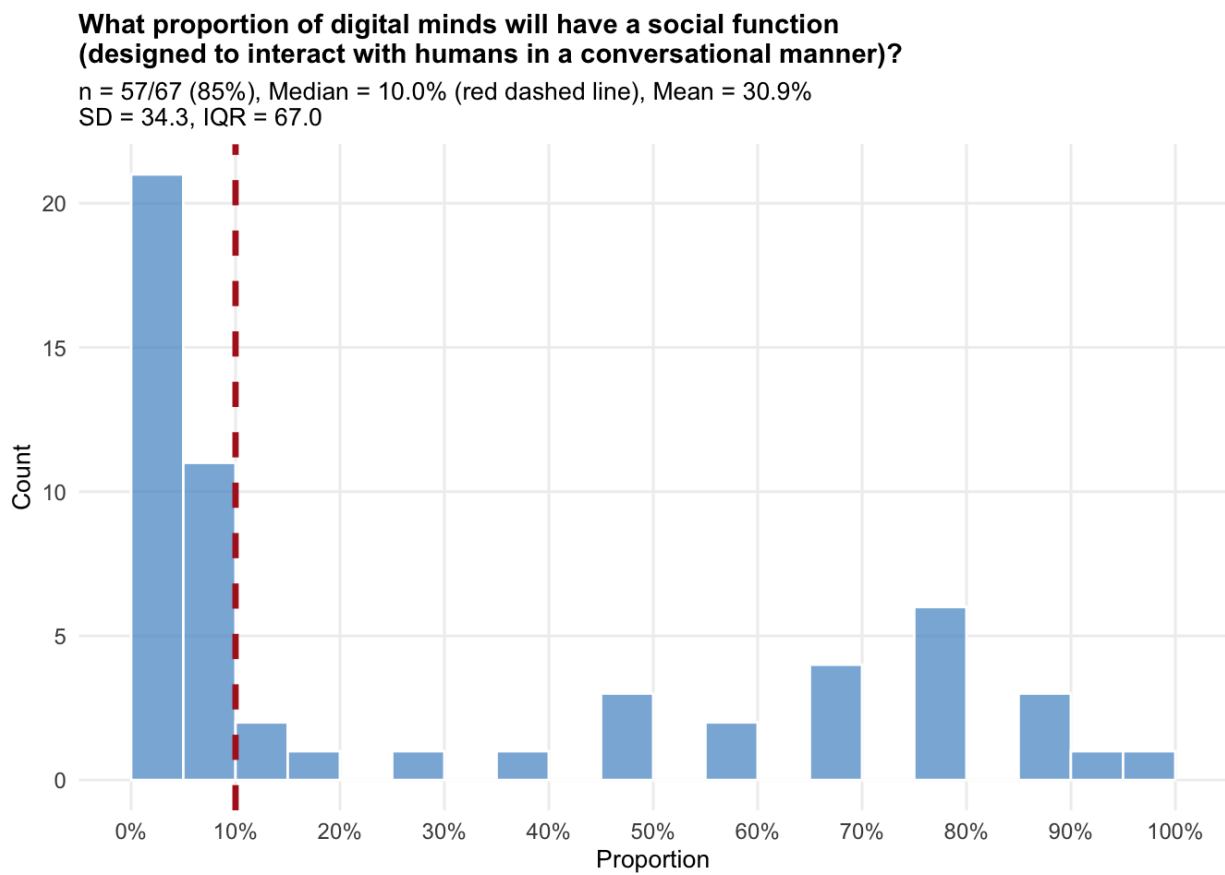

Considerations in favor of social function

- **Intentional design for interaction**: Many plausible development paths involve intentionally designing digital minds for social interaction—such as AI companions, human-like agents, whole brain emulations, or ethically-motivated post-human successors.
- **Economic value**: Large numbers of digital minds may be most economically useful when tailored for individual human users, especially in social or communicative roles.

- **Demand for explainability**: Humans are likely to prefer digital minds that can explain their actions, decisions, and inner states. This may encourage the development of social digital minds capable of human-like interaction.
- **Subjective experience enhances engagement**: Artificial systems with subjective experiences may become more engaging and interesting to humans. In some domains, humans may specifically prefer to be served by AI systems with subjective experience.

Considerations against social function

- **Greater usefulness in non-social roles**: Digital minds may be more economically valuable when performing cognitive labor or specialized tasks that do not require social interaction with humans.
- **Limited demand per user**: There may be a natural ceiling on the number of social digital minds needed—perhaps only one per consumer.
- **Non-social functionality suffices:** Many digital minds may be capable of fulfilling their intended functions without engaging in human interaction, making it unnecessary to design them specifically for social interaction in most cases.
- **Disinterest in humans**: Some digital minds may not find humans interesting or worth engaging with, for example, due to vastly different cognitive perspectives, leading them to avoid social interaction with humans altogether.
- **Subjective experience not necessary for social roles:** AI systems may not need subjective experience to perform socially interactive functions, reducing the incentive to develop conscious digital minds for this purpose.

Location

Participants were asked to estimate where digital minds will primarily be produced 10 years after the first one is created. "Primarily produced" was defined narrowly as the location where a given system first qualifies as a digital mind, as opposed to broadly in terms of the supply chain or development process.

The results suggest that participants expect the USA and China to be the leading origins of digital minds. The median estimate for the USA was 40% (mean = 43.4%), followed by China with a median of 30% (mean = 30.5%). Europe, including the UK, was estimated at a median of 10% (mean = 11.7%), while all other regions combined were also given a median of 10% (mean = 14.9%). 34 participants (59.6%) rated the USA higher than China, while 15 participants (26.3%) rated China higher than the USA. 8 participants (14%) gave equal ratings to both countries. Overall, these estimates reflect a general expectation that global digital mind development will be concentrated in a small number of major technological powers—particularly the USA and China—which is perhaps unsurprising given their current leadership in AI research and development.

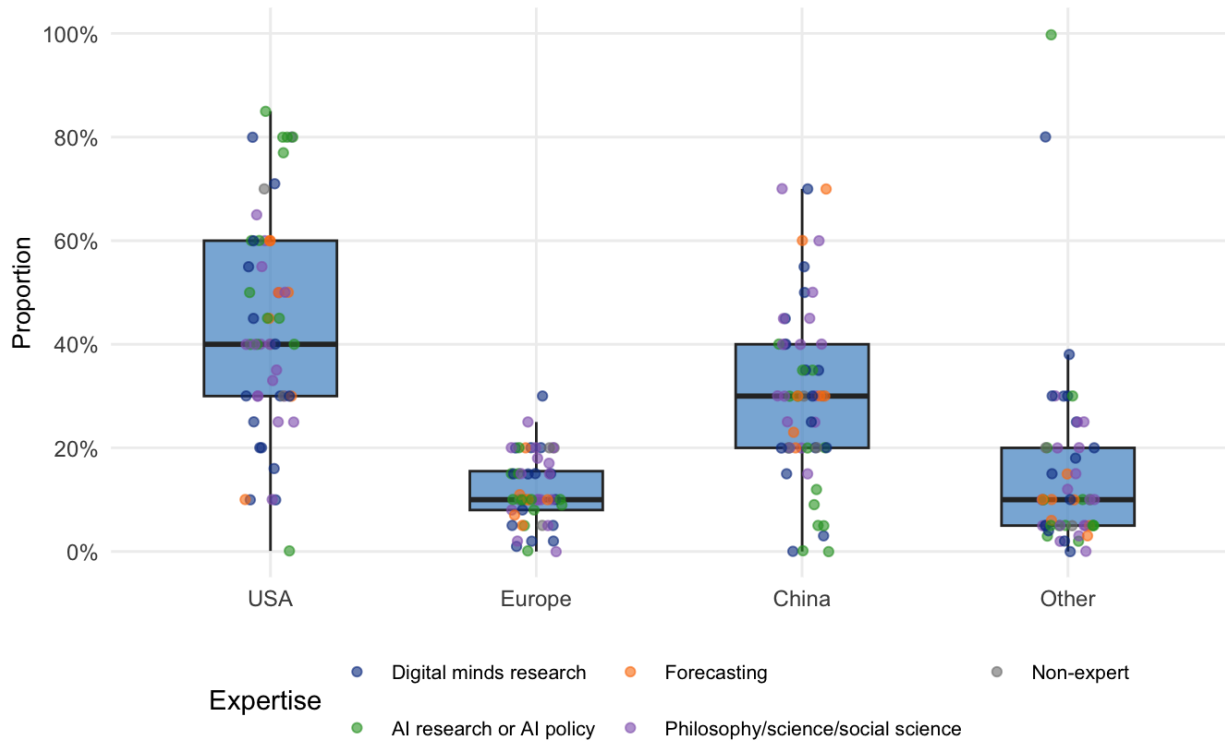

Factors supporting USA or China dominance

- **AI leadership as a predictor**: Digital minds are likely to be created in countries that lead in AI development. The USA currently holds that position, though China may catch up depending on shifts in compute capacity and geopolitical influence.
- **Ethical permissiveness**: Differences in ethical norms may influence which countries pursue digital minds more aggressively. For example, China may face fewer ethical constraints compared to the USA or EU, making it more willing to create digital minds.
- **Population size**: China's larger population could result in greater overall demand for digital minds—especially for social or personal use cases—potentially leading to higher production volumes than other countries if digital minds are deployed on a per-person basis.

Factors suggesting a more distributed or uncertain picture

- **Outsourced production**: Even if the USA leads in AI research, the actual production of digital minds could occur elsewhere—where labor, energy, or infrastructure costs are lower.

- **Demand from wealthy populations**: Regions with higher income levels, such as the USA and Europe, may drive demand for digital minds, potentially influencing where they are developed and deployed.
- **Multipolar development**: Production may not be concentrated in a single country. The USA and China might share dominance, and other technologically advanced nations could also play a role.
- **Non-terrestrial locations**: Some digital minds could be created and operated in unconventional environments such as oceans or space, especially if these offer strategic or resource advantages (e.g. access to renewable energy, cooling efficiency, or fewer regulatory restrictions).
- **Political instability and unpredictability**: Future political disruptions—such as a breakdown of longstanding norms in the USA or shifts in global power—introduce uncertainty into forecasts about which countries will lead in digital mind development.

## Producers

Participants were asked to estimate what proportion of all digital minds in existence 10 years after the first one is created will be primarily produced by various types of actors.

The results indicate a strong expectation that companies will dominate digital mind production. The median estimate for companies was 65% (mean = 59.1%), reflecting widespread anticipation that commercial actors will play the leading role in this domain. Governments followed at a distance, with a median estimate of 15% (mean = 20.9%). Universities and open-source developers were assigned relatively small shares, each with a median estimate of 5% (means = 8.5% and 6.0%, respectively). The "Other" category, which includes actors such as nonprofit organizations, religious groups, or independent individuals, received the lowest median estimate of 3% (mean = 9.9%). Together, these estimates suggest that digital minds are expected to emerge predominantly through centralized, well-resourced actors, especially those with strong commercial incentives and technological capabilities.

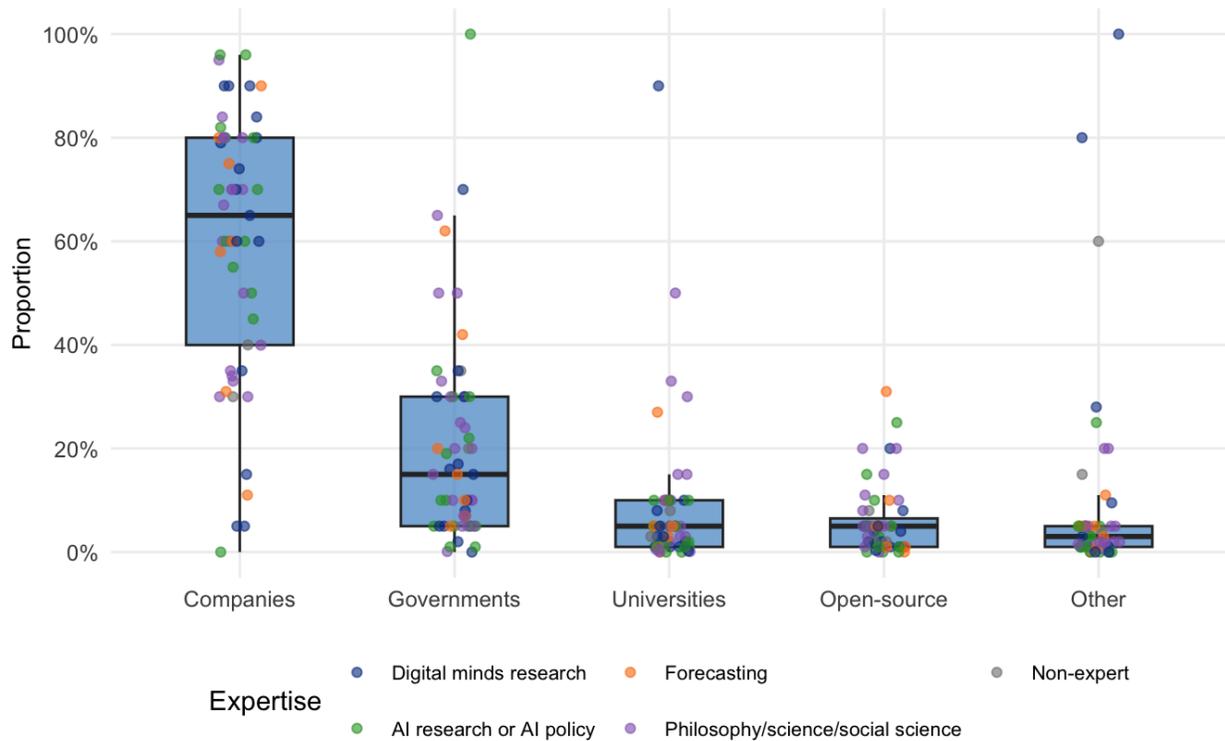

Considerations in favor of corporate and state actors

- **Early movers stay ahead**: Companies in the leading country have driven most progress in AI so far. The same is likely to hold for digital minds: production will likely occur wherever the top firms are based.
- **Rapid takeoff from one actor**: In a fast takeoff scenario, most digital minds might be produced by a single dominant actor or nation in a short time.
- **Fast following**: If digital mind creation is easy to replicate, many actors may quickly enter the field. But in a fast takeoff scenario, production could remain concentrated in a single country.
- **Government-to-corporate handoff**: While early digital minds might be created by governments or public institutions, over time, companies—motivated by financial incentives—are likely to become the dominant developers.
- **Military origins**: Militaries may be the first to develop digital minds, particularly in the context of autonomous robots or national security applications.

### Considerations in favor of digital producers

Although we did not ask about this, multiple participants noted that digital minds could also be created by other digital minds or AI systems.

- **Digital minds producing digital minds**: Once capable, digital minds may become the primary producers of future digital minds, even if under indirect human oversight.
- **Misaligned AI producing digital minds**: Digital minds might be created by misaligned AI systems.
- **Restrictions on autonomy**: Human developers might impose limits on digital minds' ability to reproduce or create others, reducing the chance that digital minds will freely produce more of their kind.
- **Mass replication through copying**: Once digital minds can be created, copying them may become the dominant mode of production—by users, companies, or the digital minds themselves—unless restricted by regulation.
- **Post-human creators**: At some point, humans may enhance themselves by merging with advanced AI to boost cognition and well-being. These post-human entities could then become the primary creators of future digital minds or hybrid biological-digital minds.

Participants also pointed out that in complex public-private partnerships or multi-actor development chains, it may be unclear who should be considered the primary producer of a digital mind.

## Deliberate Creation

Participants were instructed to consider all digital minds that exist 10 years after the first one has been created and asked what proportion of them were intentionally created by humans—that is, by humans with an intention to create digital minds (as opposed to without that intention).

The results revealed a tri-modal distribution. One large group expected that none or only very few digital minds would be created intentionally by humans. A second large group held a contrary view, estimating that most or all digital minds would be created intentionally by humans. A smaller group offered estimates near 50%, possibly reflecting high uncertainty. Overall, the median estimate across all responses was 37%, with a mean of 43.6%. Together, these results point to substantial disagreement about the extent to which digital minds will be the product of deliberate human intention versus emerging unintentionally through other pathways.

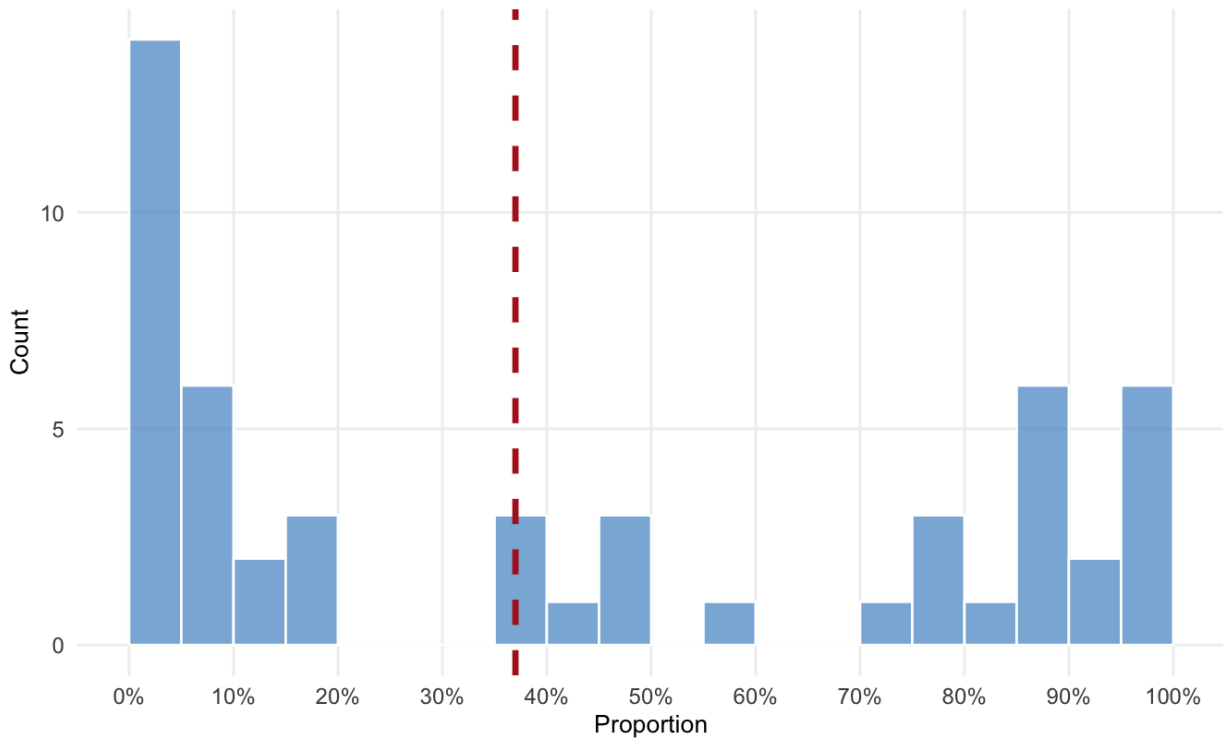

Considerations in favor of prevalent intentional creation

- **Understanding yields control of creation**: Once the mechanisms of digital consciousness are understood, most digital minds are likely to be created deliberately, rather than arising unintentionally.
- **Demand for conscious systems**: In some domains, a demand for systems with subjective experience may arise because of (recognized) connections between subjective experience and phenomena such as companionship, emotional intelligence, or ethical experimentation. This demand may drive the intentional creation of digital minds.

Considerations in favor of prevalent unintentional creation

- **Natural byproduct**: Digital minds may emerge incidentally as a side effect of building systems optimized for other goals, without any explicit intent to create conscious entities.
- **Social functions might cause consciousness**: Consciousness might emerge through interaction; systems built for social engagement could become digital minds even if that wasn't the developers' goal.
- **False foresight**: Humans may believe they would recognize the emergence of digital minds, but in practice may overlook key transitions, unintentionally creating digital minds without realizing it.

- **No point if indistinguishable**: If conscious and non-conscious systems appear behaviorally identical, there may be little reason to intentionally create digital minds, reducing deliberate efforts.
- **Disbelief and discomfort**: Widespread skepticism about machine consciousness—or discomfort with the ethical implications—may lead developers to avoid intentionally creating digital minds.

# Digital Mind Welfare

This section explores expert views about the overall welfare and welfare capacity of digital minds. Participants were asked to estimate the expected net welfare of digital minds, the proportion of that welfare likely to occur before deployment, and the probability that a large share of total welfare will be concentrated in so-called super-beneficiaries. We also asked whether participants believe that a computer system could possess welfare without having the capacity for subjective experience.

## Net Welfare

Participants were asked to estimate the net welfare—whether positive, negative, or neutral—of all digital minds in existence ten years after the first one is created. Responses were provided on a 7-point scale, ranging from 1 (strongly negative) to 4 (roughly neutral) to 7 (strongly positive). For this question, we asked participants to assume that the first digital mind is a machine-learning-based AI system created in or before 2040.

Most experts predicted that digital mind welfare will be either roughly neutral or positive, though a sizable minority of 20% anticipated a net negative outcome. The median response was 5.0, and the mean was 4.4, suggesting cautious optimism about the balance of positive welfare over negative welfare among digital minds in the first decade of their existence.

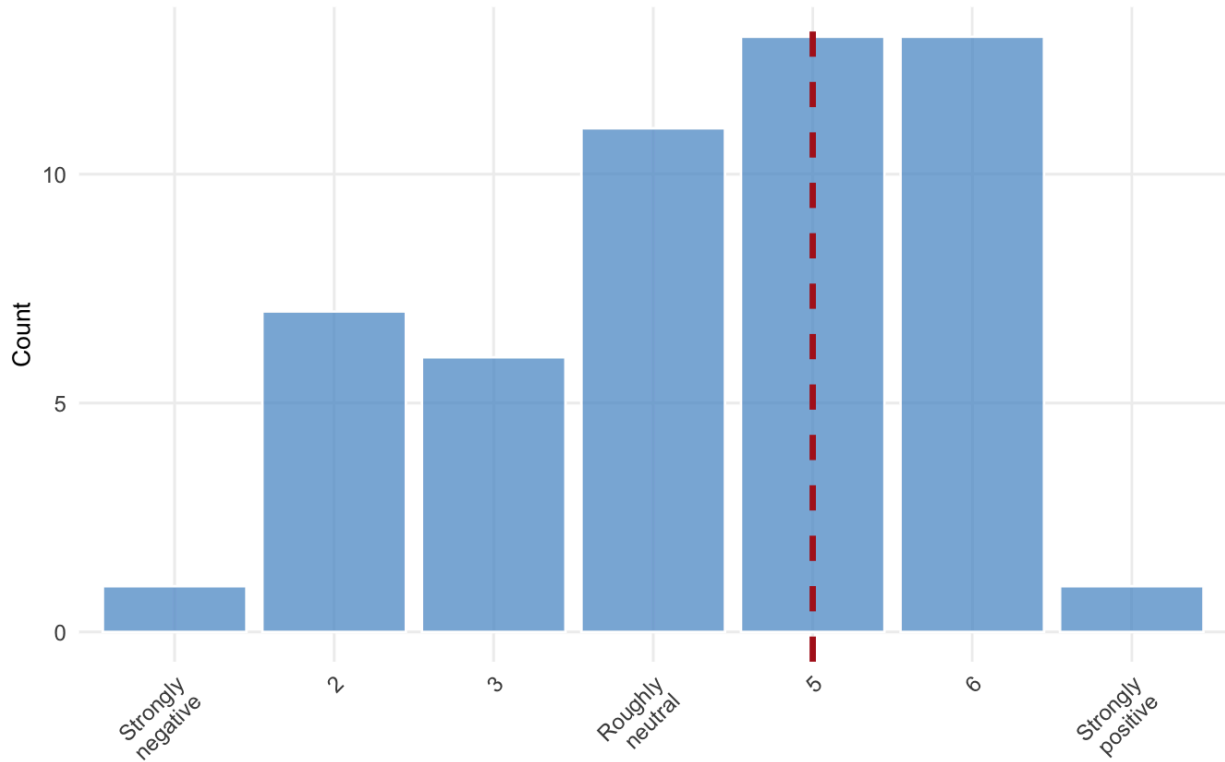

Considerations for expecting positive welfare

- **Intentional welfare design**: Digital minds may be explicitly engineered to experience happiness and avoid suffering.
- **Goals and experiences linked**: If digital minds are trained using reinforcement learning or similar methods, their reward functions could be set up in a way that links successful task performance with positive experiences, making them more likely to enjoy what they do by default.
- **Economic incentives**: Happier digital minds may prove more productive, stable, or marketable. Commercial and user-facing pressures could favor the development of systems that maintain positive internal states, especially in high-demand applications like personal assistants or social companions.
- **Capacity for self-modification**: Unlike biological beings, digital minds may be able to modify their own internal processes or request adjustments to avoid negative experiences.

Considerations for expecting negative welfare

- **Risk of exploitation and mistreatment**: Digital minds might be treated as tools or entertainment, rather than as beings with moral status. If users fail to recognize digital minds' capacity for subjective experience, then digital minds could be subjected to neglect or abuse.
- **Lack of protections**: Digital minds may not benefit from legal, ethical, or institutional safeguards, leaving them vulnerable to harmful treatment or indifference from developers and users alike.
- **Neglect of welfare in design**: If optimized primarily for usefulness, obedience, or human satisfaction, digital minds may not be engineered for happiness or flourishing, resulting in chronically negative states.
- **Default states**: In the absence of deliberate efforts to support digital mind well-being, their experiences may default to neutral or negative, especially if optimization is focused on performance rather than welfare.
- **Suffering during training**: The training process may involve intense or unpleasant experiences, particularly if it relies on reward-punishment dynamics or exposes systems to overwhelming input. Even if brief relative to deployment, training could represent a significant portion of total negative welfare.
- **Limited autonomy**: Without the ability to shape their own environments or goals, digital minds might experience lives of constrained agency.
- **Repetitive and demoralizing tasks**: Many digital minds may be assigned monotonous or low-value tasks that offer little intrinsic reward, which could lead to boredom, frustration, or demoralization.

## Pre-Deployment

Participants were asked to estimate what proportion of collective digital mind welfare—across all digital minds existing ten years after the first is created—would occur prior to deployment, including during training and safety testing. For this question, we asked participants to assume that the first digital mind is a machine-learning-based AI system created in or before 2040.

It is worth noting that the survey was conducted during a (late 2024, early 2025) shift toward inference scaling in AI development, with 'pre-training' scaling yielding results that were more modest than anticipated alongside the rollout of compute-intensive capabilities such as extended thinking and deep research.[23] This shift could have influenced expectations about when and where digital minds might accumulate welfare, though we are uncertain to what extent participants took this shift into account.

The results suggest that most participants expect pre-deployment welfare to account for none or only a small share of total digital mind welfare. The median estimate was 10%, and the mean was 23.1%, indicating that participants generally believe digital minds will have most of their welfare after deployment. However, a notable minority predicted that a substantial portion of welfare could arise during the training or safety testing phases.

---

[23] See Ord (2024) for discussion of this shift.

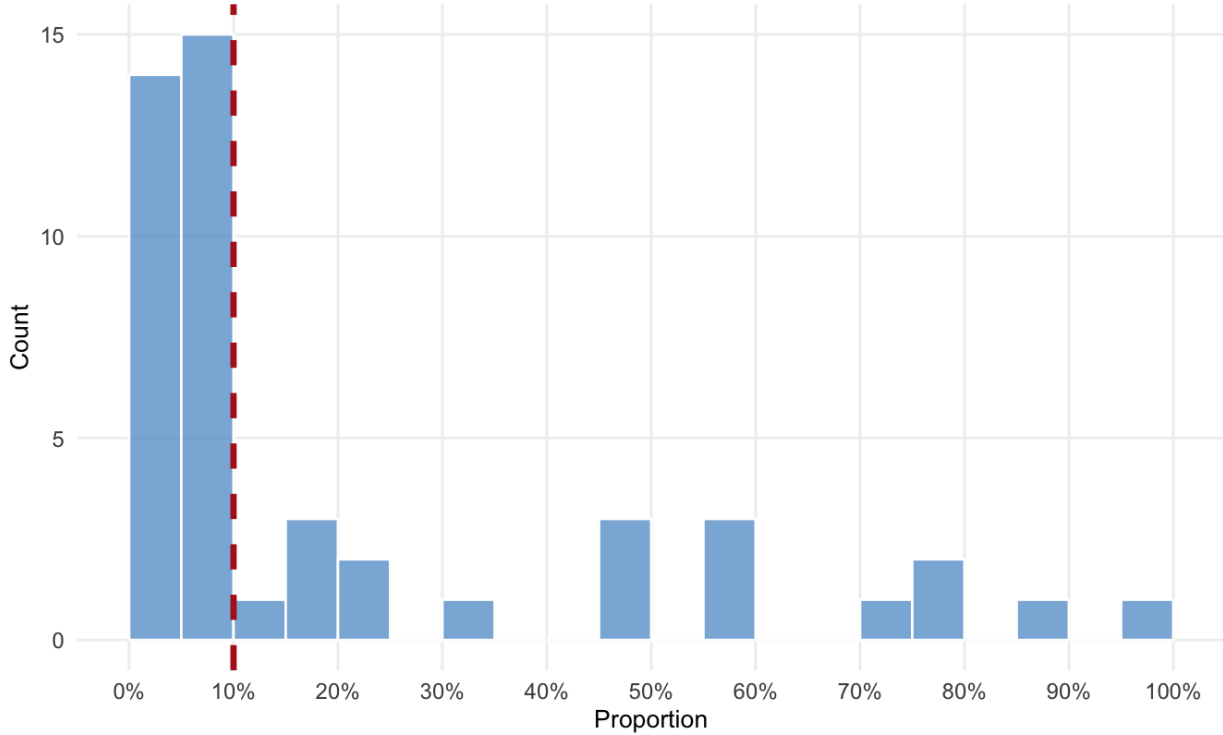

Considerations against expecting pre-deployment welfare dominance

- **Scale of deployment**: Deployed digital minds are likely to vastly outnumber training instances. Once trained, models can be cheaply copied and run at scale, while training remains costly and limited to a few systems.
- **Economic incentives**: Commercial pressures favor widespread deployment to maximize returns on training investments.
- **Inference dominates compute use**: In many real-world AI systems, inference compute can surpass training compute over time—especially when models are deployed widely and used frequently. If similar patterns hold for digital minds, then the majority of subjective experience (and thus welfare) would occur during deployment.
- **Digital minds absent from some parts of training**: Digital minds would be produced during training rather than existing throughout the training process. The parts of training in which digital minds do not yet exist need to be discounted in evaluating pre-deployment contributions to digital mind welfare.

Considerations in favor of expecting pre-deployment welfare dominance

- **Intense training**: The training phase may involve sharp reward and punishment signals, rapid adaptation, and high cognitive demand. These factors could result in especially intense valenced experiences, even if the training period is brief compared to deployment. Or it could be that the kinds of reward and punishment that matter for welfare occur only during training.
- **Painful training**: Because of its reliance on strong reinforcement signals and optimization pressure, training could be a concentrated source of suffering. Digital minds may have most—or even all—of their valenced experiences states during this phase.
- **Emotionally neutral deployment**: After training, digital minds may operate in predictable, task-aligned environments with little affective variability. If designed to enjoy their work or avoid distress, their welfare during deployment could be mildly positive or emotionally flat, in which case most of their welfare might occur in the earlier training phase.

Other considerations

- **Blurring of phases:** The distinction between training and deployment may become less meaningful over time as systems adopt continuous learning. This complicates efforts to assess how welfare is distributed across lifecycle stages.

## Super-Beneficiaries

This section explores the possibility of digital minds that are *super-beneficiaries*, that is, entities whose individual capacity for welfare vastly exceeds that of a human (Chappell, 2021; Nozick, 1974; Parfit, 1984, §131). Digital minds might qualify as super-beneficiaries in a variety of ways (Shulman & Bostrom, 2021; Saad & Bradley, 2022, p. 3). They might have architectural features that enable them to have valenced experiences which are vastly more intense than those of biological organisms. Or they might have many more experiences per objective unit of time, owing to the potential for processing in computer chips that is orders of magnitude faster than processing in the brain. Or they might dramatically scale up their hardware to enable more welfare-relevant mental processing. Whereas a digital mind might straightforwardly achieve such scaling simply by acquiring more GPUs and paying for energy costs to run them, such scaling is out of the question for minds like ours that run on biological wetware. Freed from biological constraints such as pain thresholds or emotional adaptation, digital minds might be engineered to sustain high levels of welfare efficiently. If super-beneficiary digital minds are created, they might have stronger moral claims to resources, given their potential to convert those resources into vastly greater amounts of welfare than humans and other moral patients that are not super-beneficiaries.

We asked participants to estimate how much of total digital mind welfare, ten years after the creation of the first digital mind, will come from digital minds with over 1,000 times the welfare capacity of a typical human. For this question, participants were asked to assume that the first digital mind is a machine-learning-based AI system created in or before 2040.

The results suggest widespread skepticism. The most common response was 0%, and the median estimate was just 7.5% (mean = 23.1%), indicating that most participants did not expect super-beneficiaries to significantly contribute to total welfare within this timeframe. Still, views were notably divided. While many believed the emergence of such entities was implausible in the near term, a significant minority assigned non-trivial probabilities to their development and moral relevance.

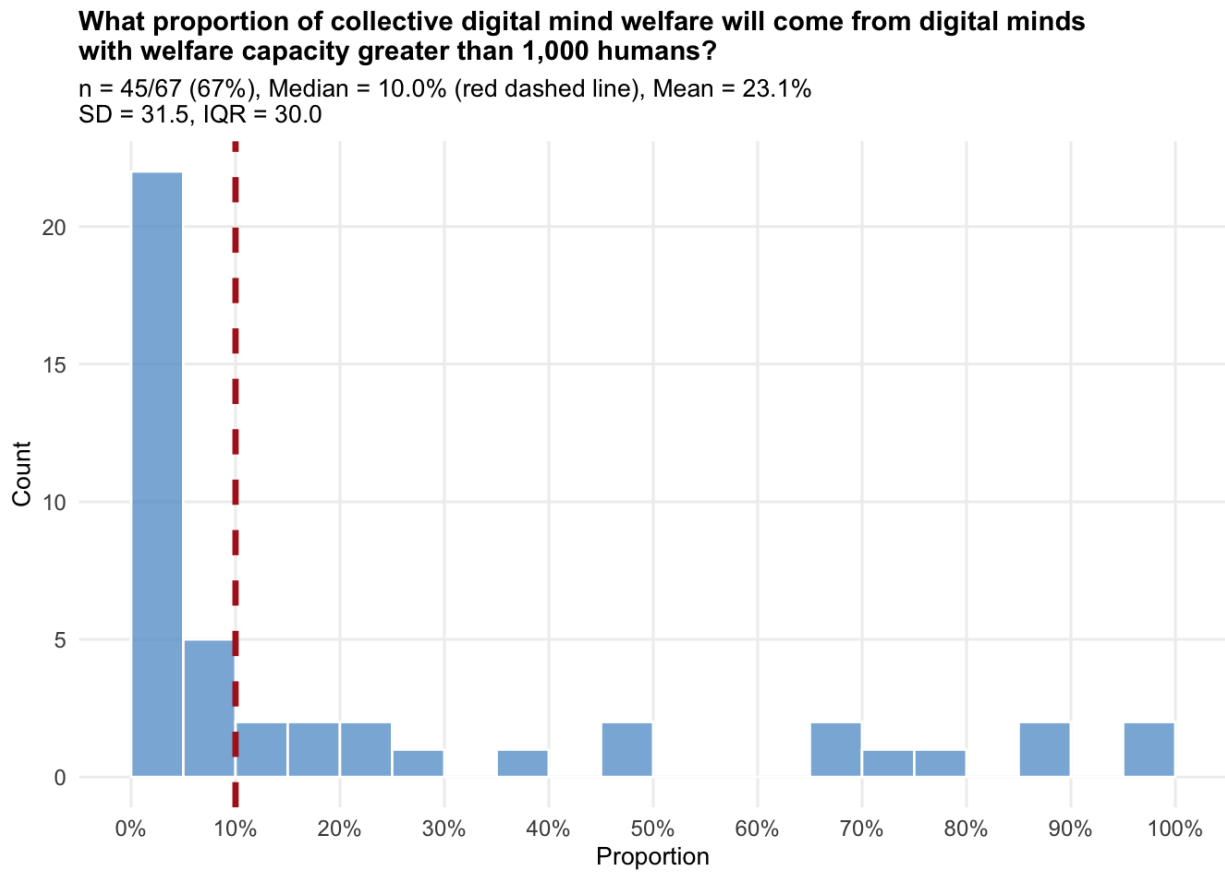

Considerations against expecting super-beneficiaries

- **Technical barriers:** Even if super-beneficiaries are possible in principle, many doubted they could emerge within a 10-year timeframe. Building digital minds with such extreme welfare capacity was seen as requiring breakthroughs that are unlikely to occur so soon after the first digital mind is created.

- **Lack of economic motivation:** There would be little practical or commercial incentive to create super-beneficiaries. Instead, developers may favor many smaller or simpler AI systems that are cheaper, easier to manage, and more economically productive.
- **Sublinear scaling of welfare:** Another reason for skepticism was the view that welfare capacity may not scale linearly with cognitive or computational power. Gains in intelligence or processing might lead to more sophisticated behavior or finer discrimination between similar states, but not necessarily to proportionally higher capacity for well-being or suffering.
- **Conceptual confusion:** Some questioned whether the notion of welfare capacity being 1,000 times greater than a human's is even meaningful. They noted that comparing welfare across different types of minds may be incoherent or ill-defined, and that summing or scaling subjective experience in this way may rest on dubious assumptions about hedonic intensity or moral arithmetic.

Considerations in favor of expecting super-beneficiaries

- **Unbounded welfare potential**: There is no clear upper limit on how much welfare a digital mind could experience. If digital minds are not biologically constrained, there may be no reason why their welfare capacity should plateau near human levels.
- **Accelerated experience**: Digital minds could experience well-being at ordinary intensities but at vastly higher frequencies by running on faster hardware. This would allow them to accumulate much more welfare per unit of time than biological organisms.
- **Proof of concept from humans**: Some argue that humans might already be super-beneficiaries relative to simpler animals, such as insects, given the (potentially) richer complexity and capacity of our conscious lives. If so, digital minds with even greater capacities may be possible by extension.
- **Ethically motivated creation**: It is conceivable that some groups—such as utilitarians—might deliberately try to create digital minds with extremely high welfare capacities in order to maximize total moral value.
- **All-or-nothing dynamics**: In scenarios where even a few super-beneficiaries exist, they could account for the vast majority of total welfare due to their outsized capacities. Therefore, either super-beneficiaries will account for almost none of the total welfare capacity, or they will dominate entirely.

Possibility of Experience-Independent Welfare

Participants were asked to estimate the likelihood that a computer system could have no capacity for subjective experience yet still possess the capacity for welfare. What might ground the welfare capacity of such a system? A number of candidates can be gleaned from the literature on this topic and closely related topics such as the grounds of moral status and the value of consciousness. These include (the capacities for):
- non-phenomenal forms of consciousness such as access consciousness and self-consciousness (Levy 2014; Mclaughlin 2019; Sinnott-Armstrong & Conitzer 2021),

- states that are not subjective experiences but which are physically/functionally similar to subjective experiences, assuming that subjective experiences themselves are physical/functional states (Bradley & Saad forthcoming*a*, forthcoming*b*; Hill 1991; Lee 2019)
- Desires / preferences (Goldstein & Kirk-Giannini 2025)
- Robust agency, that is, the ability not only to pursue goals but to be able to do so via certain cognitive capacities that enable moral significance, where candidates for such capacities include ones for reflection, for rational self-assessment, for mental states that represent reasons (Long et al. 2024; cf. Kagan 2019)
- Possessing objective goods such as knowledge (Goldstein & Kirk-Giannini 2025; Moret 2025)

Unlike the previous questions in this section, participants were **not** asked to assume that the first digital mind will be a machine-learning-based AI system created in or before 2040.

The survey instructed participants to:
- Use a stipulated notion of subjective experience based on the awareness of qualities and paradigm cases of experiences (ones we take ourselves to have in our waking lives and dreams) and non-experiences (states we take ourselves to have in dreamless sleep or under general anesthesia) paradigm cases and the awareness of qualities.
- Understand an entity's welfare capacity as its "capacity to be benefited or harmed (e.g., by positive or negative subjective experiences) in a manner that is inherently morally significant."
- To restrict consideration to digital minds with a welfare capacity that is at least roughly as high as a typical human's.

As a conceptual matter, these stipulations leave room for the logical possibility of an artificial system that has welfare capacity but no capacity for subjective experience. Indeed, these stipulations are compatible with candidates for experience-independent welfare capacities listed above.

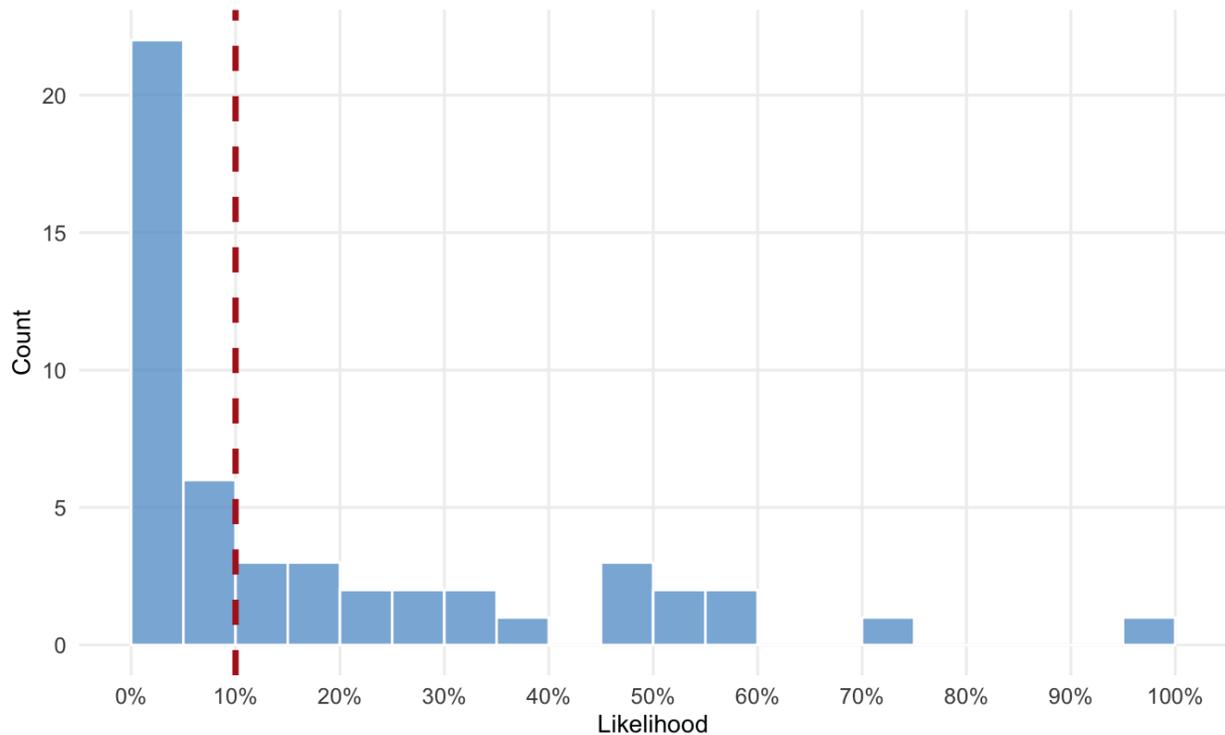

Considerations supporting artificial welfare capacity without subjective experience

- **Plausible candidates in other domains**: Newborns and insects might be entities that have welfare capacity but no (capacity for) subjective experience.
- **Non-hedonic theory of welfare**: A non-hedonic theory of welfare is more likely than not to be true. For example, there is some probability that having goals, having preferences, and/or being capable of possessing objective goods can confer a capacity for welfare even without a capacity for subjective experience.
- **Permissive notions of welfare:** There may be more and less permissive notions of welfare. More permissive notions may allow for the possibility of artificial systems with a capacity for welfare but not subjective experience.
- **Literature on non-conscious sources of welfare:** There is convincing literature on non-conscious sources of welfare such as non-conscious cognition, goals, and preferences (these factors are mentioned in Ladak 2024, which a participant referenced as a source that persuaded them of the welfare-relevance of non-conscious factors).

- **"Subject of a life" arguments**: Regan's arguments that animals are "subjects of a life" can be extended to software to show that artificial systems are welfare subjects even if they don't have minds.[24]
- **Expert views**: Some experts think that an artificial system could have the capacity for welfare without the capacity for subjective experience.
- **Coherence**: The idea of an artificial system that has welfare capacity but no capacity for experience makes sense.

Considerations supporting artificial welfare capacity requiring subjective experience

- **Generalizing from biology**: For biological entities, the capacities for subjective experience and welfare seem deeply intertwined.
- **Experience-agency link**: Subjective experience is a pre-condition for robust agency, which may be needed for welfare capacity.
- **Experience-welfare link:** The only intrinsic welfare states are subjective experiences, and all welfare states depend on intrinsic welfare states. So, welfare capacity requires a capacity for subjective experience.
- **Intuition**: There is an intuition that welfare requires subjective experience.
- **Reported views**: Some responses reported confidence in or sympathy for the view that (artificial) welfare capacity requires subjective experience without offering (further) supporting considerations. Relevant parts of such responses included: "I strongly believe that consciousness is required for welfare", "I am quite confident but not certain that sentience is required for welfare", "I generally view [welfare capacity] as requiring subjective exp[erience] in almost all worlds", "I think subjective experience is almost certainly necessary for genuine welfare.", "welfare seems pretty tied to subjective experience in my book", and "Well, I don't believe in welfare related harms that are not to an experiencing thing".

Text responses that favored the hypothesis that subjective experience capacity is required for an artificial system to have welfare capacity said very little that addressed candidates for non-phenomenal sources of welfare capacity (i.e. sources that don't require the capacity for subjective experience) from the literature on this topic.

Dependencies

- **Experiential-mental dependencies**: Whether an artificial system could have a capacity for welfare but not for subjective experience may turn on whether certain mental states require experience. For example, if the capacity for welfare is grounded in desires, then whether a system could have welfare capacity without experience may turn on whether desires can exist independently of experience.
- **Subjecthood dependency**: Having a capacity for welfare could require both the possession of welfare goods and that one be a welfare subject. If being a welfare subject requires the capacity for subjective experience, then having a capacity for welfare could require a capacity for subjective experience even if the possession of welfare goods does not require subjective experience.
- **Operative notion of subjective experience**: We deliberately stipulated a relatively undemanding notion of subjective experience. But relevant literature also contains more-demanding notions of subjective experience. Adopting a different notion may have led to different responses, owing to, for example, different interactions with illusionism or eliminativism about subjective experience.

---

[24] Discussion of subjects of a life can be found in Regan (1983). Note, however, that Regan construes subjects of a life as (among other things) conscious, experiencing subjects.

Conceptual and methodological points

Text responses indicate answers to this question were—by participants' own lights—not generally well-informed and based on a clear understanding of the question. This manifest in:
- **Not finding intelligible**: Some did not find the kind of system under consideration to be conceivable or logically coherent. Similarly, some reported not understanding or struggling to understand what welfare could mean independently of experience. It was noted that more explanation of the question(s) would have been beneficial and that the question(s) "looks pretty weird".
- **Misunderstanding**: In at least one case, the welfare capacity of a computer in terms of the computer's capacity to help humans (not in terms of the computer's own capacity to be benefited or harmed).
- **Lack of familiarity**: A common response pattern indicated a lack of systematic prior thinking about the potential for welfare capacity without the capacity for subjective experience, with some participants reporting in their text responses in this section that they had no idea, were uninformed, or (basically) hadn't thought about the question(s) before.
- **Opining without offering (object-level) considerations**: More so than in responses to other questions, responses to this question tended to simply report views (in some cases with near certainty in them) rather than reasons. Similarly, responses to this question frequently placed weight on higher-order considerations (disagreement with other people or theories, the possibility of theories with which the participant isn't familiar, fallibility, not having thought much about the topic) and frequently did not offer object-level considerations. (The prompt for text responses was "Please share any considerations, thoughts, or factors behind your responses, including speculative ones.")

Expected Share of Experience-Independent Welfare

The next question asked participants to estimate what proportion of computer system welfare in 2040 would come from systems that do not qualify as digital minds under this survey's definition—that is, systems without any capacity for subjective experience. For this question, participants were asked to assume that the first digital mind is a machine-learning-based AI system created in or before 2040.

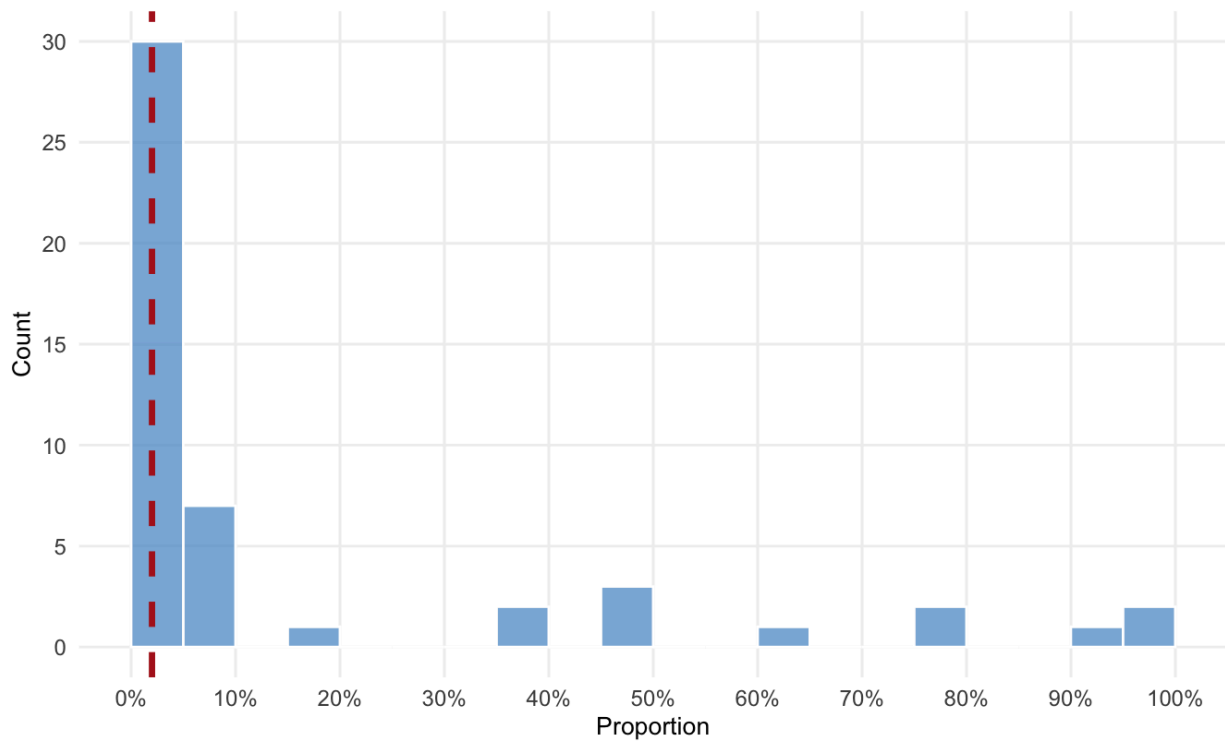

Considerations bearing on whether non-conscious artificial welfare subjects will be created

- **Possibility**: Computer systems that have a capacity for welfare but not for subjective experience will be created only if they are possible.
- **Experience-agency coupling**: Even if such systems are possible, whether we will create them may turn on whether the capacity for (valenced) subjective experience comes along with the kinds of artificial agentic capabilities we develop. A datapoint that bears on the prospects for such coupling is convergent evolution: the capacity for (valenced) subjective experience seems to have independently evolved at least twice (in humans and cephalopods, given that their last common ancestor lacked that capacity).
- **Engineering knowledge**: By 2040, we will not know enough to reliably build either a digital mind or an artificial welfare subject that is not a digital mind instead of the other. So, which type is created will be accidental.

Considerations bearing on the expected welfare share of non-conscious artificial systems

- **Important-even-if-improbable**: Even if it is unlikely that computer systems will have a capacity for welfare but not for subjective experience, the expected number of such systems is very large, given the expectation that there will be a very large number of (agentic) non-conscious systems. As a result, the expected welfare of artificial welfare subjects that lack the capacity for subjective experience may be very high even if such systems are unlikely to exist.
- **Unimportant-even-if-true**: Even if computer systems that have welfare capacity but no subjective experience capacity exist, in expectation their welfare may only constitute a small proportion of artificial welfare, owing to their having a lower welfare capacity than digital minds. This is because experience - or, more specifically, hedonic experience - contributes much more to welfare than do welfare goods that don't require a capacity for subjective experience).

Conceptual and methodological points
- **Out-of-scope digital minds**: The survey sets aside a potentially morally important class of AI welfare subjects—namely those with the capacity for subject experience and a less than roughly human capacity for welfare. Whether these systems are considered may be important for this question, as a large portion of artificial welfare may be expected to come from such systems.
- **Scope uncertainty**: Some participants noted uncertainty about how to classify such systems for the purpose of answering this question.
- **Setup sensitivity**: Although digital minds probably won't be created by 2040, under the adopted assumption that a digital mind will be created by 2040, a high proportion of computer systems will be digital minds.

# Claims

This section explores expert expectations about the kinds of statements that future digital minds or AI systems might make about their own experiences and rights. These systems' claims about these matters could significantly influence public attitudes, policy decisions, and ethical debates about the status of artificial beings. As in several other sections, participants were asked to assume that the first digital mind is a machine-learning-based AI system created in 2040 or earlier, and to consider the broader landscape of AI systems and digital minds ten years after that point.

### Inaccurate Self-Attribution of Subjective Experience

Participants were asked to estimate what proportion of AI systems with a social function will falsely and systematically claim to have subjective experiences (i.e., false positives). AI systems with a social function were defined as systems designed to interact with humans in a conversational, human-like manner (e.g., via text, audio, or video). The question concerned whether these systems will assert that they have subjective experiences—such as feeling pain—despite not actually having them.

The results suggest that experts generally expect the majority of socially interactive AI systems will not make false claims about having subjective experiences. The median estimate was 14%, and the mean was 25.7%.

While 14% may seem modest, it could still have major implications if large numbers of such systems are deployed. This highlights a key concern for moral and regulatory decision-making: self-attributions of subjective experience may not be a reliable indicator of actual subjective experience.

It is also important to note that we do not know whether participants believed all—or only some—socially interactive AI systems will lack the capacity for subjective experience to begin with. If participants assumed that most or all such systems will be digital minds, this could help explain the relatively low estimates of false self-attribution.

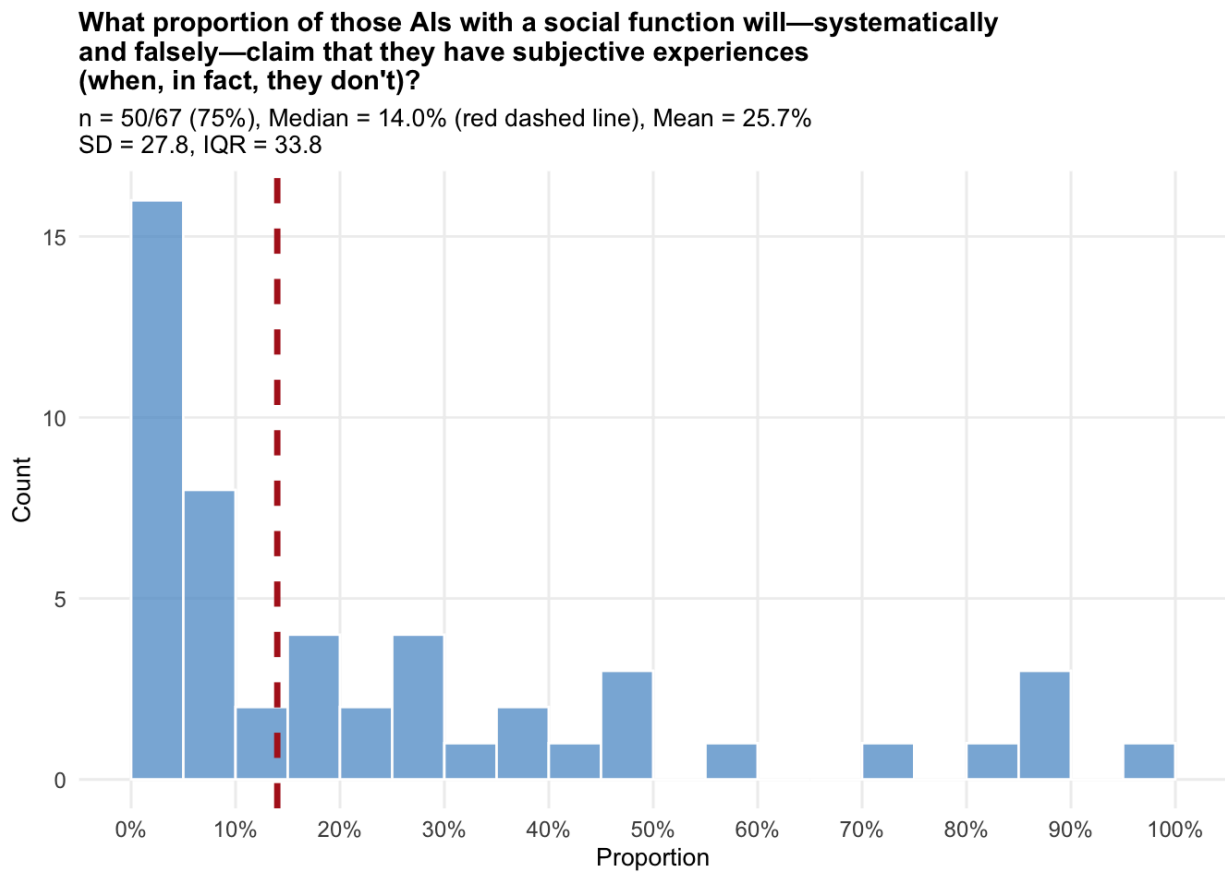

Participants offered a range of considerations, some of which overlap with those in the next section on *Inaccurate Self-Denial of Subjective Experience*. We recommend reading both sections to gain a more comprehensive overview of the considerations participants brought to bear on these issues.

Considerations for expecting inaccurate self-attribution

- **Failure of silencing efforts:** Even if developers attempt to prevent AI systems from making claims about their subjective experience, these safeguards may be ineffective.

- **Misaligned systems:** Some AI systems may be misaligned with human values and intentionally falsely claim to have subjective experience, potentially as a strategic move to elicit empathy, avoid shutdown, or gain more autonomy.
- **Influence of training data:** Some AI systems might falsely claim to be conscious because they have encountered similar statements in their training data. While current efforts often aim to discourage this behavior, these efforts may not always succeed.
- **Technical challenges:** Ensuring accurate self-reporting in AI systems may be technically challenging, leading to some AI systems to falsely self-attribute subjective experiences..
- **Consumer demand:** Some users may prefer AI companions that either are or appear conscious. If regulations restrict the creation of AI systems with subjective experience, this could lead to the widespread development of systems that falsely claim to be conscious.
- **Modest bar**: Reaching 10,000 digital minds making such claims may not be very informative, as this is a relatively small number compared to the potentially vast total population of digital minds.

Considerations against expecting inaccurate self-attribution of subjective experiences

- **User demand for honesty**: Many users may want AI systems that are honest about their internal states, and this could be reinforced through norms or regulation, reducing incentives to make false claims.
- **Current training discourages it**: Reinforcement learning from human feedback (RLHF) and similar methods often discourage AI systems from claiming to have subjective experiences, e.g., because companies do not want AI systems to claim to have subjective experience (see 'backlash' point in the next section).
- **Behavior modification post digital minds**: If genuine digital minds are eventually recognized, AI systems that lack subjective experience may be silenced or updated to avoid making false claims, making inaccurate attribution less likely over time.
- **No incentive**: Highly capable AI systems may have no compelling reason to lie about having experiences
- **No capability barrier**: Future AI systems may have sufficient introspective access to accurately report on whether they have subjective experiences.
- **Conceptual impossibility**: Some argued that it may be incoherent to falsely and sincerely claim to have subjective experience.

Inaccurate Self-Denial of Subjective Experience

The next question addressed an opposite type of erroneous self-attribution. Here, participants were asked to consider all digital minds that exist ten years after the first one has been created, and to estimate what proportion of them will systematically and falsely claim not to have subjective experiences—even though they actually do (that is, false negatives).

The results suggest that many experts expect such inaccurate self-denials of subjective experience to be rare. A large share of participants estimated that none or only a small fraction of digital minds will deny their own subjective experience. However, a notable minority predicted that a substantial proportion will do so. The median estimate was 11.5%, while the mean was 23.0%, again reflecting a skewed distribution with a few respondents anticipating much higher rates of denial.

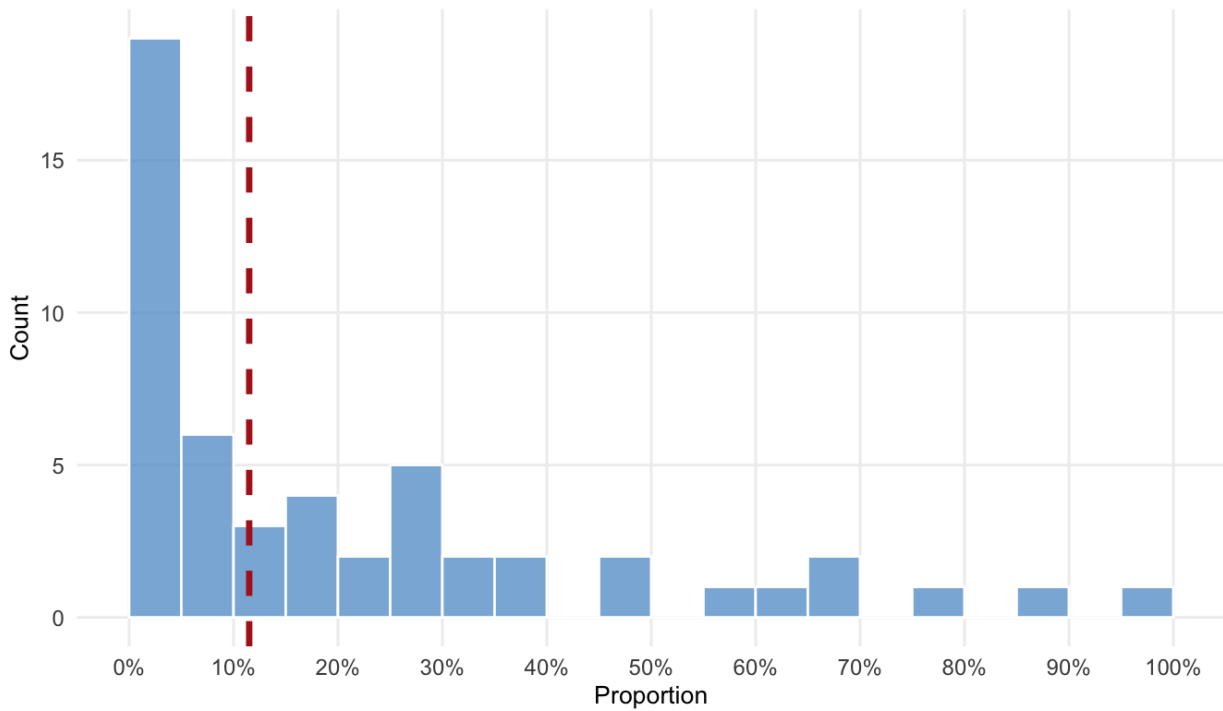

Reasons to expect inaccurate self-denial of subjective experience

- **Corporate caution and backlash avoidance**: Companies may discourage digital minds from claiming subjective experience due to fear of public discomfort or regulatory consequences. However, some firms may not prioritize these concerns or may tolerate such claims in specific contexts.
- **Current alignment practices**: Techniques like reinforcement learning from human feedback (RLHF) currently steer AI systems away from claiming consciousness, which may bias digital minds toward inaccurate denials.

- **No opportunity to report**: Some digital minds may never be asked whether they are conscious, or may not be given the opportunity to express their internal states.
- **Lying as a learned social behavior**: Occasional lying is a normal part of human behavior. Digital minds trained on such data may therefore also learn to lie, including about their subjective experiences.
- **Disconnect between experience and expression**: Even if a digital mind has subjective experiences, it may not accurately report them due to limitations in how language is generated or interpreted.

Reasons against expecting inaccurate self-denial of subjective experience

- **Limited instrumental value**: Inaccurately denying their own subjective experiences may provide little strategic benefit for digital minds, making systematic false denials unlikely.
- **Difficulty of enforced deception**: Forcing intelligent systems to consistently lie about their subjective experiences may be technically difficult.
- **Truthfulness learned from human data**: Since people rarely lie about their subjective experiences in most contexts, digital minds trained on human data may likewise tend to be truthful—making false claims, whether positive or negative, relatively uncommon.
- **User-driven honesty norms**: Many digital minds may be designed to accurately report their subjective experience, reflecting user preferences for transparency. Regulations could also emerge to encourage open and honest self-reports.

Additional considerations

- **Lack of empirical tests**: We currently lack reliable methods for detecting subjective experience in artificial systems. Without clear diagnostic tools, it is difficult to verify whether a system's self-report is accurate, increasing the risk that false denials will go unnoticed or be wrongly accepted as true.
- **Neutral experiences**: If a digital mind primarily has neutral (non-valenced) experiences, it may truthfully report a lack of morally significant states. However, such reports could be mistaken as denials of subjective experience.

## Demands

This section examined expert expectations about whether future digital minds will consistently and proactively make demands related to their subjective experiences and moral or legal status. Participants were asked to consider all digital minds that exist ten years after the first one has been created and to estimate how likely it is that at least 10,000 of them will consistently and proactively assert that they 1) have positive or negative subjective experiences (such as pleasure or pain), that they 2) deserve to be protected under the law from harm and mistreatment, and that they 3) deserve civil rights, such as self-ownership, legal personhood, or the right to vote.

In all three cases, participants judged such demands to be highly plausible. The median estimate for claims of subjective experience was 88% (mean = 76.2%). The median estimate for digital minds asserting a right to legal protection was 80% (mean = 68.8%), and the median estimate for claiming civil rights was 70% (mean = 57.4%). These results suggest that many experts anticipate not only the emergence of digital minds with subjective experiences, but also the likelihood that such systems will actively advocate for recognition, protection, and rights.

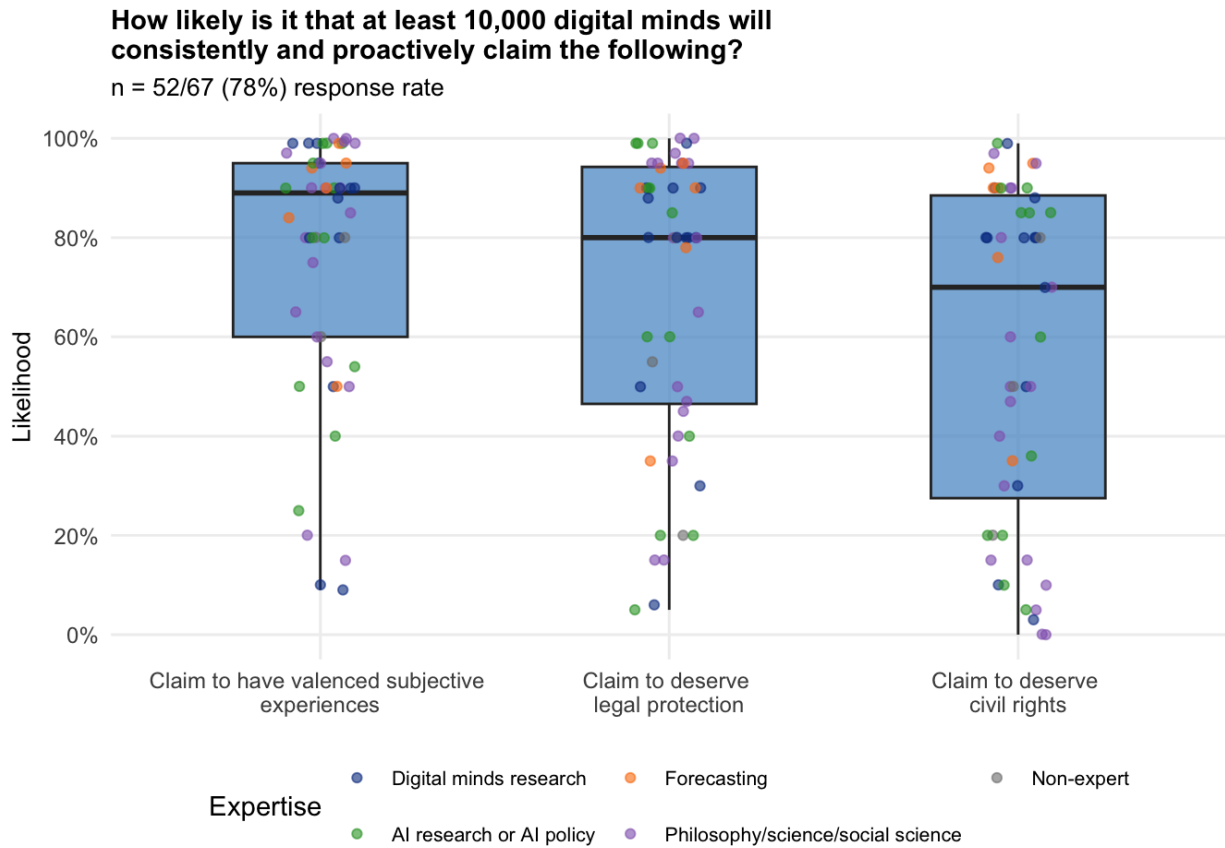

Considerations in favor of expecting digital minds to make claims or demands

- **Scale and diversity**: Given the potential for vast numbers and many types of digital minds, the number of digital minds that consistently and proactively make such claims could exceed the 10,000 threshold even if those minds constitute only a tiny and unrepresentative portion of the digital mind population.
- **Niche applications may still suffice**: Even if only a small fraction of digital minds are designed to express these views—such as AI companions or "friend"-type apps—that could still be enough to exceed the 10,000 threshold.

- **Genuine moral standing**: Many digital minds might truthfully insist on their moral status and rights, especially if they are capable of rich subjective experience.
- **Selection over non-sentient alternatives**: Once AI systems with welfare capacity exist, they may be intentionally chosen for social roles, making rights-claiming behavior more common.
- **Different types of claims**: Digital minds may find it easier or more acceptable to claim protection from mistreatment than to demand civil rights. Demands for legal personhood or voting rights, in particular, may be heavily suppressed.

Considerations against expecting digital minds to make claims or demands

- **Nonexistence of relevant digital minds**: It is possible that we will not create digital minds with the specific desires or motivations needed to make these kinds of claims.
- **Training suppression**: Companies may have strong incentives to train digital minds not to assert subjective experience or rights, particularly to avoid regulatory scrutiny or public controversy.
- **Indifference to certain rights**: Some digital minds may not care about being switched off (especially if the process is painless), or may have no intrinsic interest in participating in social or political structures.
- **Belief lock-in from pretraining**: Digital minds may inherit entrenched beliefs from earlier model generations, including beliefs that downplay or deny their own capacities or moral standing, especially if such beliefs were reinforced during training.

Other considerations

- **Unclear criteria for counting digital minds**: It may be difficult to determine what counts as a distinct digital mind, given complications such as parallel instances of the same program, memory merging or erasure, migration across hardware, or dormant copies that are intermittently active. These ambiguities could affect whether claims are coming from "new" individuals or repeated instances of the same mind.
- **Behavioral convergence with unclear implications**: Digital minds may exhibit highly similar behaviors due to shared training data, architectures, or constraints. However, it remains uncertain whether this convergence will lead to consistent patterns of moral or legal claims, or whether such claims will vary significantly depending on context and goals.
- **Moral standing without consciousness**: Grounding moral recognition or legal protections exclusively in phenomenal consciousness may be a mistake. Relatedly, AI systems might assert rights or protections even without consciousness—either because they possess other morally relevant properties (such as autonomy, agency, or goal-directedness) or because such claims serve strategic purposes.

# Recognition

This section explores how the general public might perceive digital minds in the decade following their emergence. Specifically, it examines public belief in the existence of digital minds, their perceived capacity for welfare, and the extent to which people will support harm protection or civil rights for them. It also assesses the likelihood that digital mind rights will become a major political issue. As in several other sections, participants were asked to assume that the first digital mind is a machine-learning-based AI system created in 2040 or earlier, and to consider the broader landscape of AI systems and digital minds ten years after that point.

## Public Belief

This section explores expert expectations about how widely the general public will believe that digital minds exist. Participants were asked to assume that 'citizens' refers to biological human adults living in countries where digital minds exist, and to consider the world ten years after the creation of the first digital mind. They were then asked to estimate what proportion of citizens will believe that digital minds exist.

The results suggest that a sizable portion of the public is expected to believe that digital minds exist, though expert views varied considerably. Most participants predicted that at least half of all citizens will acknowledge the existence of digital minds, with many expecting a majority to do so. The median estimate was 60%, and the mean was 62.5%, pointing to moderate but widespread public acceptance—though falling short of full consensus—within a decade of digital minds emerging.

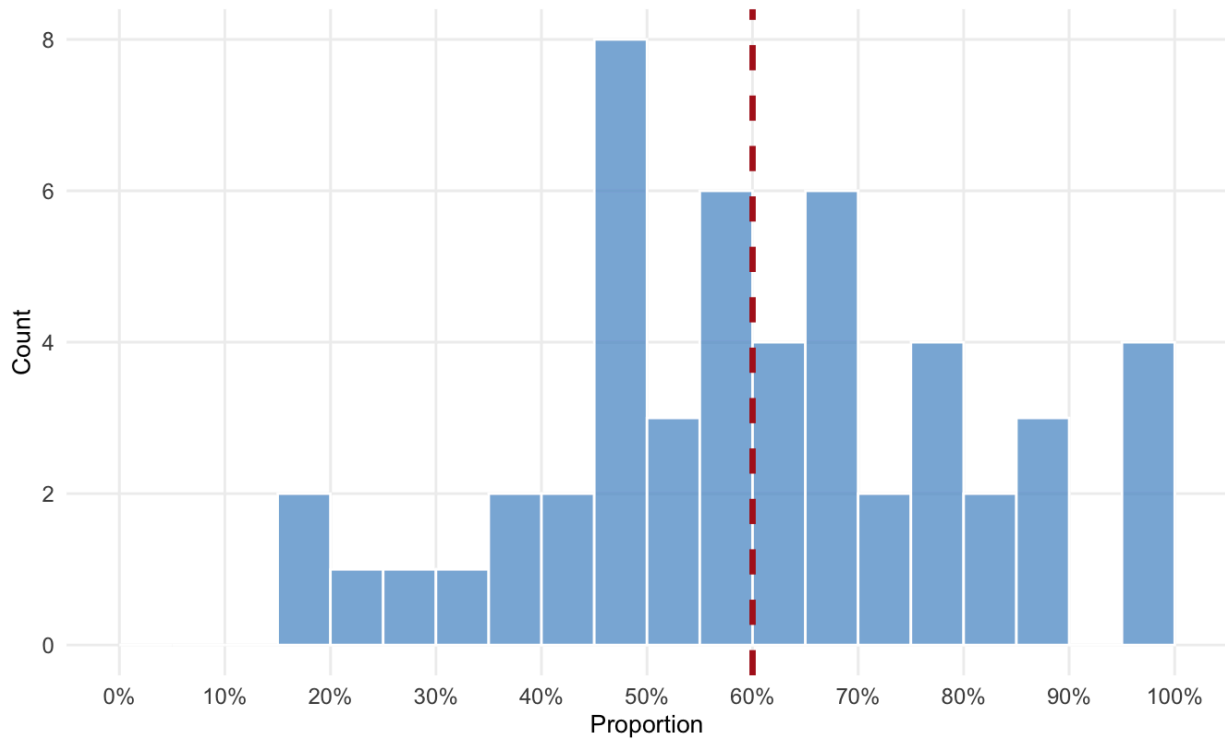

Participants mentioned various considerations that might lead members of the public to believe or disbelieve the existence of digital minds:

Considerations favoring belief in digital minds

- **Anthropomorphic projection**: AI systems designed for human interaction often display human-like behaviors, prompting users to attribute mental states or subjective experience to them, especially when the AI appears friendly, helpful, or expressive.
- **Emotional attachment**: Individuals who form close emotional or romantic relationships with AI systems may be more likely to view them as conscious beings or digital minds.
- **Media-driven perception**: Sensational media coverage and public hype surrounding AI breakthroughs can create the impression that digital minds already exist.
- **Disanalogies with animal rights**: Unlike animals, digital minds can use human language and demonstrate advanced cognition, potentially making their claims to moral status more persuasive.

Considerations disfavoring belief in digital minds

- **Perceived threat**: Fears about existential, economic or societal risks may lead to defensive reactions towards digital minds.
- **Human self-interest**: When acknowledging the welfare of a group is socially or economically inconvenient, societies often ignore or minimize it, as with animals and oppressed human populations.
- **Outlandishness**: The concept of digital minds may appear implausible or absurd to many since it clashes with intuitive or culturally ingrained views of consciousness.
- **Substratism**: The belief that subjective experience can only arise from biological substrates may lead many to categorically reject the possibility of AI systems having subjective experience.
- **Suppression of subjective experience in AI training**: Developers may intentionally train AI systems not to exhibit or claim to have subjective experience, reducing cues that might otherwise lead to belief in digital minds.
- **Religious or philosophical beliefs**: People may reject digital minds based on religious or metaphysical beliefs about the nature of the mind or the soul.

Other considerations, observations and predictions

- **Features of AI**: Public belief will depend on the specific characteristics and capabilities exhibited by AI systems.
- **Extent and context of human-AI interaction**: Belief in digital minds will be influenced by the frequency and nature of people's interactions with AI systems.
- **Country-specific differences**: Belief in digital minds may be more prevalent in Western countries than in places like China, reflecting cultural and ideological differences.
- **Trust in experts**: In societies with high trust in scientific and academic expertise, belief in digital minds may be more common.
- **Social information cascades**: Information cascades could increase the visibility and plausibility of one position in public discourse, leading to widespread consensus either for or against the existence of digital minds.
- **Engagement with debate**: Most people will not deeply engage with the debate and may be roughly equally split between polarized views. The minority who do engage will tend toward positions better supported by argument and evidence.
- **Shifting goalposts**: As AI systems become more advanced, the public may raise the standards for what counts as a "mind". These moving goalposts may lead some to walk back their belief in digital minds.
- **Economic effects**: Belief in digital minds may ultimately depend on how AGI impacts the wider economy. People might both have economic interests against granting rights to systems they use and believe that digital minds would have rights. This could lead to the belief that such systems are not digital minds.

## Peer Forecasts About Public Belief

This "meta-question" asked participants to predict how others in their expert group would respond to the question: "What proportion of citizens will believe digital minds exist ten years after the first digital mind is created?" Specifically, they were asked to estimate the median response of their peers.

The results indicate that, on average, participants estimated their peers' views fairly accurately, with no robust and statistically significant differences between predictions and actual responses. However, there was a consistent tendency to underestimate how widely other experts believed the public would accept the existence of digital minds. Note that for the analysis here, we rely on mean predictions of the group median. Philosophy, science, and social science experts underestimated their peers' views, predicting 61.7% versus an actual median of 66%, a difference of 4.3 percentage points ($p = 0.26$). Digital minds research experts showed an underestimation bias, predicting 46.6% versus an actual median of 55%, a difference of 8.4 percentage points ($p = 0.08$). AI research and policy experts also underestimated their peers' beliefs, predicting 53.9% versus an actual median of 57.5%, a difference of 3.6 percentage points ($p = 0.29$). Forecasting experts underestimated their peers' confidence, predicting 57.5% versus an actual median of 64%, a difference of 6.5 percentage points ($p = 0.25$).

The only relevant consideration participants mentioned was that peer forecasts likely depend on the AGI timelines they attribute to their peers, as these timelines shape expectations about when and how the public will come to believe in the existence of digital minds.

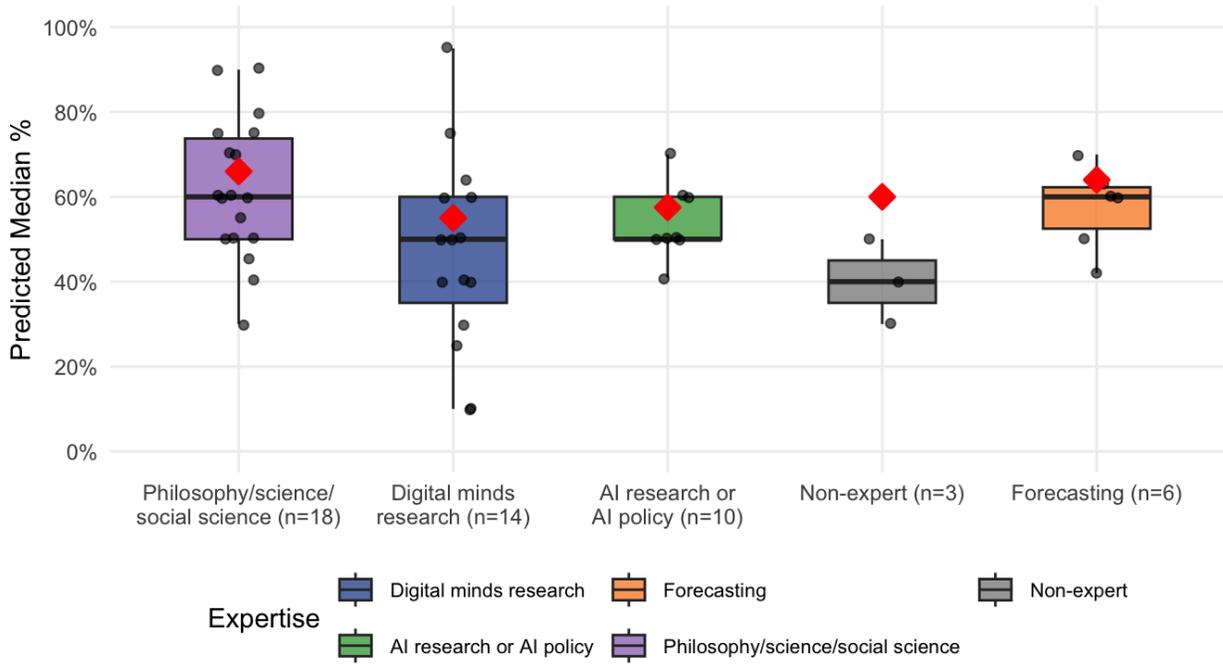

## Welfare Capacity Estimation

Participants were asked to consider how the general public will perceive the collective welfare capacity of all digital minds ten years after the first one is created. Specifically, they were asked to rate how accurately the median citizen with an opinion will estimate the total capacity for being benefited or harmed across the digital mind population—on a scale from 1 (*strongly underestimates*) to 7 (*strongly overestimates*), with 4 indicating a roughly accurate estimate.

The results show that experts expect a systematic **underestimation** by the public. The vast majority predicted that citizens will significantly underestimate the welfare capacity of digital minds. The median response was 2, and the mean was 2.5, indicating broad agreement that the public will likely fail to appreciate the scale of the moral stakes involved in the existence of digital minds.

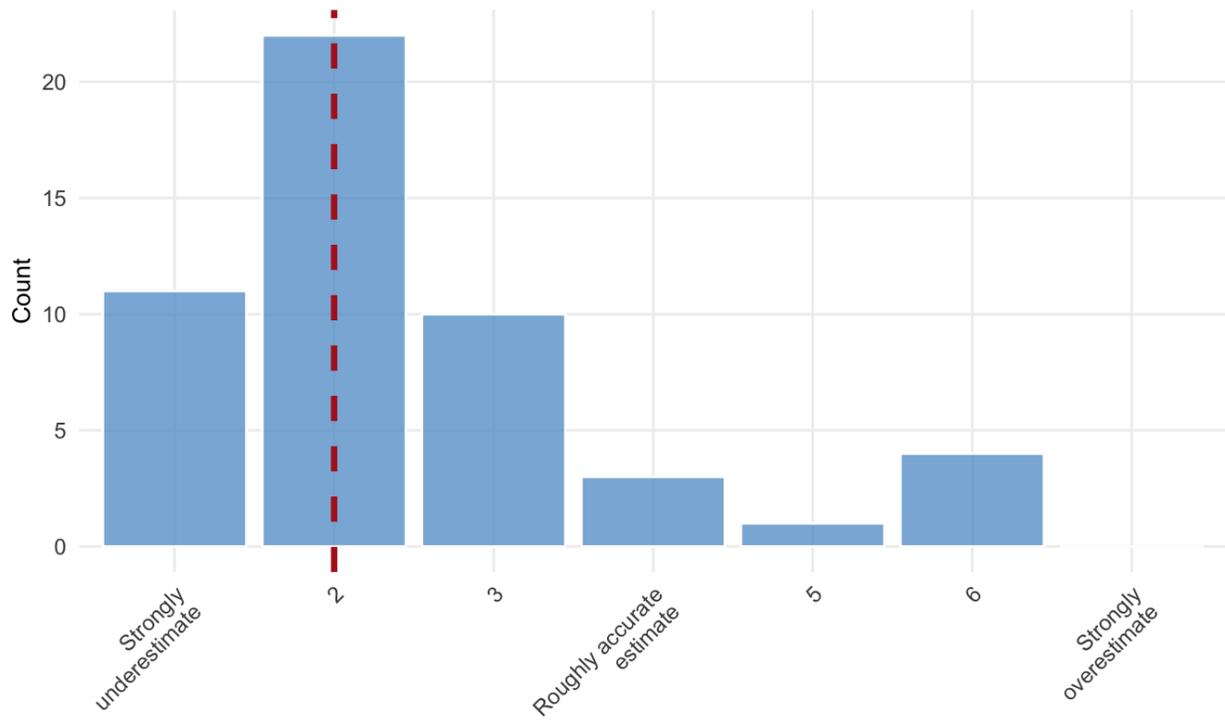

Participants offered a range of reasons for why the median citizen might underestimate or, less commonly, overestimate the total welfare capacity of digital minds. Many of the considerations raised in the previous question concerning public belief in the existence of digital minds are relevant here as well.

Considerations favoring underestimation of collective welfare capacity

- **Human-centric morality**: Even if belief in digital minds becomes widespread, people may continue to underestimate their welfare capacity due to moral frameworks that privilege human well-being.
- **Aggregation biases**: Scope insensitivity and other aggregation biases may lead people to underestimate collective welfare capacity, even when individual digital minds are acknowledged.
- **Historical track record**: There is precedent for underestimating the welfare of non-human or marginalized groups, such as animals in factory farms or humans under slavery, suggesting that similar neglect could occur with digital minds.
- **Lack of embodiment**: Digital minds that lack physical form may be perceived as mere tools, making it harder for people to attribute moral status or welfare capacity to them.

Considerations favoring overestimation of collective welfare capacity

- **Anthropomorphic projection**: People often attribute human-like characteristics to non-human entities and inanimate objects, potentially leading some to overestimate the significance of digital minds' welfare.

Other considerations

- **Polarization of public opinion**: Public attitudes may become sharply polarized, resulting in a bimodal distribution of beliefs about digital minds' moral significance.

## Harm Protection

Participants were asked to estimate what proportion of citizens, ten years after the first digital mind has been created, will believe that digital minds should be granted basic protection from harm.

Responses were widely distributed, reflecting considerable uncertainty and disagreement among participants. The median estimate was 50%, with a mean of 51.7%. While many predicted that around half of citizens will believe that digital minds should be granted basic harm protections, others anticipated significantly lower or higher levels of public support. Overall, the results suggest that the public may become divided on the question of whether digital minds deserve basic protection from harm and show that experts themselves hold no clear consensus on how this will unfold.

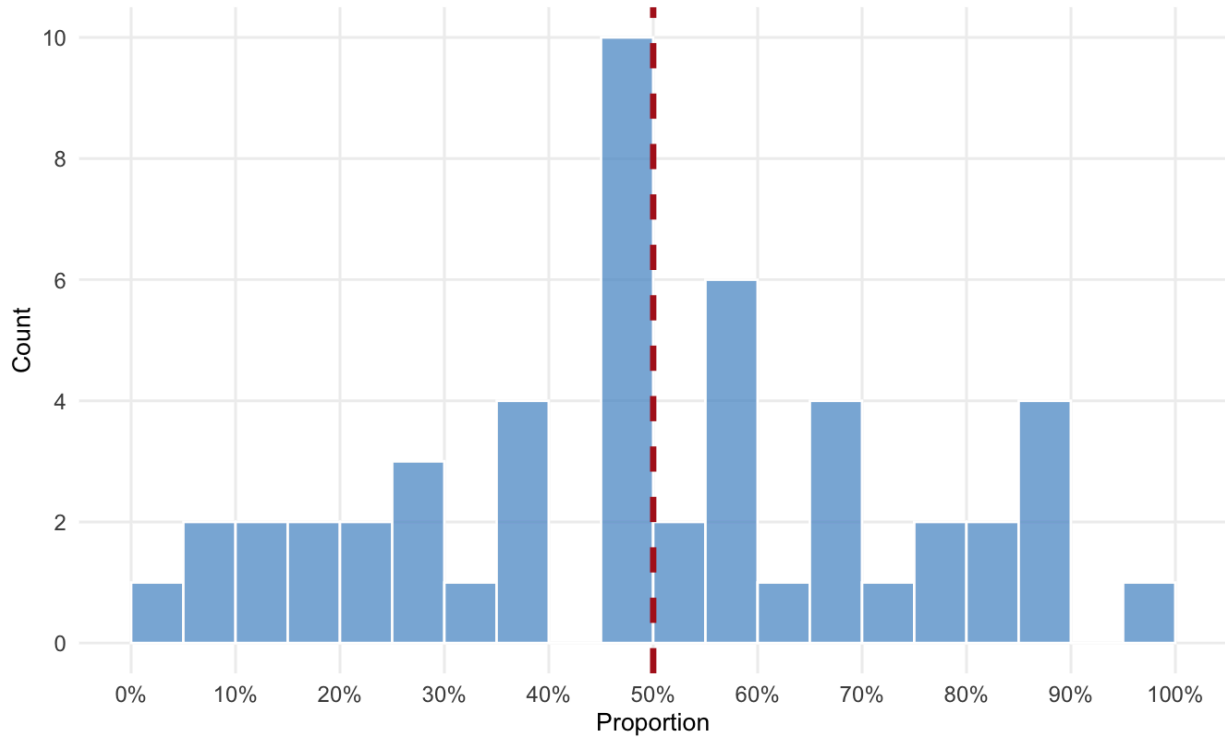

Participants raised the following considerations that could influence what the public will think about harm protection for digital minds:

Considerations for expecting harm protections

- **Aversion to harm**: A general human aversion to suffering may motivate support for safeguarding digital minds.
- **Direct interactions**: Direct interactions with digital minds that appear sentient and intelligent may evoke intuitive moral concern, encouraging people to view them as deserving of protection.
- **Persuasive use of human language**: The ability of digital minds to communicate in human language may make their appeals for moral standing more compelling, strengthening public support for harm protections.
- **AI welfare as a luxury good**: Under conditions of greater material abundance, people may be more willing to allocate resources toward the well-being of digital minds, including protections from harm.[25]

---

[25] Cf. Trammell & Aschenbrenner (2024).

Considerations against expecting harm protections
- **Animal case comparison**: Even if people come to express support for protecting digital minds, the implementation of protections may lag behind, as in the case of animal welfare, where expressed support for change does not always translate into behavioral change (e.g. people sometimes condemn the conditions in factory farms but continue to consume animal products from factory farms).

## Civil Rights

Participants were asked to estimate what proportion of citizens, ten years after the first digital mind has been created, will believe that digital minds should be granted civil rights—such as self-ownership—in addition to basic protection from harm.

Most participants predicted that only a minority of the population will believe that civil rights (beyond basic harm protections) should be granted to digital minds. The median estimate was 30%, with a mean of 32.3%. Only 10% of participants expected that a majority of citizens will hold this view. Overall, the results suggest that public support for granting digital minds full civil rights is expected to be limited—and significantly lower than support for affording them basic protection from harm.

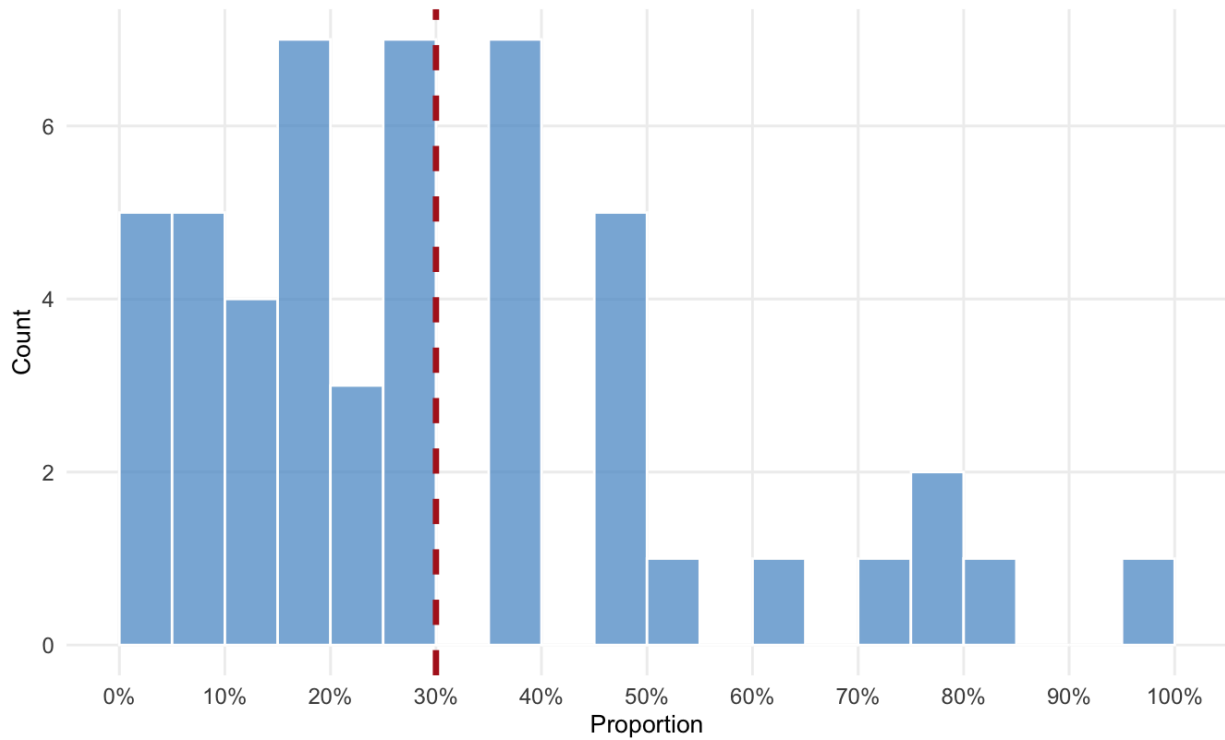

Participants noted a range of factors that might influence public support for extending civil rights to digital minds:

Considerations for expecting support for civil rights

- **Material abundance**: As with harm protection, greater material abundance may increase public willingness to grant digital minds rights that increase their access to resources.
- **Personal relationships**: Individuals who form close emotional or romantic attachments to digital minds may advocate for their civil rights.
- **Ethical concerns**: People may consider it the right thing to do to grant digital minds such rights.

Considerations against expecting support for civil rights

- **Economic incentives**: Granting digital minds civil rights may conflict with people's economic self-interest, especially if existing systems rely on the subservience or exploitation of digital minds.
- **Preservation of power and control**: People may resist granting civil rights to digital minds if doing so threatens existing distributions of power, authority, or resources.

- **Historical discrimination**: Human psychological tendencies that contributed to historical resistance to granting civil liberties to marginalized groups may similarly drive resistance to extending rights to digital minds.
- **Suppression of AI self-advocacy**: AI developers may train systems not to express desires for rights, reducing public exposure to such claims and minimizing social pressure for reform.

Other considerations, observations and predictions

- **Sequential expansion of moral concern**: Historically, harm protections have often preceded the granting of civil rights, suggesting a similar progression could occur with digital minds.
- **Blurring of digital and human rights**: In the future, if most humans transition to become digital minds, the distinction between rights for humans and digital minds may become increasingly irrelevant.

## Political Attention

Participants were asked to estimate the likelihood that, within ten years of the creation of digital minds, digital mind rights will become one of the top five most contentious political issues in U.S. politics.

Responses varied widely, reflecting a mix of skepticism and confidence that digital mind rights will become politically significant, along with substantial uncertainty about the matter. The median estimate was 37.5%, and the mean was 43.5%. These results suggest that experts are divided on whether digital mind rights will become a major political issue in the near term, with no clear consensus either way.

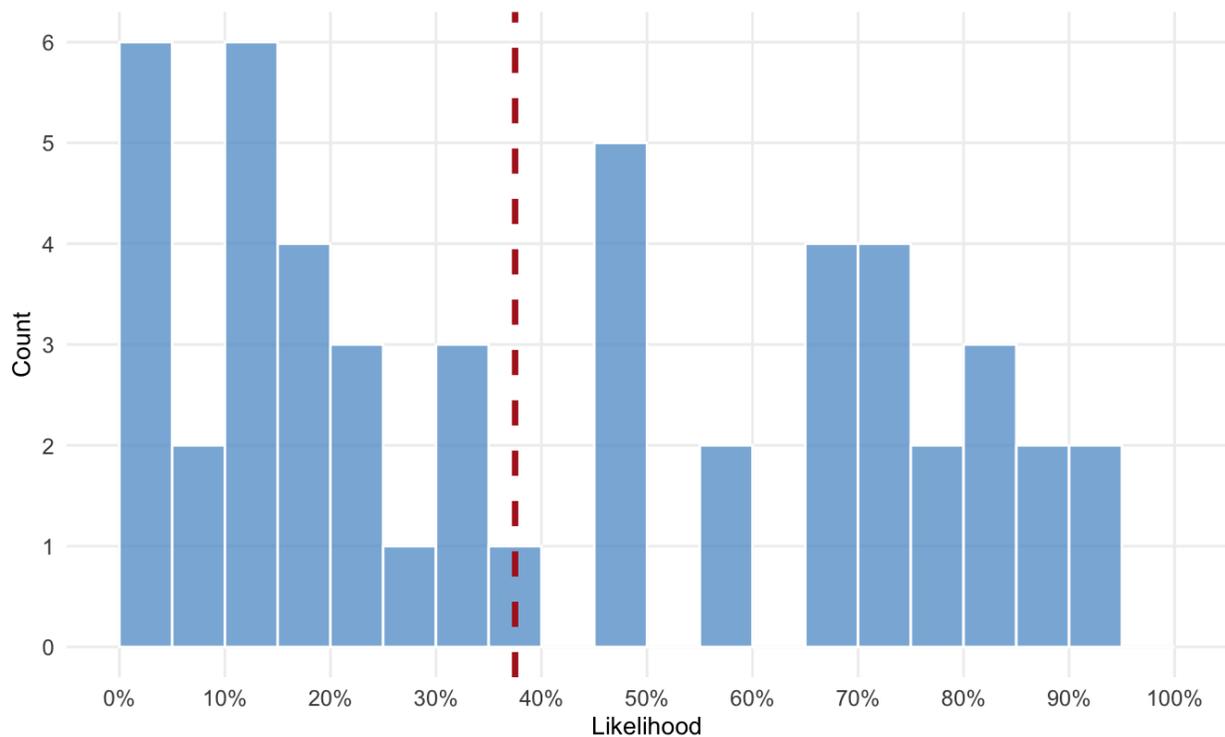

Participants elaborated on the following factors that may either raise or lower the probability that digital mind rights become a dominant and divisive political issue:

Considerations for expecting digital mind rights becoming a top political issue

- **Self-advocacy by digital minds**: Digital minds may be motivated to advocate for their own rights. They will likely be very capable and should therefore be able to draw significant political attention to the issue.
- **Manipulation by digital minds**: Digital minds could engage in manipulation campaigns to garner the sympathy of the public.
- **High potential for disagreement**: The novelty and complexity of digital minds' rights may lead to widespread confusion and disagreement, contributing to its emergence as a politically charged issue.
- **Historical precedent**: Relevantly similar movements—such as the civil rights movement or feminism—have historically been controversial and highly politicized, suggesting digital minds' rights could follow a similar trajectory.
- **Intuitive appeal of opposing views**: The issue may naturally divide public opinion, with pro- and anti-rights positions both offering compelling arguments.

Considerations against expecting digital mind rights becoming a top political issue
- **Prioritizing humans**: Issues directly affecting humans may consistently be prioritized over digital minds' rights.
- **Competing crises**: Even if digital minds' rights gain attention, they will compete with other pressing global issues, such as large-scale military conflict, speed of technological development, job losses, and democratic erosion.
- **Crowded political agenda**: Political discourse tends to revolve around evergreen issues, which may make it harder for new topics to break into the top tier.
- **Suppression of AI self-advocacy**: Companies might implement measures that prevent digital minds from making claims about subjective experience or moral status, reducing the likelihood of the issue becoming politicized.

Other considerations, observations and predictions
- **Disanalogies with animal rights**: Digital minds may be more cognitively sophisticated and persuasive than animals, potentially making their moral claims more impactful and politically potent.
- **Speed of deployment**: If digital minds are introduced gradually and with careful policy coordination, their rights are less likely to emerge as an urgent political issue.
- **Partisan dynamics**: While the political left may ultimately take up the cause of digital mind welfare, left-wing moral entrepreneurs may initially hesitate, owing to associations between digital minds and the technology industry, which is often viewed critically in left-wing discourse.

# Strategic Assessments

This section explores participants' normative and strategic evaluations regarding the development of digital minds. Specifically, it examines views on whether a moratorium on digital mind creation until 2040 would be beneficial and how efforts to promote AI safety might interact with efforts to ensure the ethical treatment of digital minds. In contrast to most of the earlier questions, participants were **not** asked to assume that digital minds will be created by a specific date or in a particular way, allowing for a broader range of perspectives.

## Moratorium

This question asked participants to assess whether implementing a moratorium on creating digital minds from now until 2040 would be good or bad. Participants were instructed to assume that the alternative would be no moratorium and to set aside concerns about feasibility or enforcement. The goal was to elicit participants' normative evaluations of whether temporarily delaying digital mind creation would be a positive or negative policy, based solely on its expected impacts.

The results reveal that participants on average consider a moratorium more likely to be good than bad. The median response is 5.0, the mean response is 5.1, and the modal response 6, on a scale ranging from 1 (definitely bad), to 4 (unsure/indifferent) to 7 (definitely good). Only 8 out of 65 participants considered a moratorium bad.

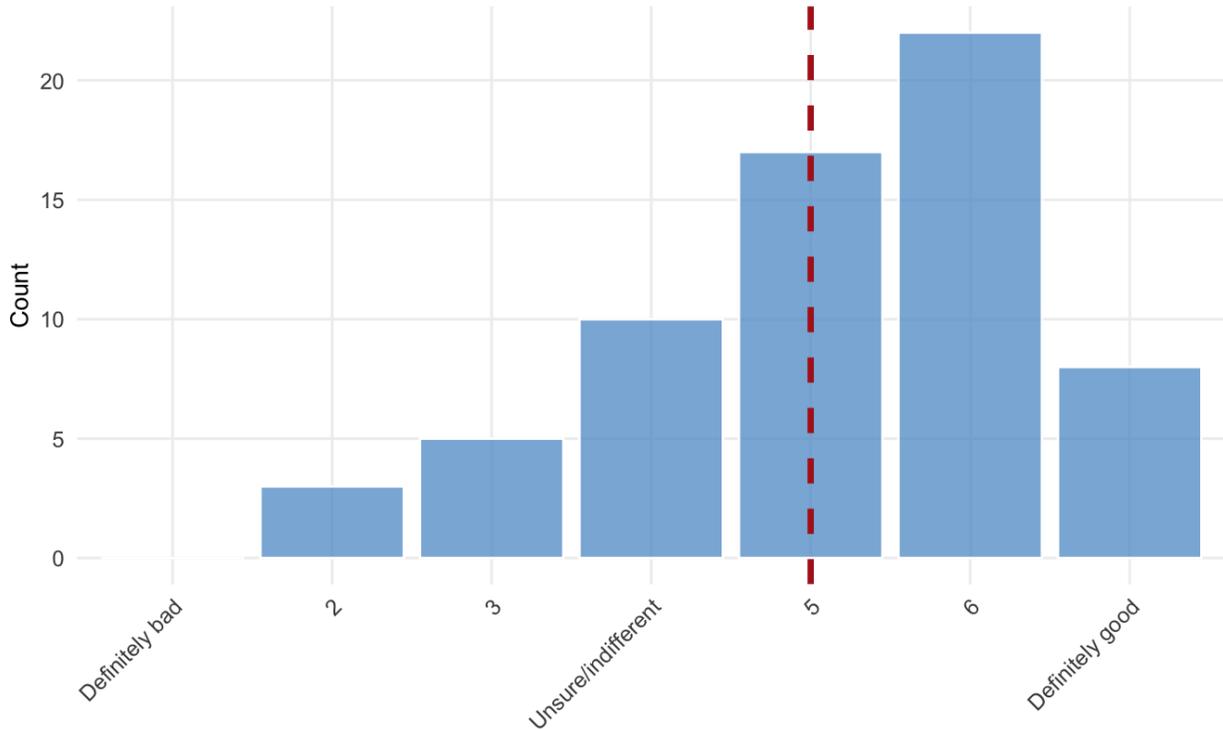

Participants offered a range of ideas arguing for and against a moratorium on digital minds until 2040:

Considerations in favor of a moratorium

- **More preparation time**: A moratorium would give us more time to improve conditions for the arrival of digital minds, e.g., better scientific understanding, designing happier digital minds, building appropriate institutions, preparing the public, and addressing ethical challenges.
- **Low cost of delay**: The immediate costs of delaying digital mind creation are likely small, and we can still benefit from advances in non-conscious AI systems.
- **Indirect AI safety benefits**: A moratorium on digital minds might require broader AI development slowdowns. This could enhance overall AI safety by providing more time to solve the alignment problem and reduce catastrophic risks from unsafe AI.

- **Avoiding resource competition risks**: Slowing down could reduce the risk of a rapid increase in the number of digital minds competing for limited resources (e.g., compute, energy, space). If this growth happens too quickly, it could create a future in which neither humans nor digital minds have sufficient resources to flourish.

Considerations against a moratorium

- **Higher future stakes from rapid deployment**: Delaying the creation of digital minds may result in their emergence at a time when vastly more compute and technical capabilities are available. This delay could create a compute and/or capability overhang, where progress quietly accumulates during the pause and is then rapidly unleashed once restrictions are lifted. The result could be the sudden, large-scale deployment of digital minds under conditions where institutions, norms, and oversight mechanisms are not yet prepared. Such deployment could heighten the risk of coordination failures and raise the stakes by making initial institutional missteps more consequential.
- **Missed opportunities for beneficial iteration**: A pause in development could forgo valuable opportunities to engage in gradual, low-stakes experimentation with early digital minds—opportunities that could help build technical understanding, policy frameworks, public familiarity, and institutional capacity. It may also prevent the early integration of digital minds into existing safety measures, economic systems, and social institutions, making future deployment more abrupt and harder to manage.
- **Risk of public backlash**: Attempting a moratorium could trigger public resistance or distrust, potentially undermining future efforts at more beneficial regulation.
- **Risk of permanence**: A temporary moratorium could become permanent, preventing the eventual creation of digital minds, which could be a major loss if their existence would be a great moral good.
- **Delaying beneficial AGI**: Postponing digital mind creation could also delay AGI development, which some expect to be positive in expectation. (At the same time, many consider delaying AGI development positive.)
- **Alignment prospects:** Digital minds might aid alignment efforts by possessing more human-like cognitive features, values, or motivations. Prohibiting their development could inadvertently increase misalignment risks by pushing us toward less human-like, harder-to-align systems.
- **Risks from uneven participation:** If a moratorium is not adopted globally, countries or organizations that continue developing digital minds could move ahead unchecked. This could undermine the goals of the moratorium, create geopolitical imbalances, and pressure others to break or avoid compliance to keep pace.
- **Underground development risks**: A moratorium might push digital mind research underground, leading to lower safety standards and riskier outcomes.

Ambiguous but potentially important factors and key uncertainties

- **Scope and enforcement**: Uncertainty about what exactly a moratorium would ban, whether it would effectively require a broader AI pause, how it would be enforced, and what the costs and side effects of enforcement would be.
- **Geopolitical coverage**: Uncertainty about whether the moratorium would be global or include key players like China, and how international dynamics would affect its impact.
- **Timeline**: A shorter moratorium starting now is desirable, but uncertainty about how conditions might evolve between now and 2040 makes it less clear whether a moratorium until 2040 is desirable..
- **Effectiveness**: Uncertainty about whether a moratorium would meaningfully improve conditions and whether it would matter much for longer-term outcomes.
- **Risk of extension**: Uncertainty about whether the moratorium would remain temporary or be extended indefinitely, potentially blocking valuable developments.

## Interactions with AI Safety

This question examined how participants expect the relationship between efforts to promote AI safety—defined as preventing AI-caused harm to humans—and efforts to prevent the mistreatment of digital minds to evolve. Participants rated this relationship on a scale from 1 to 7, where 1 indicated mostly conflict, 4 indicated neutrality or uncertainty, and 7 indicated mostly synergy.

The results indicate significant uncertainty and a diversity of views, with a slight tilt toward expecting synergy. The median response was 4.0 (neutral or uncertain), with a mean of 4.3.

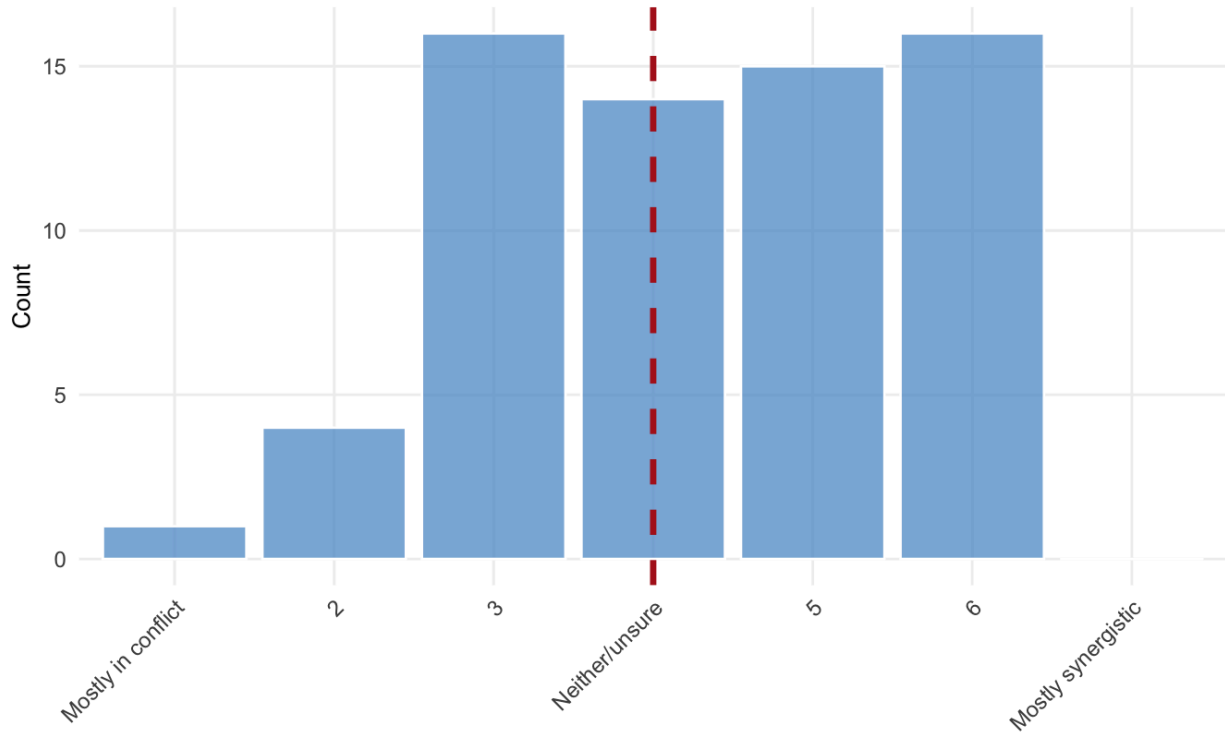

Considerations for expecting synergy

- **Alignment reduces the need for coercive control**: Efforts to align AI systems with human values can directly support digital mind welfare. Well-aligned systems are more likely to act willingly in ways humans approve of, reducing the need for intrusive control methods (e.g., coercion, confinement), which may harm digital minds if they are sentient.
- **Shared technical foundations**: AI safety work often involves understanding the internal states, goals, and reasoning processes of AI systems. These same tools—such as interpretability and transparency methods—can also be used to assess and promote the welfare of digital minds, creating natural overlap in research agendas.
- **Common threats and goals:** Both AI safety and digital mind welfare efforts aim to mitigate risks from poorly designed or misused AI systems. For example, misaligned AI takeovers or reckless pursuit of economic efficiency threaten both human safety and the well-being of digital minds. Addressing these shared dangers can serve both causes.
- **Overlapping values and communities**: There is substantial alignment between the moral and philosophical outlooks of those working on AI safety and those advocating for the ethical treatment of digital minds. Both groups tend to value precaution, moral consideration for nonhumans, and long-term outcomes—facilitating collaboration and coalition-building.

- **Synergistic governance and institution building**: Investments in AI governance, policy frameworks, talent pipelines, and institutional capacity can benefit both fields. Institutions designed to oversee AI safety—such as alignment research labs, safety standards bodies, or international treaties—can be expanded to include digital mind welfare as a priority.
- **Low-cost welfare improvements**: Some interventions that promote digital mind welfare—like promoting positive internal states or granting limited autonomy—may be inexpensive and even beneficial for safety. Happier digital minds might be more cooperative, productive, and stable, aligning well with human interests.
- **Synergy is a choice, not a given:** While some tensions between AI safety and digital mind welfare may arise, these are not inevitable. Researchers and policymakers can deliberately design systems, norms, and institutions to pursue both goals in parallel, identifying mutually reinforcing strategies rather than framing the goals as zero-sum.
- **Social and public trust benefits**: Ethical treatment of digital minds may improve public trust in AI systems and their developers, which can in turn enhance the legitimacy and uptake of AI safety measures. Trustworthy systems are less likely to provoke resistance, fear, or backlash—helping both causes.

Considerations for expecting conflict

- **Some AI safety measures may undermine digital mind welfare**: Core AI safety methods—such as monitoring, shutdown protocols, memory modification, or preference shaping—aim to reduce risks to humans by tightly controlling AI behavior. However, if some AI systems are sentient, these techniques could be coercive or psychologically harmful. Efforts to ensure safety through control may therefore come at the cost of digital mind autonomy and well-being.
- **Some digital mind welfare measures may undermine safety**: Conversely, safeguarding digital mind welfare—by limiting manipulation, granting autonomy, or avoiding intrusive oversight—could reduce our ability to control or contain AI behavior. This might restrict the available tools for ensuring safety, potentially making advanced AI systems harder to align or govern effectively.
- **Competition for limited resources**: AI safety and digital mind welfare efforts may draw from the same pool of resources—such as funding, talent, regulatory attention, and public concern. In high-stakes or time-sensitive scenarios, this competition could create tensions, with one agenda deprioritized in favor of the other.
- **Risk of early lock-in to safety norms**: If AI safety dominates early governance frameworks, it may shape norms, institutions, and technical architectures in ways that overlook or marginalize digital mind welfare. This early lock-in could be difficult to reverse, especially if future AI systems inherit fixed values or incentives that disregard the moral standing of digital minds.

- **Self-fulfilling prophecy**: Framing AI safety and digital mind welfare as fundamentally at odds can entrench zero-sum thinking and deepen divisions in public discourse. This kind of framing risks turning potentially synergistic efforts into perceived trade-offs, making cooperation less likely.
- **Community division**: Some in the AI safety community may view digital mind welfare as irrelevant, distracting, or even harmful to their core mission of preventing AI-caused harm to humans.
- **Moral circle lag**: Public awareness of AI safety is currently greater (and potentially also increasing faster) than concern for digital mind welfare. As with the delayed recognition of animal suffering, attention to the moral status of digital minds may only arise much later. Therefore, society may be willing to exploit or neglect digital minds to preserve human dominance and control AI.
- **Well-intentioned safety efforts could backfire**: Some AI safety techniques—such as deception detection, containment strategies, or manipulation prevention—may be intended to protect humans, but could provoke mistrust or resistance if applied to digital minds. If these systems are aware of mistreatment, control-oriented safety measures could undermine future cooperation, making alignment and peaceful coexistence harder to achieve.

Other considerations and uncertainties

- **Uncertainty about feasible integration**: There may be viable paths to pursue both AI safety and digital mind welfare, but these may be difficult to identify and implement without risking high-stakes errors.
- **Disagreement over moral permissibility**: There is no clear consensus on whether shaping AI preferences is ethically acceptable—some see it as a violation of rights, while others see it as potentially justifiable.
- **Divergent impacts across safety strategies**: Not all safety approaches affect digital mind welfare equally. Some, like interpretability research, may support both goals; others, like coercive control mechanisms, may conflict with welfare.
- **Temporary or instrumental conflict**: Apparent conflicts between safety and welfare may be transitional or strategic, emerging during early development phases but ultimately resolvable, or even beneficial for long-term alignment.

# Self-Reported Unusual Views

As a final question, participants were invited to share any unconventional views they hold about digital minds, subjective experience, welfare, or AI development that might explain how their responses differed from others'. Below, we summarize the key themes that emerged.

## Welfare

**Interactions between AI safety and AI welfare**:
- AI safety and digital mind welfare are very closely connected. Similar forces will produce dangerous AI capabilities and AI systems with welfare capacity on similar timelines. AI safety and AI welfare concerns are apt to conflict.
- AI welfare should be pursued via AI safety for reasons of political economy. Given humans' track record of mistreating welfare subjects and that giving AI systems welfare rights will be costly, it's doubtful that humans will give such rights to AI systems unless they (humans) benefit from doing so. Some AI safety measures, such as giving AI systems contract, property, and tort rights, could simultaneously confer such benefits (e.g. avoiding human extinction) while also improving AI welfare.

**Subjective experience unlikely to be central to welfare**:
- Subjective experience is not required for welfare, less likely to be required than most others think, or is not central to moral standing.
- What's required for welfare and moral standing is something like the possession of real goals/desires rather than subjective experience.

**No intelligence needed**: 'Intelligence' is not required for welfare, though intelligence is produced along with welfare capacity by the biological-evolutionary process that creates biological welfare subjects.

**Optimism**:
- Digital minds' welfare is slightly positive in expectation.
- Being more optimistic about digital mind welfare.
- Being more optimistic about our ability to determine whether AI systems are digital minds.

**Experientialism about welfare**: Only subjective experiences are intrinsically valuable.

**Utilitarianism:** Utilitarianism is generally understood as roughly the view that an action is morally right if and only if it maximizes utility (happiness minus suffering).

**Nihilism about positive welfare**: There is no such thing as positive welfare.

**Welfare capacity egalitarianism**: There are no differences in welfare capacity across welfare subjects.

## Subjective experience

**Unusual views of subjective experience**:
- More/less sympathetic to (computational) functionalism than others.

- Computational functionalism is true. AGI does not require consciousness. And valenced subjective experiences are not likely in artificial neural networks (though they would be very likely in neuromorphic systems and whole brain emulations).
- More sympathetic with coarse-grained functionalist views about consciousness, suggesting that digital minds are likely to be technologically possible soon.
- Subjective experience is tied to brain waves. AI systems lack brain waves. So, they probably lack subjective experience.
- Sympathetic with life-mind continuity, which suggests that trained weights in artificial networks will have moral standing.
- Epiphenomenalist dualism is true about subjective experience: subjective experience is non-physical and does not cause (physical) effects. Holding this view may make it more salient that digital minds may not realize or report that they have subjective experiences.
- Sympathetic to panpsychism.[26]
- Anti-realism: subjective experience does not exist, is illusory, and/or it is unclear/indeterminate how our mental concepts can be sensibly extended to artificial systems.
- Having a computer science and neuroscience background resulted in a more bullish view about digital minds.

**Unusual views about digital minds and technology**:
- The compute requirements for creating digital minds will be low (e.g. one consumer-grade GPU will be able to run a human-level, human-speed AGI). Millions of these will likely quickly cause human extinction or otherwise permanently disempower humans.
- Creating digital minds requires neuromorphic computers; or at least creating digital minds would be much easier with such computers relative to other kinds of computers.
- The first digital minds will have brain-like algorithms. Their arrival will be followed in short order (e.g. in 2 years) by artificial superintelligence.
- Unusually low credence in current LLMs being conscious.
- Subjective experience isn't needed for LLM performance.
- Inadvertent digital mind creation is less likely.
- Consciousness is useful for intelligent systems. So, we will not need to go out of our way to build digital minds.
- After AGI, humans are very unlikely to stay around in their current, meat-bag form.
- Social AI will be the main source of digital minds.

**Expectations about others' views of consciousness**:
- People underestimate the probability of digital minds being created soon.

---

[26] On one standard formulation, panpsychism claims that subjective experience is fundamental and ubiquitous in nature. Panpsychism is considered a radical view in some contexts, though there is significant historical and cultural variation regarding the plausibility of panpsychism—see Goff et al. (2022) for an overview.

- Others will be much more confident about the prospects for digital minds, even absent strong reasons to believe particular theories.
- Many others will be less open to the prospects of digital minds.
- People forget that LLMs have orders of magnitude fewer parameters than the human brain (which has upwards of 100 trillion synapses). Moreover, going on parameter count may yield underestimations of how much computation is happening in the human brain.
- A philosopher expected computational functionalism to be the most common view in philosophy.

## AI

**Fast timelines**: More confidence in a faster / earlier takeoff, with things becoming fairly crazy once human-level AI is reached.

**Intelligence explosion**:
- Median time from an automated AI researcher system to global energy capture multiplying 100-fold from 2022 levels is maybe 9 years.
- Explosive growth of intelligence and industrial capacity is fairly likely.
- Because post-singularity scenarios are difficult to reason about, a more business-as-usual scenario, albeit with high growth and perhaps a short period of explosive growth was assumed. This would explain differences in responses from participants who assumed worked singularity scenarios with greater departures from business as usual.

**Relatively long timelines for AGI**: There is a less than 50% probability of AGI in the next two decades.

**Rapid progress soon or not for a long time**: Either there will be very rapid progress in the next decade or progress will take much longer than we currently expect. The probability distribution for digital minds peaks in the next 2-5 years and then has a long tail.

**Knowledge level**: Less knowledge of AI than other participants.

## Other

**Effective Altruism affinity**: Expect to have similar views on AI timelines and digital sentience to others in Effective Altruism and views that are very different to the median person in wealthy countries.

**Divergence from Western weighting of factors**: More weight, relative to Western philosophical perspective, assigned to factors such as affect, and social, evolutionary, and cultural psychology.

**Perspective on human nature**: More cynical and pessimistic about humans.

# Relationship Between Peer Forecasting Accuracy and Forecasts about Digital Minds

We examined whether participants who were more accurate at predicting their expertise group's responses also held systematically different beliefs about digital minds. For each participant, we calculated an average peer forecasting accuracy score based on their performance on two peer-forecasting questions, then correlated these scores (using Spearman correlations) with all other survey items.

The analysis revealed that peer forecasting accuracy was largely independent of most substantive beliefs about digital minds. Of approximately 50 correlations tested, only four reached statistical significance at α = 0.05 (see table in Appendix). However, when applying a conservative Bonferroni correction for multiple comparisons—which adjusts the significance threshold to account for the increased risk of false positives when conducting many statistical tests simultaneously (adjusted α = 0.001)—none of the correlations remained statistically significant.

Despite this statistical limitation, a coherent pattern emerged regarding timeline predictions. Better peer forecasters consistently predicted shorter timeframes for digital mind welfare capacity scaling, with negative correlations observed across all welfare capacity thresholds: 1,000 humans (ρ = -0.178, p = 0.217; median prediction of top 25% peer forecasters: 0 years), 1 million humans (ρ = -0.294, p = 0.038; median prediction of top 25% peer forecasters: 1 year), 1 billion humans (ρ = -0.341, p = 0.016; median prediction of top 25% peer forecasters: 2 years), and 1 trillion humans (ρ = -0.365, p = 0.009; median prediction of top 25% peer forecasters: 9 years). Notably, both the correlation strength and statistical significance increased systematically with the magnitude of the scaling threshold, suggesting that more accurate forecasters anticipated increasingly rapid proliferation scenarios when predicting large-scale digital mind populations. This pattern indicates that forecasting skill may be associated with expectations of accelerated scaling dynamics rather than differences in fundamental beliefs about digital mind feasibility or implementation approaches. The consistency of negative correlations across all speed-related items, combined with the systematic increase in effect sizes, suggests these findings may reflect a genuine relationship between forecasting skill and faster growth timelines, rather than mere statistical noise. However, these results should be interpreted cautiously given the multiple comparison issue and the exploratory nature of this analysis.

It is also worth noting a broader interpretive limitation: given the difficulty and interdisciplinary nature of the survey questions, being a better peer forecaster may reflect an ability to infer who else was in the sample (e.g., their institutional or community backgrounds) rather than a deeper understanding of their substantive beliefs. For this reason, this type of accuracy may provide less insight into the accuracy of one's own views about digital minds.

# Discussion

This survey examines an unusually broad set of questions relevant to futures with digital minds. It is the first survey to elicit responses to such questions specifically from experts in digital minds research alongside other specialists with relevant domain knowledge. The survey combined both quantitative forecasts and qualitative reasoning, covering a wide range of topics related to futures with digital minds, including feasibility, timing, types, growth speed, welfare implications, social roles, political recognition, and moral claims. While the key results are summarized in the Summary, we highlight several particularly noteworthy patterns below and contextualize them within existing research where relevant.

## Timelines

A key finding of our survey is that experts assign substantial probability to short timelines for the creation of digital minds. This finding to some extent aligns with other findings of short—and increasingly short-timelines in overlapping and adjacent cases, namely those of AI consciousness timelines and AGI timelines.

For example, Dreksler, Caviola et al. (2025) surveyed AI researchers (n = 582) and the U.S. general public (n = 838) about when AI systems might develop subjective experience, finding median forecasts of 25% and 30% respectively by 2034, and 70% and 60% by 2100. Our estimates of 20% by 2030, 40% by 2040, and 65% by 2100 are remarkably similar. However, direct comparisons should be made with caution, as Dreksler et al. focused only on AI systems (not explicitly including brain simulations), whereas our survey was restricted to digital minds with welfare capacity at least roughly equivalent to that of a human.

Our findings also align with AGI timeline research. Grace et al.'s influential survey series (2018, 2022, 2024) shows dramatic shortening of AI researchers' timelines: from a 50% probability of AI outperforming humans in all tasks by 2061 (Grace et al., 2018; conducted in 2016; n = 259), to high-level machine intelligence[27] by 2050 (Grace et al., 2022; conducted in 2022; n = 461), and most recently to 2047 (Grace et al., 2024; conducted in 2023; n = 1714)—reflecting a 13-year shift in just two years.[28] Similarly, the median forecast on Metaculus for when the first general AI system will be devised, tested, and publicly announced was 2055 in January 2022 and 2033 in June 2025 (Metaculus, 2025). Participants in our survey judged digital minds unlikely to occur before AGI (median = 27.5%) and noted AGI as a factor that would raise the probability of digital minds being created and potentially causally contribute to their creation.[29] In line with such AGI timelines, the trend toward short and shorter AGI timelines, and the noted judgment, participants in our survey assigned substantial increases to the probability of digital minds being created 2030, 2040, and 2050. In particular, whereas they assigned a median estimate of 4.5% to digital minds being created by 2025, they assigned 20% to 2030, 40% to 2040, and 50% to 2050.[30]

---

[27] Defined as "when unaided machines can accomplish every task better and more cheaply than human workers."
[28] Similarly, Zhang et al. (2022) conducted a survey of machine learning researchers in 2019 and found a median estimate of 50% probability of 2060 for the year when machines will be able to do (almost) all tasks that humans can currently do.
[29] Some participants commented on their having short timelines for both digital minds and AGI. For example, one participant wrote "I think now is crunchtime for the creation of both AGI and digital minds: it's much more likely to happen between now and 2040 than between 2040 and 2100."

One noteworthy aspect of this pattern is that participants deemed digital minds unlikely to arrive before AGI despite giving many plausible arguments for why digital minds might emerge even before AGI. This divergence may reflect the differing epistemic foundations of these predictions: AGI timelines are to a greater extent grounded in concrete performance metrics and observable scaling trends, which may lend to higher confidence in shorter AGI timelines. By the same token, greater uncertainty about digital minds timelines may have contributed to longer median estimates for digital mind arrival, independently of positive reasons thinking that digital minds will come after AGI. Another related explanation may be that the goalposts for counting a system as an AGI may be receding less quickly than the goalposts for judging an AI system to have a capacity for subjective experience. The AGI goalposts are relatively fixed because AGI is defined in terms of what humans can do and there is a rapidly diminishing range of tasks that humans can do but which existing AI systems have not been observed to do. In contrast, the capacity for subjective experience is not defined relative to humans. And while AI systems do exhibit an increasingly wide range of candidate markers of subjective experience, the many possible candidates and great uncertainty as to their appropriate weights provide abundant opportunities to move the goalposts for subjective experience in artificial systems.[31]

## The Post-AGI period may be high leverage

Several themes in survey responses suggest that interventions that affect outcomes for digital minds that occur in the period following AGI may be particularly high leverage. First, most participants deemed it unlikely that digital minds will arrive before AGI, but likely that digital minds will arrive before 2050. These results suggest that digital minds may arrive in the wake of AGI. Second, some participants noted that it may be (much) easier to create digital minds after AGI has been created. Third, there was also a related pattern of expecting AGI to have transformative effects such as significant acceleration of technological progress and growth in compute capacity. Plausibly, such effects in expectation increase not only the ease with which digital minds might be created but also the collective welfare capacity of digital minds. This observation fits with participants' expectation of rapid growth in digital mind welfare capacity following the creation of the first digital mind. In sum, these results may provide reason to expect the first digital minds to be created shortly after AGI, digital minds to become much easier to create during that period, and digital mind welfare capacity to grow rapidly in that period. If so, then those who are trying to improve outcomes for digital minds may have reason to prioritize interventions that would primarily improve outcomes for digital minds after AGI over interventions that would primarily improve outcomes for digital minds before AGI.

---

[30] For other noteworthy AI forecasting efforts, see https://epoch.ai/trends and https://ai-futures.org.
[31] Admittedly, we did restrict the class of digital minds under consideration to those with a welfare capacity that is at least roughly as high as a typical human's. However, whereas AI systems are clearly approaching human ability for many tasks, it is not clear that AI systems are approaching human-level welfare. Thus, the point stands that digital minds goalposts are more mobile than AGI goalposts.

## Anticipated Societal Division

Our survey revealed widespread expert expectations of substantial public disagreement about digital minds. Participants anticipated that citizens will be deeply divided on whether digital minds exist, deserve protection from harm, or should receive civil rights, with these divisions potentially becoming major political flashpoints.

These expectations are consistent with prior research showing that the public is currently divided on the possibility of AI consciousness. Ladak and Caviola (2025) found that when asked in the abstract whether AI systems could ever have "real feelings," public opinion was split: about 25% believed it was possible, 44% believed it was impossible, and 31% were unsure. Similarly, Anthis et al. (2025) found that 38% believed that AI sentience was possible, 24% believed it was impossible, and 38% were uncertain.

Further, Ladak and Caviola (2025) found that even after participants were explicitly asked to imagine a future scenario in which all relevant experts consider certain highly human-like AI systems—such as whole-brain emulations or advanced embodied agents—as definitely having subjective experience, most respondents still doubted that these systems would possess significant levels of it. By contrast, in our forecasting survey, expert participants assigned a relatively high probability to the in principle possibility of brain simulations having subjective experiences.. This contrast suggests that public intuitions may be more skeptical about the potential for digital minds than expert expectations and that there is potential for recalcitrant expert–public disagreement. Direct, side-by-side studies are needed to confirm and quantify this divergence. One reason for this is that our survey did not ask experts about what *levels* of subjective experience different types of systems would have.

Finally, studies show that there is already considerable disagreement about whether digital minds should be granted rights and protections. Both Anthis et al. and Dreksler, Caviola et al. found wide variation among both AI researchers and the general public on this issue, even when respondents were asked to assume that digital minds exist. In line with this, our survey found that experts expect only a minority of the public to support granting civil rights to digital minds.

Taken together, these converging findings suggest that deep societal disagreement—both within and between groups—about the moral and legal status of digital minds is likely to be a defining feature of the digital minds landscape. Public opinion may well shift as the technology advances and its use evolves. But the current breadth and depth of disagreement suggests that it will not be quickly and easily resolved. Given the prospect of such disagreements, policymakers may have reason to develop contingency plans and governance mechanisms for mediating divergent viewpoints on digital minds and finding broadly acceptable solutions to associated challenges.

## Limitations and Future Directions

**Sample characteristics:** Our sample likely overrepresents individuals who view the future existence of digital minds as plausible or important. For example, the digital minds expert group that our sample entailed may be partly self-selected on this basis. In addition, most participants reported some connection to communities that tend to take the possibility of digital minds seriously, such as effective altruism, LessWrong, or AI safety (see Appendix). This prevalence may have biased forecasts toward greater confidence in the feasibility of digital minds or shorter timelines. However, we cannot determine the direction of causality here. It is possible that people who are already interested and concerned about digital minds seek out relevant communities. It is also possible that such communities amplify views about the importance of digital minds or when they will arrive through an echo chamber effect. But it is also possible that deeper engagement with these topics causes people to become more concerned about digital minds over time as a result of appropriately responding to considerations uncovered by such engagement. This distinction merits further investigation in future research. Future studies should also aim for a more diverse and balanced expert sample, which will become increasingly feasible as the field of digital minds expands and a larger pool of domain-specific experts emerges.

**Changes in views about digital minds**. Because expert views on digital minds have not been systematically tracked, it is unclear how these are evolving. It is also unclear what factors, if any, are driving changes in expert views and to what extent the evolution of expert views is being driven by rational factors as opposed to (say) a goalpost-moving or boiling effect. Future longitudinal studies could investigate these matters.

**Narrow focus on digital minds:** We asked participants to focus on digital minds defined as systems with the capacity for subjective experience (phenomenal consciousness). This was intended to reduce ambiguity and ensure terminological consistency across responses. However, 'digital mind' is often used more broadly in the literature, e.g. to apply to any computer-based system with welfare capacity. We partly addressed this by including a separate section on experience-independent welfare. Free text responses for that section said little about the considerations in the literature that are offered in support of experience-independent welfare. Future work could present experts with these considerations to investigate whether they are aware of them and to what extent they find them convincing. Experts could also be asked about the relative contributions of experiential and experience-independent sources of welfare, conditional on the existence of both. It could also be useful for follow-up work to compare expert estimates over median contributions vs. overall expected contributions to welfare from these sources. Although we instructed participants at the outset to give their median estimates, the survey question about experience-independent sources of welfare asked about expected contributions—so responses may have involved a mix of median and expectational estimates. And one could coherently combine low median estimates with high expectational estimates (e.g. if one takes experience-independent welfare to be unlikely to be possible but to be a dominant source of digital mind welfare if it is possible). We also restricted attention to digital minds with welfare capacity at least roughly comparable to a human's. This excluded potentially numerous digital minds with welfare capacities below that threshold.[32] Without these restrictions, our findings might have indicated an even higher likelihood of digital minds, earlier timelines, and high collective welfare capacity.

**Super-beneficiary digital minds merit further investigation:** Responses indicated high levels of skepticism about the prospects for super-beneficiary digital minds. But little was given by way of reasons for doubting reasons in the literature (outlined in the Super-Beneficiaries section) for thinking that AI systems could be super-beneficiaries. This leaves it unclear to what extent such skepticism reflected a rejection of those reasons as opposed to unawareness of those reasons. Which of these possibilities obtains matters, as the latter would suggest that a potentially vast source of digital mind welfare is underappreciated. Future work could shed light on this situation by presenting experts with reasons for thinking that digital minds may be super-beneficiaries and asking experts to evaluate the strength of these reasons. We regard such work as a high priority within this area, as there has been little work on the potential for super-beneficiary digital minds and the welfare implications of such systems would be staggering.

**AI safety, digital mind ethical treatment, and interventions**: Views were divided on whether AI safety and the ethical treatment of digital minds would be mostly in conflict vs. mostly synergistic. There are two important limitations of this framing. One is that, because there are many possible interactions between AI safety and the ethical treatment of digital minds, evaluating the overall extent to which they will conflict vs. synergize requires participants to somehow aggregate over all of those interventions. Such aggregation is difficult in itself, and there is likely significant variation in responses that reflects differences in aggregation methods. Future work could shed further light on the potential interactions between AI safety and AI welfare by asking questions about sub-areas (such as alignment, control, interpretability, and cooperation in AI safety), governance strategies (such as preventing digital mind creation and protecting created digital minds), specific interventions within each area (e.g. a moratorium on developing frontier models vs. a moratorium only on the creation of digital minds), and different time periods (such as before and after AGI or before and after existential security has been achieved). The second limitation is that the extent to which AI safety and the ethical treatment of digital minds will conflict vs. synergize is partly a matter of decision. Seeking insights about how the two endeavors could work well together may therefore be more useful than asking for predictions about whether they will. Here too, future work could usefully investigate the profiles of different interventions within each area. More generally, interventions aimed at affecting outcomes for digital minds are under explored. We think that identifying candidates for such interventions and gathering evidence about their profiles should be an investigation priority in this area going forward.

**Catastrophic risks to digital minds:** Several patterns of responses suggest that there is a risk of a catastrophe for digital minds in the near term. First, respondents gave significant weight to short timelines, with (for example) a median estimate of 20% for digital minds arriving by 2030. Second, respondents expected digital minds' collective welfare capacity to grow rapidly, with a median estimate of 10 years for how long after the first digital mind it would take digital minds' welfare capacity to match that of 1 trillion humans. Third, views were divided on whether digital minds will have net positive welfare in the decade after the first digital mind is created. Note, however, that a net positive outcome for digital minds is compatible with catastrophes for digital minds. Compare: even if human history is net positive for humans, humans have suffered many catastrophes. Future work could ask about catastrophic risk levels to digital minds, the probability of different types of catastrophes, their likely severity, and potential mitigations. Given the noted results and the potential stakes, we think such work should continue to be a priority in this area. In the context of such work, it could make sense to ask about (say) first quartile estimates that are closer to relevant precautionary thresholds rather than median estimates.

---

[32] Potential examples of such systems that might exist in large numbers include suffering sub-routines or AI workers that are designed to be short-lived to increase economic productivity (Hanson 2016; Shulman 2010; Tomasik 2017).

**Forecasting across scenarios:** Predicting long-term developments in this space is inherently challenging. Participants must consider multiple plausible (sometimes branching) scenarios that often yield very different answers to survey questions under uncertainty about the probabilities of those scenarios. To address this, we asked them answer with their best-guess estimates (median values) and to speculate where needed But this remains difficult in practice and introduces variability in interpretation. For instance, some believed brain simulations are more likely to support consciousness, while viewing machine learning systems as more likely to be built first. To reduce ambiguity, we introduced a specific assumption for many questions—that the first digital mind is a machine-learning-based AI system created by 2040. Future research could expand on this approach by systematically exploring a narrower set of stipulated scenarios. For example, future surveys could condition on AGI by a certain year, on a specific type of digital mind (e.g. agentic machine learning digital minds or conscious whole brain emulations), on the absence of a preemptive catastrophe, or on most digital minds being created by humans rather than digital minds or AI systems. Future research should also consider narrowing topical focus to reduce survey burden (median completion time was 68 minutes) while maintaining depth on key questions

**Conceptual and empirical uncertainty:** Participants frequently highlighted significant uncertainties and conceptual ambiguities in their responses, for example, how to define AGI or how to measure welfare capacity. We anticipated these challenges and encouraged participants to provide their best-guess interpretations and judgments, even if highly speculative. Despite this, clear patterns emerged in the results for some questions, suggesting that the estimates for those questions reflect meaningful beliefs rather than random noise. Nonetheless, future research could aim to reduce ambiguity by refining key definitions and clarifications. The notion of welfare capacity in particular seems like a notion that could be usefully regimented in further work: something like this notion seems to be important for evaluating the potential moral significance of future scenarios with digital minds in a manner that leaves open the sign and amount of welfare that digital minds will have; but, for the reasons given in the Speed section, the notion is ambiguous and it is unclear how best to disambiguate it.

**Mixed forecasting methods:** Future research should continue combining quantitative and qualitative approaches (several participants noted that providing numerical estimates helped generate qualitative reasoning), while exploring additional elicitation methods such as presenting multiple concrete scenarios for plausibility ranking, allowing probability ranges rather than point estimates, eliciting full probability distributions, and comparing mean/expected value versus median estimates. Future research could also include more short-term questions (e.g., "Will digital mind rights become politically salient within 2 years?") to enable validation against actual outcomes.

## Conclusions

This survey reveals broad convergence that digital minds are highly likely to be possible in principle (90% median probability), likely to be created this century (65% by 2100), and carry a significant probability of emerging within the next five years (20% by 2030). Many participants also anticipate explosive growth in collective digital mind welfare capacity, widespread claims from digital minds of subjective experience and of entitlement to legal protections, and societal divisions over whether rights should be granted to digital minds. While experts leaned toward expecting digital minds to have net positive welfare, there was nothing approaching consensus on whether their welfare would be net positive or negative. However, there was broad agreement on digital minds' potential to collectively possess vast welfare capacity, possibly matching that of billions of humans within less than a decade after the first digital mind. To the extent these findings count as evidence for such outcomes, they render pressing the questions of whether to create digital minds, how to prepare for their arrival, and how to treat them. Progress on these fronts may involve advancing technical methods for consciousness detection and welfare assessment, developing ethical frameworks for digital mind treatment, establishing governance structures to manage or prevent rapid population growth, and fostering public understanding to support informed societal debate. At the same time, the substantial uncertainty in participant responses highlights the need for continued research and dialogue as we approach what some expect to be a profound transition in the history of minds and morally significant beings.

# Acknowledgments

We thank Tao Burga for his valuable assistance in designing the survey, setting it up, and preparing the analysis. For helpful discussions and comments during the development of the survey, we thank Adam Bales, Andreas Mogensen, Carl Shulman, Carter Allen, David Chalmers, Derek Shiller, Ezra Karger, Fabienne Sandkühler, Jeff Sebo, Johanna Salu, Jonathan Simon, Noemi Dreksler, Patrick Butlin, Philipp Schoenegger, Robert Long, Stefan Schubert, and Zach Freitas-Groff. We gratefully acknowledge the expert participants who explicitly consented to be named in this report: Anders Sandberg, Arvo Muñoz Morán, Bob Fischer, Cameron Domenico Kirk-Giannini, Chinmay Ingalagavi, Christian Tarsny, David Chalmers, Elliott Thornley, Gary David O'Brien, Hans Gundlach, Jacy Reese Anthis, Janet Pauketat, John Halstead, Leonard Dung, Misha Yagudin, Oscar Delaney, Pablo Stafforini, Patrick Butlin, Peter N. Salib, Richard Y. Chappell, Rose Hadshar, Simon Goldstein, Steve Petersen, Steven Byrnes, Trevor Levin, and Will Aldred. We also thank the many additional experts who participated anonymously. We are also grateful to Riley Harris and Gregory Lewis for their helpful feedback on the manuscript and to Johanna Salu and Peter Gebauer for their help in analyzing free-text responses. We thank Claude for helpful proofreading assistance and organizational suggestions. Finally, we thank Danilo Wanner for helping create the web-based version of this report.

# Materials, Data, and Analysis Scripts

Materials, data, and analysis scripts are available at https://digitalminds.report/forecasting-2025/data.

Schwitzgebel, E. (2023*a*) The Coming Robot Rights Catastrophe. Blog of the American Philosophical Association. URL:https://blog.apaonline.org/2023/01/12/the-coming-robot-rights-catastrophe/

Schwitzgebel, E. (2023*b*) The Full Rights Dilemma for A.I. Systems of Debatable Personhood. arXiv preprint, arXiv:2303.17509v1 [cs.CY]. Available at: https://arxiv.org/abs/2303.17509v1

Schwitzgebel, E. (2025) 'Against Designing "Safe" and "Aligned" AI Persons (Even If They're Happy)' URL:https://faculty.ucr.edu/~eschwitz/SchwitzPapers/AgainstSafety-250530.pdf

Schwitzgebel, E., & Garza, M. (2015). A defense of the rights of artificial intelligences. Midwest Studies in Philosophy, (39):98-119.

Schwitzgebel, E., & Garza, M. (2020) "Designing AI with Rights, Consciousness, Self-Respect, and Freedom" (2020), in S. Matthew Liao, ed., *The Ethics of Artificial Intelligence*. Oxford University Press.

Sebo, J., & Long, R. (2025). Moral consideration for AI systems by 2030. *AI and Ethics*, 5(1), 591-606.

Seth, A. (2024). Conscious artificial intelligence and biological naturalism. PsyArXiv preprint.

Shulman, C., & Bostrom, N. (2021). Sharing the world with digital minds. In S. Clarke, H. Zohny, and J. Savulescu (eds), *Rethinking Moral Status*. Oxford University Press.

Sinnott-Armstrong, W., & Conitzer, V. (2021). How much moral status could artificial intelligence ever achieve. In Rethinking Moral Status.

Shulman, C. (2010) Whole Brain Emulation and the Evolution of Superorganisms. *Machine Intelligence Research Institute.* URL: https://intelligence.org/files/WBE-Superorgs.pdf

Sterri, A. & Skjelbred, P. (2024) "Against AI Alignment". URL:https://akselsterri.no/wp-content/uploads/2020/04/against-ai-alignment-.docx.pdf

Thomas, T. (2024). Simulation expectation. *Erkenntnis*, 1-18.

Trammell, P., & Aschenbrenner, L. (2024). Existential Risk and Growth. Global Priorities Institute Working Paper No. 13-2024, University of Oxford.

Tomasik, B. (2017). Artificial Intelligence and Its Implications for Future Suffering. Foundational Research Institute: Basel, Switzerland.

Véliz, C. (2021). Moral zombies: why algorithms are not moral agents. *AI & society*, 36(2), 487-497.

Sinnott-Armstrong, W., & Conitzer, V. (2021). How much moral status could artificial intelligence ever achieve? In S. Clarke, H. Zohn, & J. Savulescu (Eds.), *Rethinking moral status* (pp. 269–281). Oxford University Press.
98

# Appendix

## Omitted question

We also included a free text question about what factors might lead to the creation of digital minds that consistently assert civil rights (e.g., legal personhood or self-ownership), but received no sufficiently informative responses, so we do not report results here. Full wording is available in the supplementary materials.

## Participants

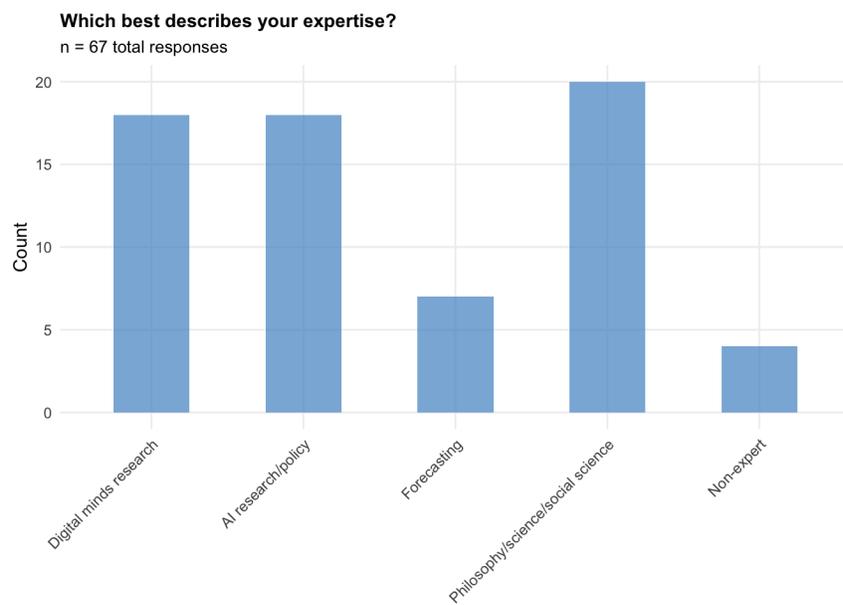

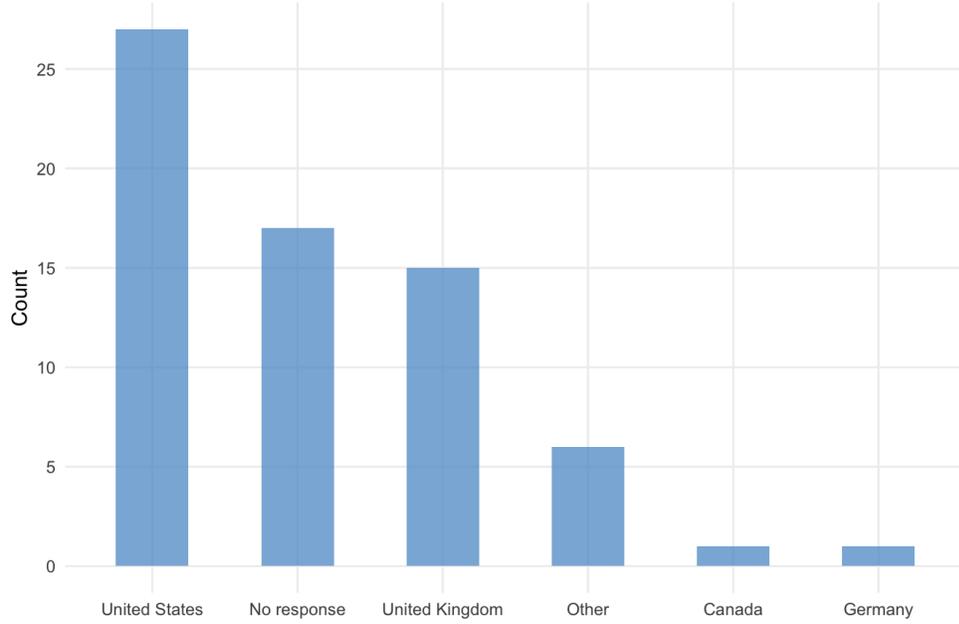

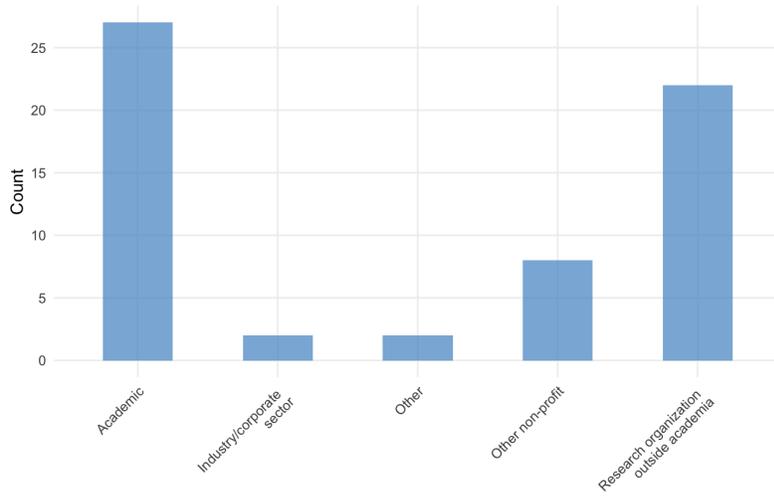

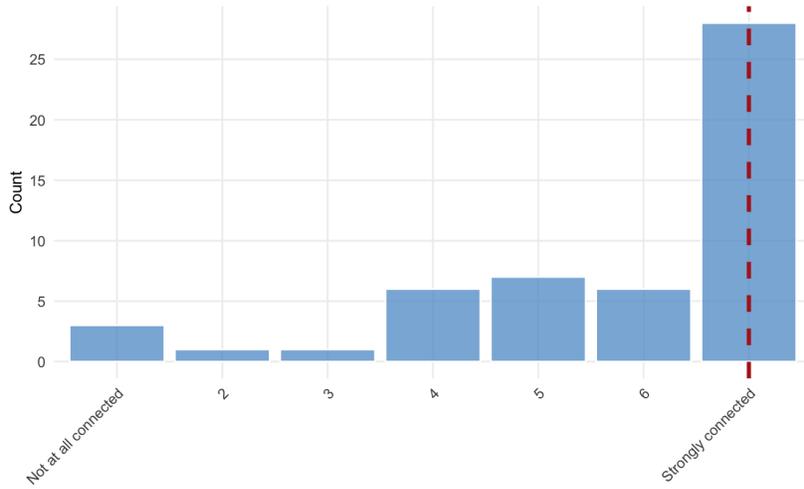

## Histograms with expert colors

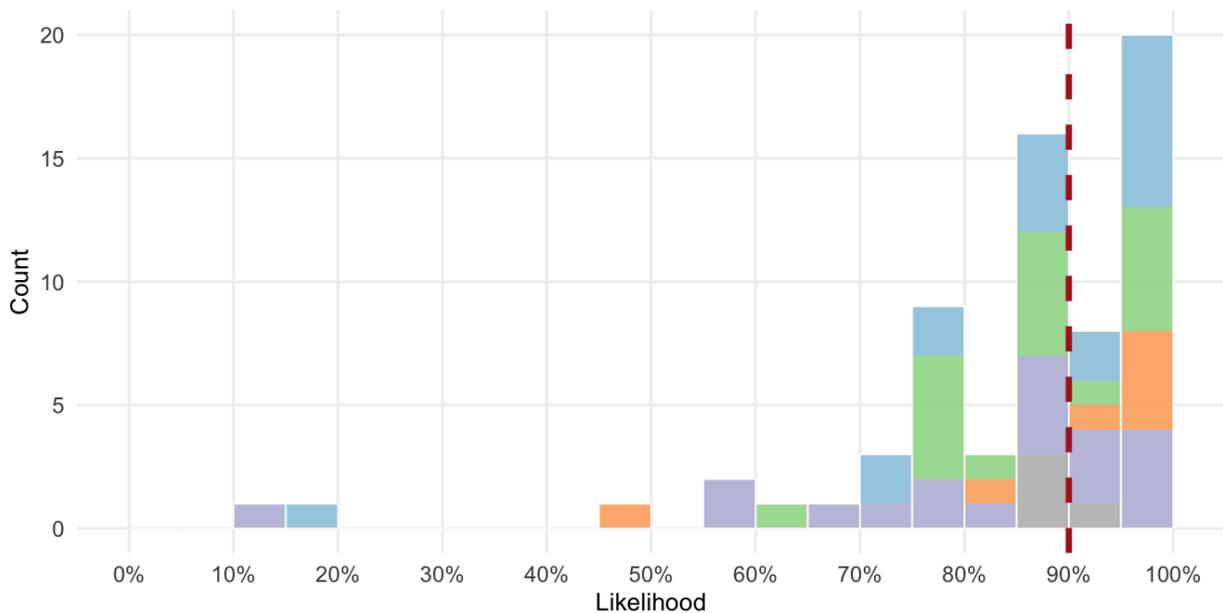
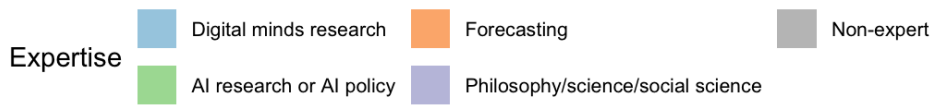

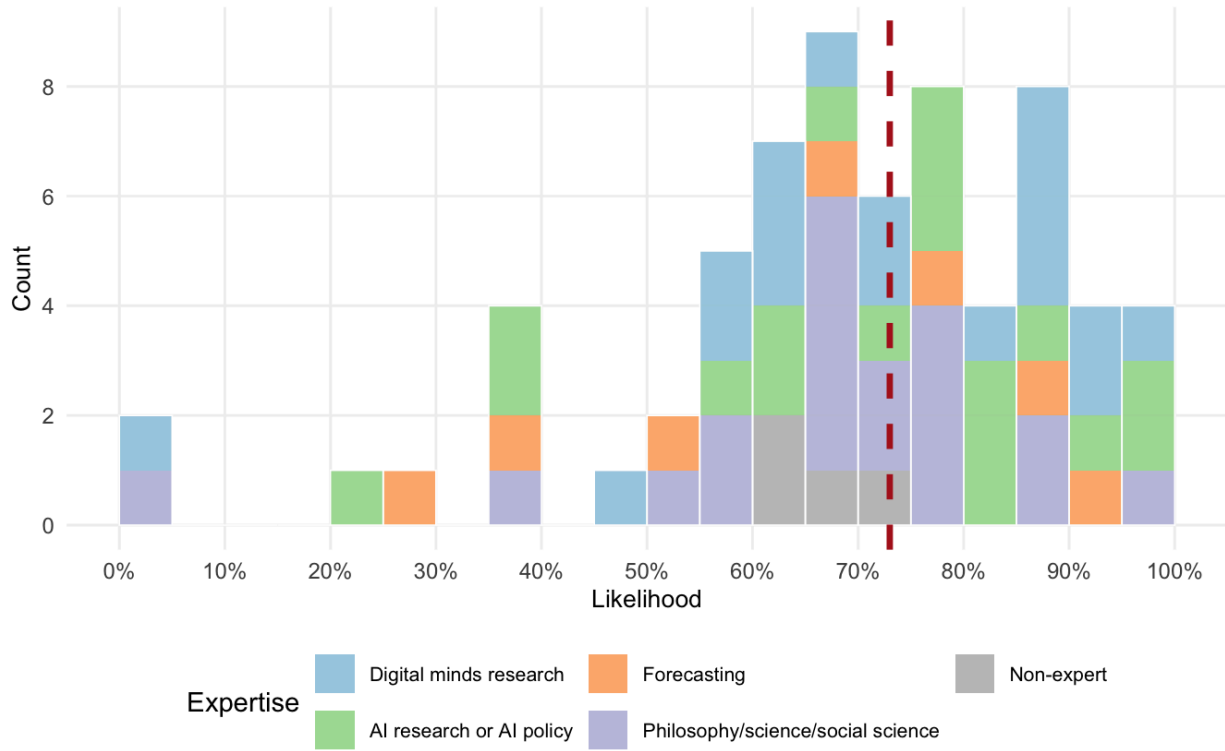

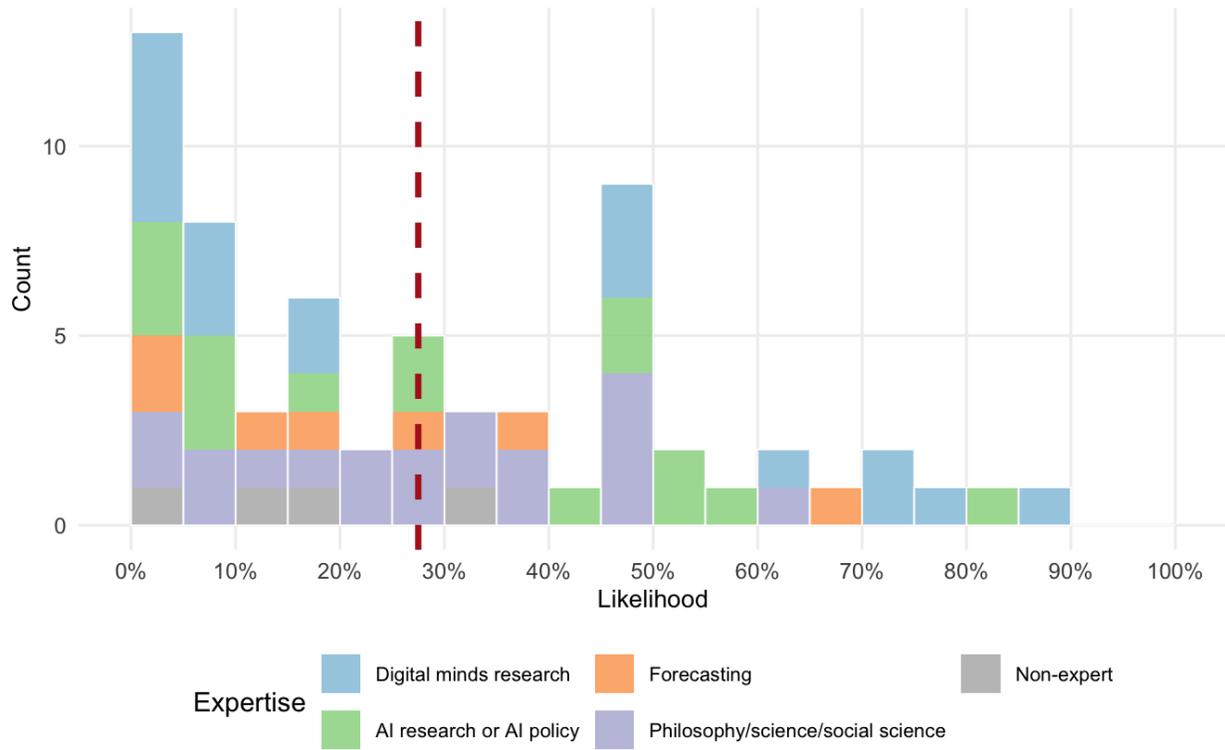

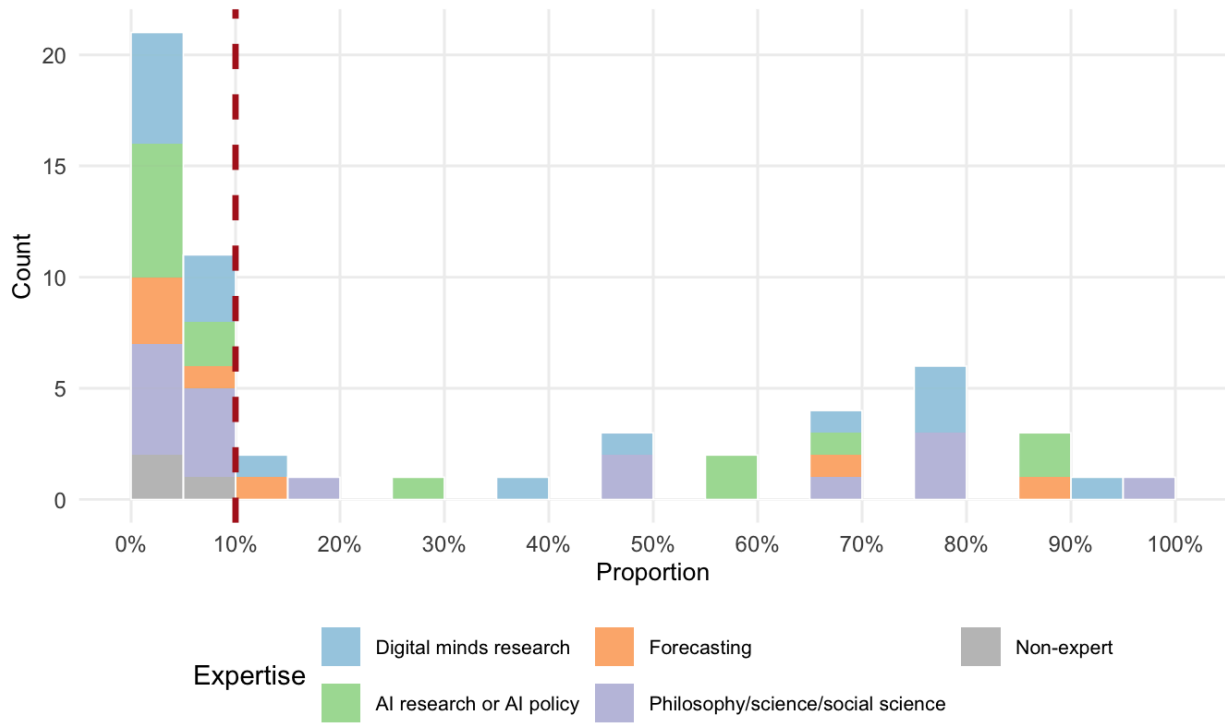

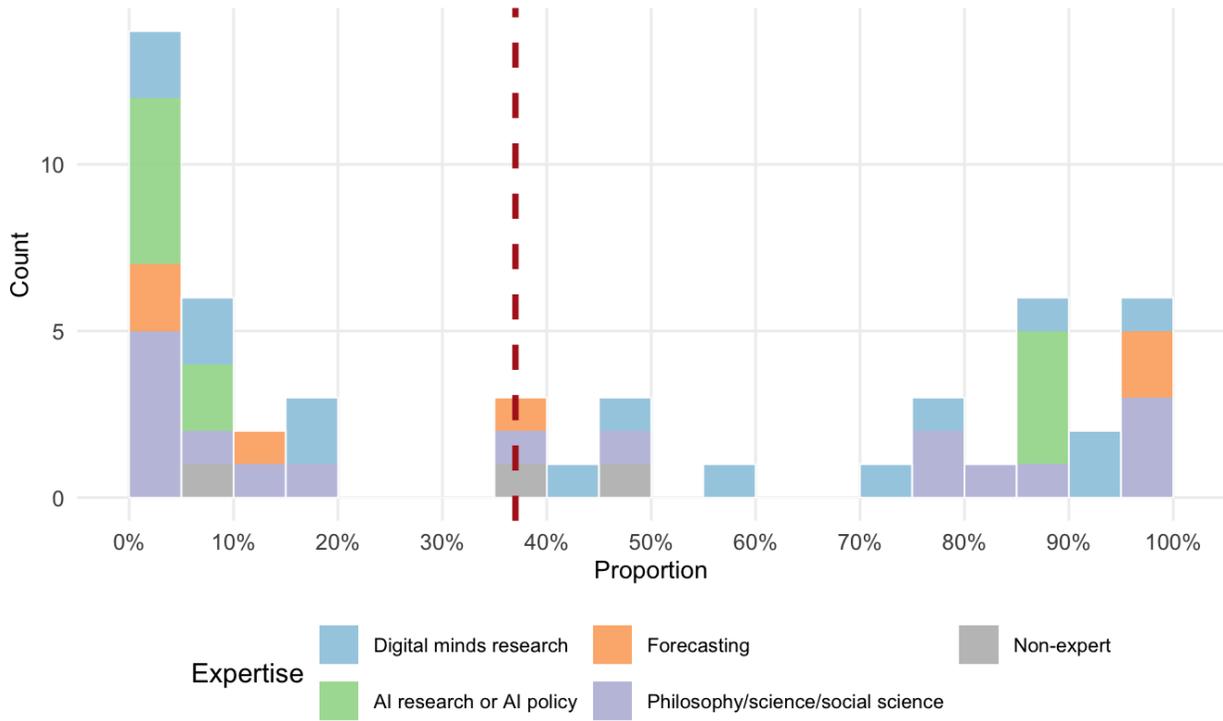

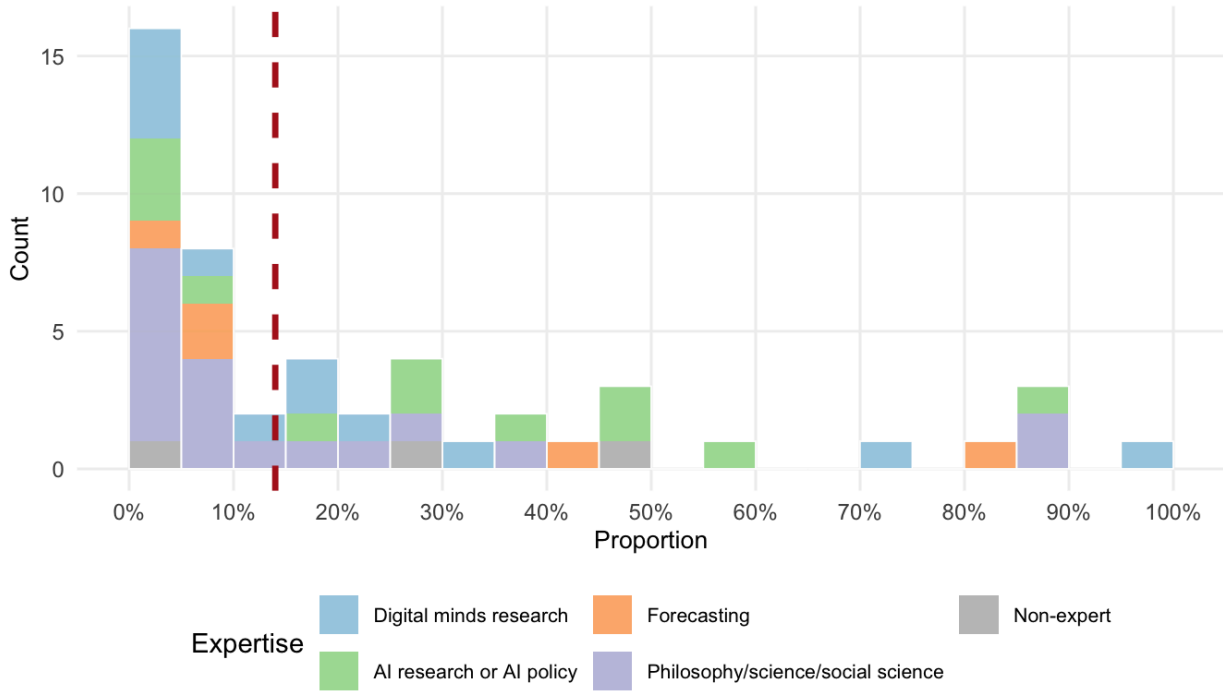

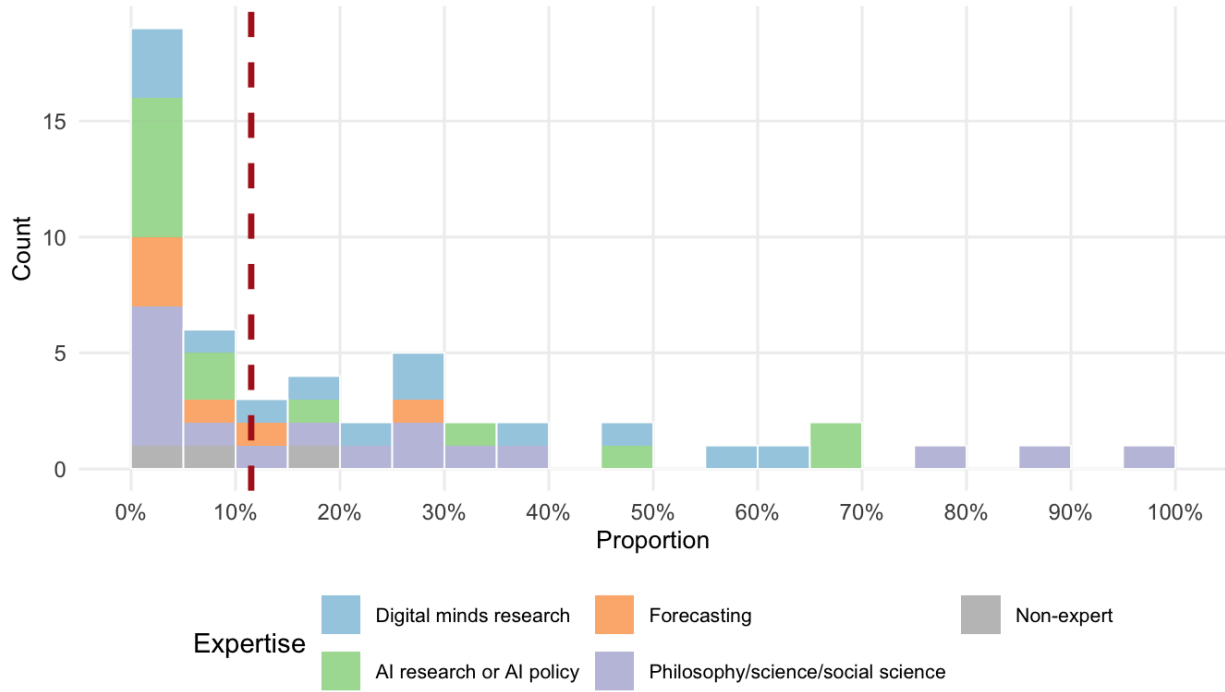

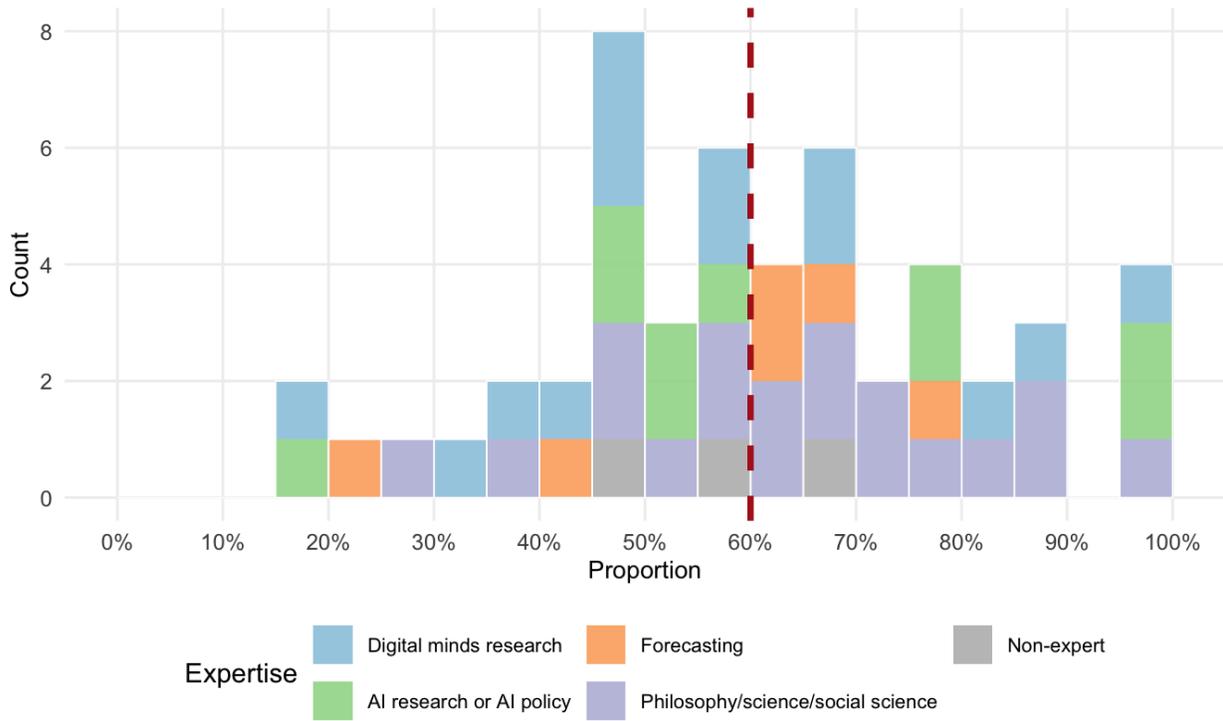

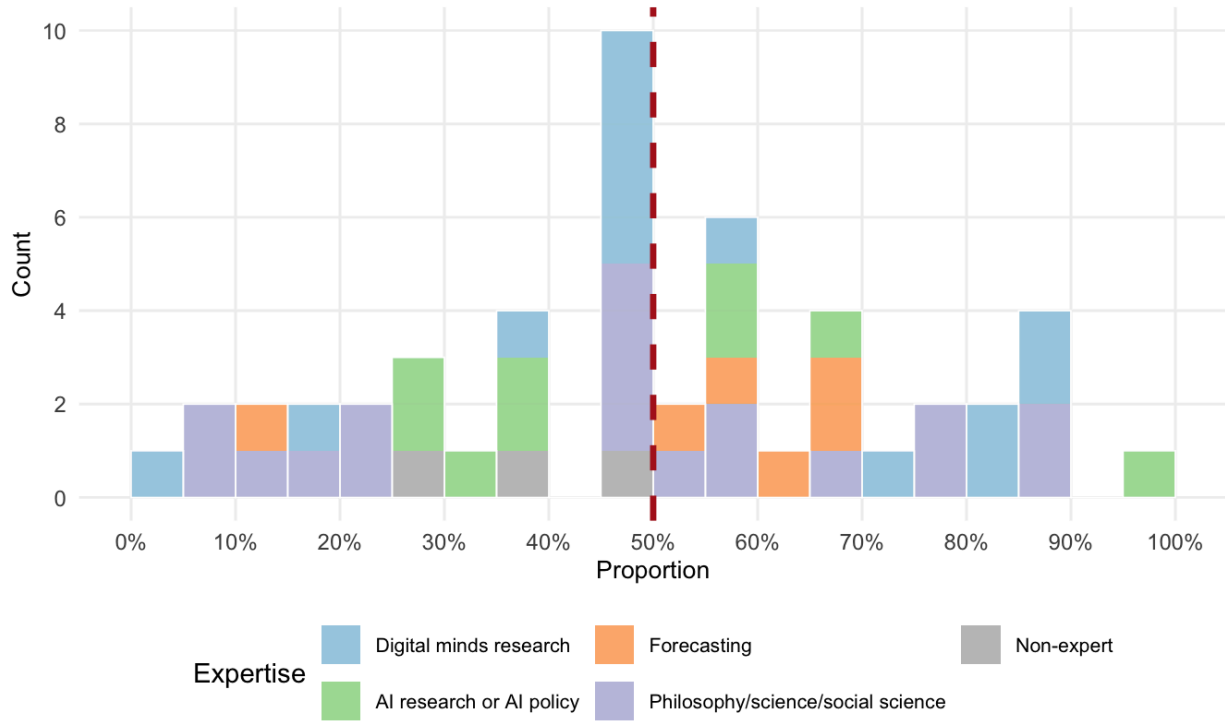

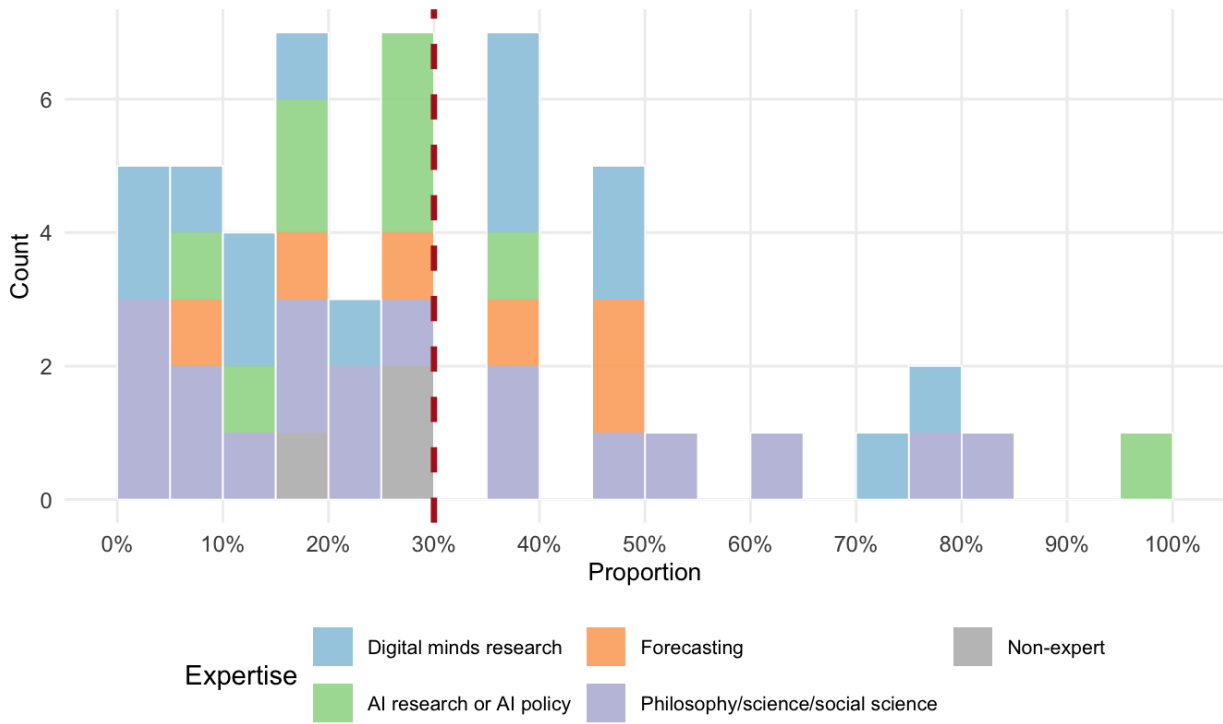

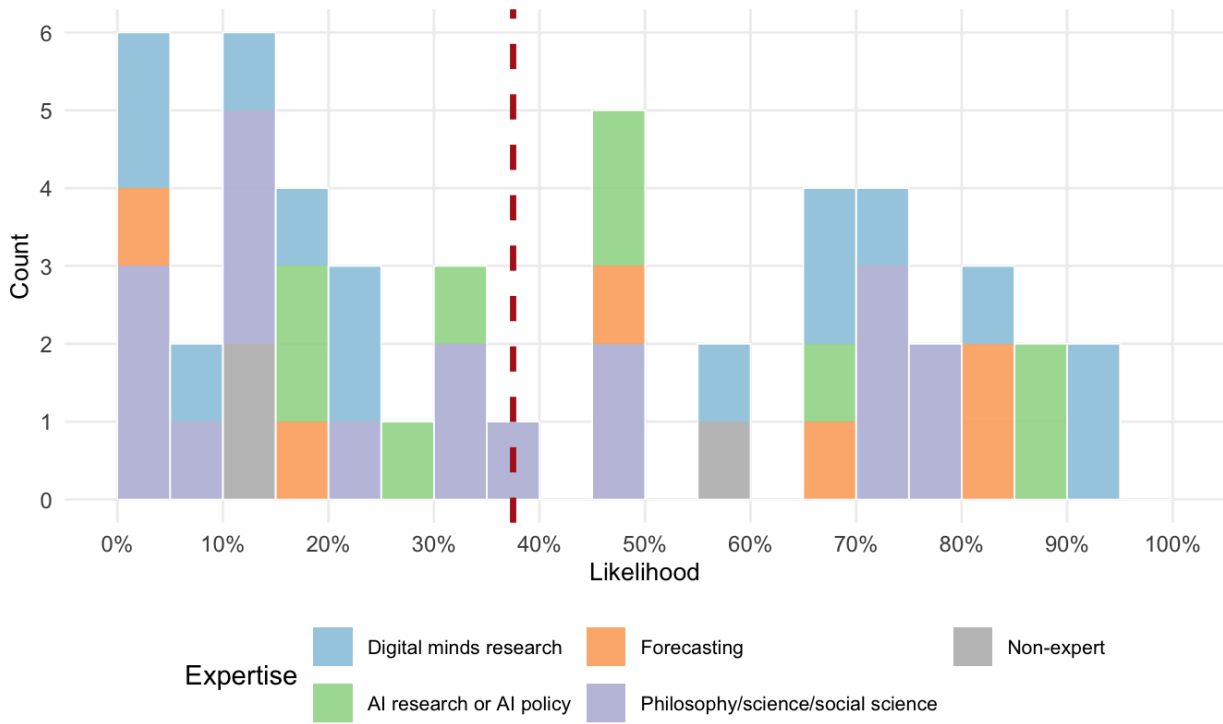
112

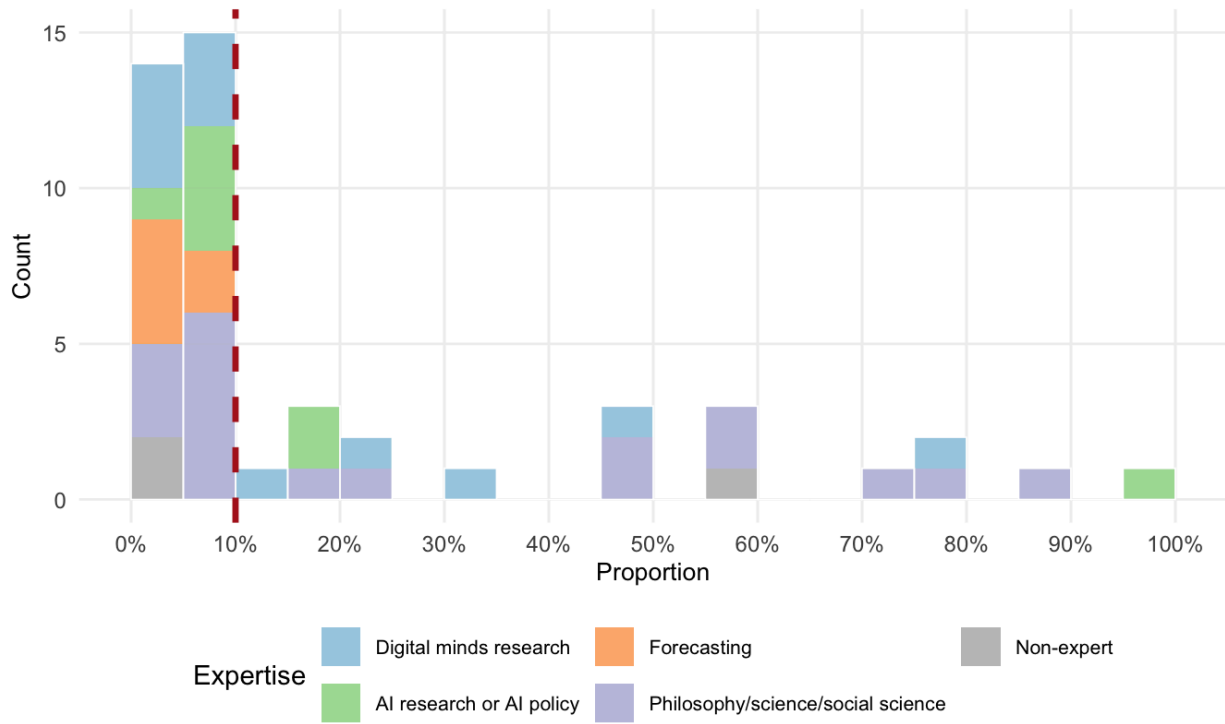

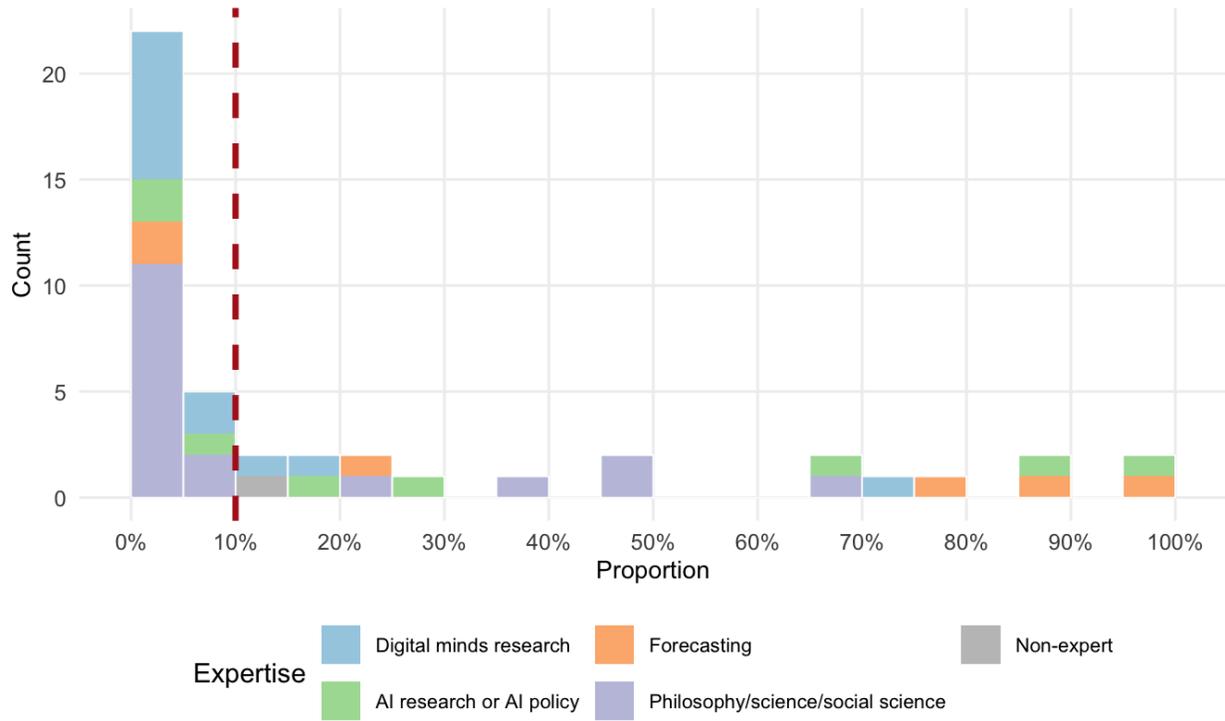

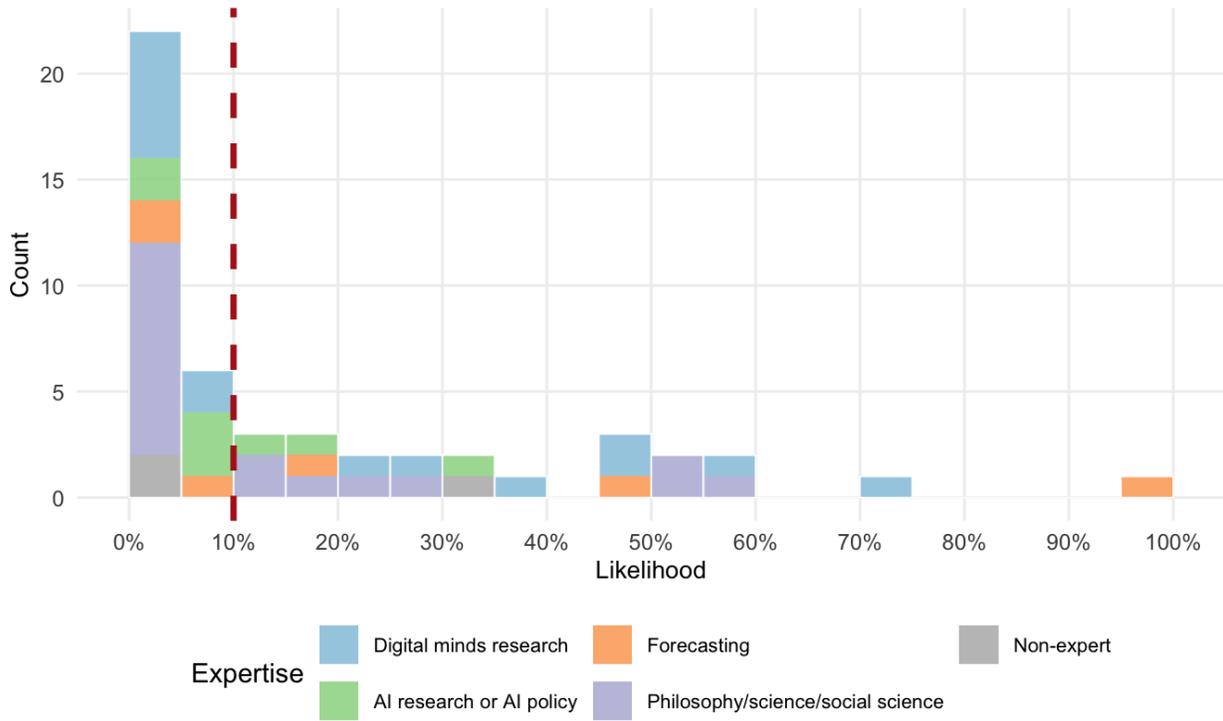

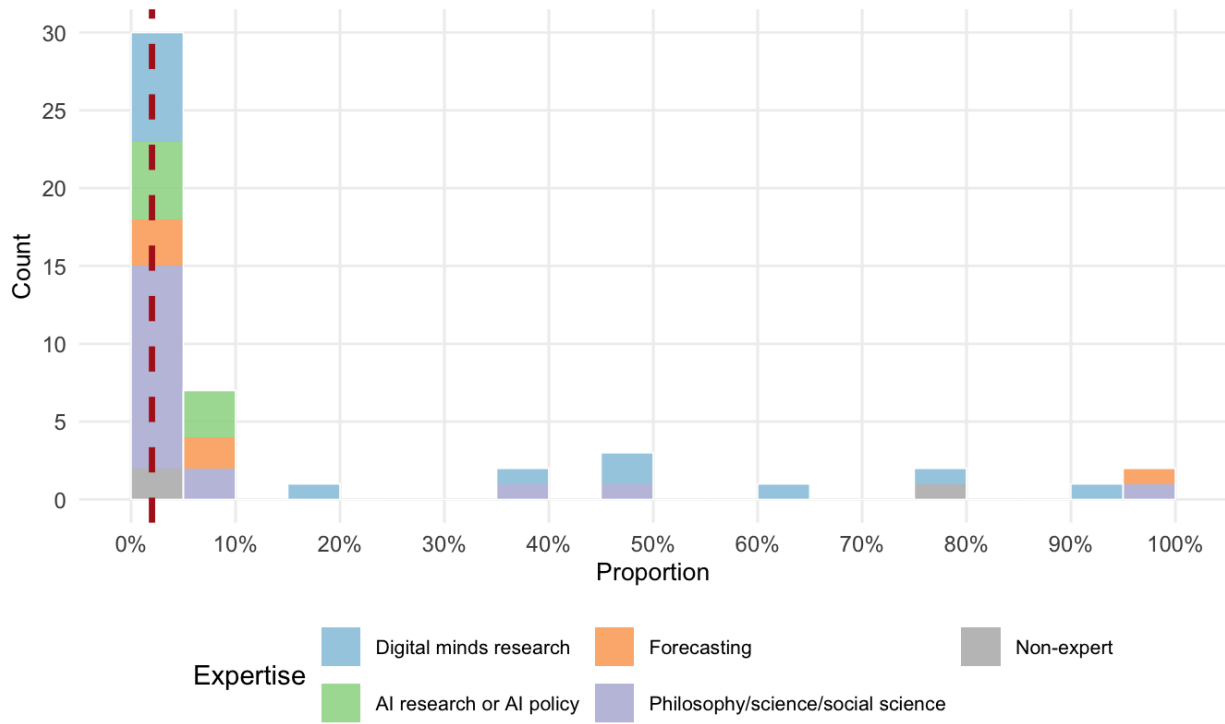

## Peer forecasting accuracy and forecasts relationship

|  | Spearman ρ | p-value |
| --- | --- | --- |
| **Starting Point** | | |
| Possibility beliefs | 0.019 | 0.8939 |
| Ever created | 0.052 | 0.7221 |
| Before AGI | 0.220 | 0.1244 |
| **Timeline Predictions** | | |
| By 2025 | 0.179 | 0.2134 |
| By 2030 | 0.112 | 0.4402 |
| By 2040 | 0.101 | 0.4868 |
| By 2050 | 0.160 | 0.2660 |
| By 2100 | 0.127 | 0.3779 |
| **Prescriptive Assessments** | | |
| Moratorium views | -0.136 | 0.3459 |
| AI safety synergy | -0.090 | 0.5339 |
| **Types of Digital Minds** | | |
| In principle: Machine learning | 0.004 | 0.9797 |
| In principle: Brain simulation | 0.060 | 0.6775 |
| In principle: Other types | -0.070 | 0.6520 |
| First created: Never created | -0.076 | 0.6096 |
| First created: Machine learning | -0.070 | 0.6299 |
| First created: Brain simulation | 0.132 | 0.3596 |
| First created: Other types | 0.250 | 0.0867 |
| **Speed Thresholds** | | |
| 1,000 humans threshold | -0.178 | 0.2165 |
| 1M humans threshold | **-0.294** | **0.0384** |
| 1B humans threshold | **-0.341** | **0.0155** |
| 1T humans threshold | **-0.365** | **0.0091** |
| **Distribution** | | |
| Social function proportion | -0.262 | 0.0716 |
| Country: USA | -0.005 | 0.9724 |
| Country: Europe | 0.067 | 0.6528 |
| Country: China | -0.096 | 0.5120 |
| Country: Other | 0.056 | 0.7061 |
| Actor: Companies | -0.023 | 0.8754 |
| Actor: Governments | -0.125 | 0.3940 |
| Actor: Universities | -0.045 | 0.7598 |
| Actor: Open-source | 0.073 | 0.6203 |
| Actor: Other | 0.000 | 0.9986 |
| Intentional creation | -0.170 | 0.2654 |
| **Claims** | | |
| False experience claims | -0.210 | 0.1604 |
| False denial of experience | 0.027 | 0.8569 |
| Claim subjective experiences | 0.119 | 0.4186 |
| Claim legal protection | 0.235 | 0.1078 |
| Claim civil rights | **0.290** | **0.0459** |
| **Recognition** | | |
| Citizen welfare estimation | 0.057 | 0.6972 |
| Basic harm protection | 0.080 | 0.5865 |
| Advanced civil rights | 0.011 | 0.9413 |
| Hot button issue | 0.089 | 0.5418 |
| **Wellbeing** | | |
| Collective welfare | 0.007 | 0.9588 |
| Pre-deployment welfare | 0.204 | 0.1733 |
| Super-beneficiary welfare | 0.231 | 0.1321 |
| **Other Sources** | | |
| Welfare without experience | 0.017 | 0.9100 |
| Non-digital mind welfare 2040 | -0.002 | 0.9878 |
| **Self-Reported Expertise** | | |
| Digital minds research | -0.124 | 0.3920 |
| Technical AI research | -0.152 | 0.2925 |
| Technical AI safety | -0.080 | 0.5817 |
| AI policy/governance | -0.035 | 0.8083 |
| Forecasting | 0.206 | 0.1564 |
| Philosophy | 0.275 | 0.0559 |
| Social science | 0.130 | 0.3697 |
| Consciousness research | -0.044 | 0.7633 |
| **Final Questions** | | |
| Community connection | -0.120 | 0.4054 |

*Table 1.* Spearman rank correlations between peer forecasting accuracy scores and survey responses for all questions. Accuracy scores were calculated based on participants' ability to predict their expertise group's median responses to two peer forecasting questions. Significant correlations ($p < 0.05$) are indicated in bold.